\documentclass[structabstract]{aa}  
\usepackage{booktabs}
\usepackage{calrsfs}
\usepackage{enumerate}
\usepackage{footmisc}
\usepackage{graphicx}
\graphicspath{{graphics/}{../graphics/}}   
\usepackage{longtable, pdflscape}
\usepackage{natbib}
\bibpunct{(}{)}{;}{a}{}{,} 
\usepackage{txfonts, color}
\usepackage{soul}
\usepackage[switch]{lineno}
\usepackage{xfrac} 

\usepackage{multirow}

\usepackage[draft=False,colorlinks=true,linkcolor=black,citecolor=blue,urlcolor=blue,linktoc=all]{hyperref}

\usepackage{etoolbox}
\makeatletter
\patchcmd\@combinedblfloats{\box\@outputbox}{\unvbox\@outputbox}{}{%
	\errmessage{\noexpand\@combinedblfloats could not be patched}%
}%
\makeatother

\makeatletter
\def\LongtableFooter{%
  \multicolumn{\LT@cols}{r}{\framebox[1.1\width]{\textbf{continued on the following page}}}\\}
\makeatother

\newcommand{\cdbox}[1]{%
  {\color{blue}%
    \dbox{\color{black}#1}}%
}
\usepackage{dashbox,framed,color,ocg-p}
\newcommand{\ToggleLayer}[2]{%
  \leavevmode
  \pdfstartlink user {
    /Subtype /Link
    /Border [0 0 0]%
    /A <<
      /S/JavaScript
      /JS (
         var aOCGs = this.getOCGs(), Layer;
         var Layers = "#1".split(","), Active = -1, i, l;
         for (l=0; l<Layers.length; l++) {
           Layer = Layers[l];
           for (i=0; aOCGs && i<aOCGs.length; i++) {
             if (aOCGs[i].state && aOCGs[i].name == Layer) {
               Active = l;
               aOCGs[i].state = false;
             }
           }
           if (Active >= 0) break;
         }
         if (Active == -1) {
           for (l=0; l<Layers.length; l++) {
             if (Layers[l] == "") Active = l;
           }
         }
         Active = Active + 1;
         if (Active == Layers.length) Active = 0;
         Layer = Layers[Active];
         for (i=0; aOCGs && i<aOCGs.length; i++) {
           if (aOCGs[i].name == Layer) aOCGs[i].state = true;
         }
      )
    >>
  }#2%
  \pdfendlink
}

\DeclareMathAlphabet{\pazocal}{OMS}{zplm}{m}{n}

\newcommand{\co}[2]{^{#1}\mathrm{C}^{#2}\mathrm{O}}
\newcommand{\transition}[2]{\,({#1}\text{--}{#2})}

\newcommand{\hi}{H\,\textsc{i}}
\newcommand{\hii}{H\,\textsc{ii}}
\newcommand{\Ncomp}{N_{\mathrm{comp}}}
\newcommand{\Nsrc}{N_{\mathrm{src}}}
\newcommand{\sect}{Sect.$\,$}
\newcommand{\fig}{Fig.$\,$}
\newcommand{\Tb}{\mathrm{T}_{\mathrm{MB}}}
\newcommand{\wco}{W_{\mathrm{CO}}}

\newcommand{\veldisp}{\sigma_{v}}
\newcommand{\vlsr}{v_{\mathrm{LSR}}}
\newcommand{\weight}[2]{w_{\mathrm{#1}}^{\mathrm{#2}}}

\newcommand{\prob}[2]{\mathrm{P}_{\mathrm{#1}}^{\mathrm{#2}}}

\newcommand{\veldispexp}{\sigma_{v}^{\text{exp}}}
\newcommand{\zgal}{\mathrm{z}_{\mathrm{gal}}}
\newcommand{\zoff}{\mathrm{z}_{\mathrm{offset}}}
\newcommand{\rgal}{\mathrm{R}_{\mathrm{gal}}}
\newcommand{\dsun}{\mathrm{d}_{\odot}}

\newcommand{\kms}{km$\,$s$^{-1}$}

\newcommand{\msun}{M$_{\odot}$}

\newcommand\gausspyplus{\textsc{GaussPy+}}

\begin{document}
    \title{Autonomous Gaussian decomposition of the Galactic Ring Survey} 
    \subtitle{II. The Galactic distribution of \textsuperscript{13}CO}
   
   \author{M. Riener
          \inst{1, }{\thanks{Member of the International Max-Planck Research School for Astronomy and Cosmic Physics at the University of Heidelberg (IMPRS-HD), Germany}}
          \and
    J. Kainulainen
          \inst{2}
          \and
    J. D. Henshaw
         \inst{1}
         \and
    H. Beuther
         \inst{1}
          }

   \institute{Max Planck Institute for Astronomy, K\"onigstuhl 17, 69117 Heidelberg, Germany
         \and
             Onsala Space Observatory, SE-439 92 Onsala, Sweden
             }

   \date{Received 23 May, 2020; accepted 04 June, 2020}

 
  \abstract
  {Knowledge about the distribution of CO emission in the Milky Way is essential to understand the impact of Galactic environment on the formation and evolution of structures in the interstellar medium.
However, currently our insight about the fraction of CO in spiral arm and interarm regions is still limited by large uncertainties in assumed rotation curve models or distance determination techniques.
In this work we use the Bayesian approach from \citet{Reid2016, Reid2019} that is based on our presently most precise knowledge about the structure and kinematics of the Milky Way to obtain the current best assessment of the Galactic distribution of $\co{13}{}$ from the Galactic Ring Survey.
We performed two different distance estimates that either included (Run A) or excluded (Run B) a model for Galactic features, such as spiral arms or spurs. 
We also include a prior for the solution of the kinematic distance ambiguity that was determined from a compilation of literature distances and an assumed size-linewidth relationship.
Even though the two distance runs show strong differences due to the prior for Galactic features for Run A and larger uncertainties due to kinematic distances in Run B, the majority of their distance results are consistent with each other within the uncertainties.
We find that the fraction of $\co{13}{}$ emission associated with spiral arm features varies from $76\%$ to $84\%$ between the two distance runs.
The vertical distribution of the gas is concentrated around the Galactic midplane showing full-width at half-maximum values of $\sim 75$~pc.
We do not find any significant difference between gas emission properties associated with spiral arm and interarm features.
In particular the distribution of velocity dispersion values of gas emission in spurs and spiral arms is very similar.
We detect a trend of higher velocity dispersion values with increasing heliocentric distance, which we however attribute to beam averaging effects caused by differences in spatial resolution.
We argue that the true distribution of the gas emission is likely more similar to a combination of the two discussed distance results, and highlight the importance of using complementary distance estimations to safeguard against the pitfalls of any single approach.
We conclude that the methodology presented in this work is a promising way to determine distances to gas emission features in Galactic plane surveys.}


   \maketitle
%


\section{Introduction}

A long-standing problem in astrophysics is how molecular gas, in particular the isotopologues of carbon monoxide (CO), is distributed in the Milky Way \citep[for reviews see e.g.][]{Combes1991review, Heyer2015review}.
Knowledge about the location of the molecular gas in our Galaxy is essential to answer important open questions in interstellar medium (ISM) research, such as the impact and importance of different Galactic environments (e.g. spiral arm and interarm regions) on star formation and the origin and evolution of ISM structures.

Addressing these scientific questions in an unbiased and systematic way requires the detailed analysis of CO emission line surveys of the Galactic plane, which usually consist of hundreds of thousands to millions of spectra \citep[e.g.][]{Dame2001, Jackson2006, Umemoto2017, Su2019}.
Many studies have focussed on extracting structures from these surveys, which have been compiled into catalogues of physical objects such as molecular clouds and clumps \citep[e.g.][]{Solomon1987, Rathborne2009grs, Rice2016, Miville-Deschenes2017, Rigby2019}.
Alternative approaches \citep[e.g.][]{Sawada2012, RomanDuval2016, Riener2020} have focussed on an analysis of these data sets without a segmentation into pre-defined physical objects, which bypasses the step of classifying the fundamentally continuous nature of the ISM into discrete objects.
However, both approaches require the determination of distances to the gas emission to permit a homogeneous analysis and comparison across different Galactic environments that accounts for differences in spatial resolution introduced by our vantage point inside the Galactic disk.

Molecular gas observations entail additional information about the radial velocity of the gas emission along the line of sight--the velocity difference to the local standard of rest or $\vlsr$--which many Galactic plane studies have used in conjunction with an assumed model for the rotation curve of our Galaxy to estimate distances via the kinematic distance (KD) method \citep[e.g.][]{Dame1986, RomanDuval2009grs, RomanDuval2016, Elia2017higal, Miville-Deschenes2017}.
However, the KD method is based on a model for the Galactic rotation curve and thus assumes the gas to be in rotational equilibrium, whereas the Milky Way is characterised by streaming motions \citep[e.g.][]{Combes1991review, Reid2009, Lopez2018gaia, Reid2019}.
Especially around spiral arms we expect strong deviations from purely circular rotation that can reach values of up to $10$~\kms\ and can lead to large kinematic distance uncertainties of up to $2-3$~kpc \citep[e.g][]{Burton1971, Liszt1981, Stark1989, Gomez2006, Reid2009, Ramon-Fox2018}.
Moreover, the non-axisymmetric potential introduced by the Galactic bar causes large non-circular motions in the gas within Galactocentric distances of $\sim\! 5$~ kpc \citep[e.g.][]{Reid2019}.
Towards the Galactic centre and close to the Sun the observed gas velocity has also almost no radial component which yields large distance uncertainties \citep[see e.g. the kinematic distance avoidance zones in Fig.~1 of][]{Ellsworth-Bowers2015}.

Another big problem of the KD method is that it always yields two possible distance solutions in the inner Galaxy (i.e. for emission within the solar orbit), which has been termed the kinematic distance ambiguity (KDA). 
Additional information is needed to resolve the KDA and previous studies have utilised an abundance of methods to solve for it by using, for example, \hi\ self absorption \citep[e.g.][]{Jackson2002, AndersonBania2009, RomanDuval2009grs, Wienen2012, Urquhart2018}, \hi\ absorption against ultracompact \hii\ regions \citep{Fish2003}, \hi\ emission/absorption \citep[e.g.][]{AndersonBania2009}, association with infrared dark clouds \citep[e.g.][Duarte-Cabral, in prep.]{Simon2006b}, or the use of scaling relationships \citep[e.g.][]{Rice2016, Miville-Deschenes2017}.

Notwithstanding all these issues in establishing reliable distances, many studies of molecular clouds extracted from $\co{12}{}\transition{1}{0}$ or $\co{13}{}\transition{1}{0}$ surveys tried to identify their position within the Galaxy \citep{Combes1991review, Heyer2015review, Rice2016, Miville-Deschenes2017} and found large variations in how well the clouds trace the gaseous spiral arm structure and the fraction of clouds located in interarm regions.
In terms of star formation, we expect an enhancement in spiral arms due to effects of gravitational instabilities, cloud collisions, and orbit crowding \citep[e.g.][]{Elmegreen2009}.
Even though sites of massive star formation seem to be predominantly associated with spiral arms \citep[e.g.][]{Urquhart2018}, recent studies have found no significant impact of Galactic structure on the clump or star formation efficiency of dense clumps \citep[e.g.][]{Moore2012, Eden2013, Eden2015, Ragan2016, Ragan2018}, or the physical properties of filaments \citep{Schisano2019higal} and molecular clumps \citep{Rigby2019}.
However, the last study reported differences in the linewidths between clumps located in interarm and spiral arm structures.
A recent study by \citet{Wang2020} also found clear differences in the ratio of atomic to molecular gas in arm and interarm regions.

We note that many of these studies used different Galactic rotation curve models and rotation parameters \citep[e.g.][]{Clemens1985, Brand1993, Reid2014} in their distance estimation; also different spiral arm models \citep[e.g.][]{Taylor1993, Vallee1995, Reid2014} have been used as a comparison.
The exact number and precise locations of the spiral arms in our Milky Way is debated, even though recent years have seen huge progress in our understanding of Galactic structure \citep[see e.g. the recent review from][]{Xu2018review}.
In particular, advances have been made by precise parallax measurements of masers associated with high-mass star-forming regions \citep[e.g.][]{Reid2009, Reid2014, Reid2019, Vera2020}.
New distance estimation approaches have emerged that use a Bayesian approach to combine these parallax measurements with additional information from CO and \hi\ surveys \citep{Reid2016, Reid2019}, which has already been used in the distance estimation to molecular clouds and clumps \citep[e.g.][Duarte-Cabral, in prep.]{Rice2016, Urquhart2018, Rigby2019}.

Our main motivation with this work is to use the currently most precise model for the structure and rotation curve of the Milky Way from \citet{Reid2019} in conjunction with the Bayesian approach presented in \citet{Reid2016, Reid2019} to analyse the distribution of molecular gas within the Galactic disk.
With the distance results we further can discuss variations of the gas emission properties with Galactic environment or Galactocentric distance.
By using additional priors based on literature resolutions of the KDA for molecular clouds and clumps and considerations based on a size-linewidth relationship, we derive distance estimates to all Gaussian components we fitted to the data set of a large $\co{13}{}\transition{1}{0}$ Galactic plane survey in the first quadrant \citep{Riener2020}.
In this work we present the results of two distance runs, one including and one excluding a prior for Galactic features.
This approach allows us to determine lower and upper limits for the fraction of emission within spiral arm and interarm locations, and enables us to discuss the robustness of our results in terms of how much the gas emission varies with Galactocentric distance and Galactic features.


\section{Data \& Methods}
\label{cha:data}

\subsection{Gaussian decomposition of the GRS}
\label{sec:grs}

In this work we use the Gaussian decomposition results of the entire Boston University–Five College Radio Astronomy Observatory Galactic Ring Survey (GRS; \citealt{Jackson2006}) data set as presented in \citet{Riener2020}\footnote{\url{http://cdsarc.u-strasbg.fr/viz-bin/cat/J/A+A/633/A14}\label{foot:grsdeco}}. 
The GRS is a $\co{13}{}\transition{1}{0}$ emission line survey that spans ranges\footnote{For $\ell < 18^\circ$ the latitude coverage is incomplete and for $\ell > 40^\circ$ the velocity range is limited to $-5$ to $85$~\kms} %
in Galactic longitude, Galactic latitude, and velocity of $14^\circ < \ell < 55.7^\circ$, $\left|b\right| < 1.1^\circ$, and $-5 < \vlsr < 135$~\kms. 
The GRS consists of about 2.28 million spectra; the data has an angular resolution of $46^{\prime\prime}$, a pixel sampling of $22^{\prime\prime}$, and a spectral resolution of $0.21$~\kms.
\citet{Riener2020} used the fully automated Gaussian decomposition package \gausspyplus\ \citep{Riener2019gausspyplus}\footnote{\url{https://ascl.net/1907.020}} to fit all GRS spectra in their native spatial and spectral resolution, which resulted in about $4.65$ million velocity fit components. 
They estimate that the decomposition was able to recover about $87.5\%$ of the flux from the GRS data set, with the remaining fraction of flux being due to diffuse emission or spectra with elevated noise levels that made the extraction of signal very challenging.
\citet{Riener2020} made the entire decomposition results available, and also provide quality metrics (such as the number of strongly blended components) for the fit results\footref{foot:grsdeco}. 

\subsection{Bayesian distance calculator}
\label{sec:bdc}

For the distance estimation we used the Bayesian distance calculator (BDC) tool \citep{Reid2016, Reid2019} that was designed for the distance calculation of spiral arm sources. 
For a given ($\ell$, $b$, $\vlsr$) coordinate, the BDC calculates a distance probability density function (PDF) based on multiple priors that can be selected by the user. 
In the current version of the BDC (v$2.4$, \citealt{Reid2019}) this includes the following priors: 
\begin{description}
	\item KD: the kinematic distance;
	\item GL: the Galactic latitude value or displacement from the Galactic midplane;
	\item PS: the proximity to parallax sources; these are high mass star-forming regions, whose trigonometric parallaxes have been determined as part of the Bar and Spiral Structure Legacy (BeSSeL) Survey\footnote{\url{http://bessel.vlbi-astrometry.org}} and the Japanese VLBI Exploration of Radio Astrometry (VERA)\footnote{\url{http://veraserver.mtk.nao.ac.jp}};
	\item SA: the proximity to features from an assumed spiral arm model; these features (such as spiral arms and spurs) have been inferred from combining information from the parallax sources with archival CO and \hi\ Galactic plane surveys;
	\item PM: the proper motion of the source.
\end{description}

\noindent The BDC allows users to set weights for these priors ($\prob{KD}{}$, $\prob{GL}{}$, $\prob{PS}{}$, $\prob{SA}{}$, $\prob{PM}{}$) that can range from $0$ to $1$.
If the weight of a prior is set to $0$ it is neglected in the distance estimation.
In the default settings of the BDC, all prior weights are set to $0.85$, with the exception of $\prob{PS}{}$, which receives a lower weight of $0.15$.
In addition, users can also supply a prior for the resolution of the KDA, that means they can provide information on whether the source location is expected to be on the near or far side of the Galactic disk.
The weight $\prob{far}{}$ for this prior is by default set to $0.5$, so that the near and far solutions of the KD prior receive equal weight. 
In this work we introduce two additional priors based on literature solutions of the KDA (\sect\ref{sec:prior_kda}) and a size-linewidth relationship (\sect\ref{sec:prior_linewidth}) that inform the $\prob{far}{}$ value for individual sources.


\section{Distance estimation}

Here we describe our method for the distance estimation.
Since the BDC was designed as a distance estimator for spiral arm sources, its default settings have an inherent bias of associating $(\ell, b, \vlsr)$ coordinates with the assumed spiral arm model (see Fig.~6 in \citealt{Reid2016}).
To better characterise the impact of this bias, we decided to perform and compare distance calculations with and without the SA prior, which we refer to further on as Run~A and Run~B, respectively.
The two distance results represent two very useful extremes in the parameter space of the distance estimation.
In one case we intentionally bias the emission towards our currently best knowledge of spiral arm features or overdensities of \hi\ and CO, which we would expect to also coincide with overdensities in $\co{13}{}$.
In the other case we obtain a picture that is unbiased by an assumed spiral arm model, but is much more dominated by the chosen Galactic rotation curve and suffers more from kinematic distance uncertainties and errors introduced by streaming motions.
In the following, we present our settings for these two BDC runs, detail how we incorporated additional prior information based on literature KDA information and the fitted linewidths, and discuss how we choose the final distance results.

\subsection{Modification of the BDC and setting of prior weights}
\label{sec:bdc_settings}

\begin{table}
    \caption[Galactic rotation curve parameters used in the BDC runs]{Galactic rotation curve parameters used in the BDC runs.}
    \centering
    \small
    \renewcommand{\arraystretch}{1.3}
    \begin{tabular}{lr|lr}
    \hline\hline
    Parameter & Value & Parameter & Value\\
    \hline
    $R_0$ [kpc] & $8.15 \pm 0.15$ & $U_\sun$ [\kms] & $10.6$ \\
    $\Theta_0$ (a$_{1}$) [\kms] & $236 \pm 7$ & $V_\sun$ [\kms] & $10.7$ \\
    a$_{2}$ & $0.96$ & $W_\sun$ [\kms] & $7.6$ \\
    a$_{3}$ & $1.62$ & & \\
    \hline
    \end{tabular}
    \label{tbl:rotcurve}
\end{table}

For the distance calculation we use the most recent version of the BDC tool (\sect\ref{sec:bdc}) with the default Galactic rotation curve parameters as determined by \citet{Reid2019}; Table~\ref{tbl:rotcurve} lists the most important parameters. 
R$_0$ denotes the distance to the Galactic centre and $\Theta_0$ (or a$_1$) is the estimated circular rotation speed at the position of the Sun; both values are in very good agreement with independent observations and measurements \citep{Abuter2019, Kawata2019}.
The a$_1$, a$_2$, and a$_3$ values are parameters used in the “universal” form of the rotation curve from \citet{Persic1996} that was adopted in the BDC \citep{Reid2014, Reid2016, Reid2019}.
$U_\sun$, $V_\sun$, and $W_\sun$ denote the solar peculiar motions towards the Galactic centre, in the direction of Galactic rotation, and towards the north Galactic pole, respectively.

\begin{figure}
    \centering
    \includegraphics[width=\columnwidth]{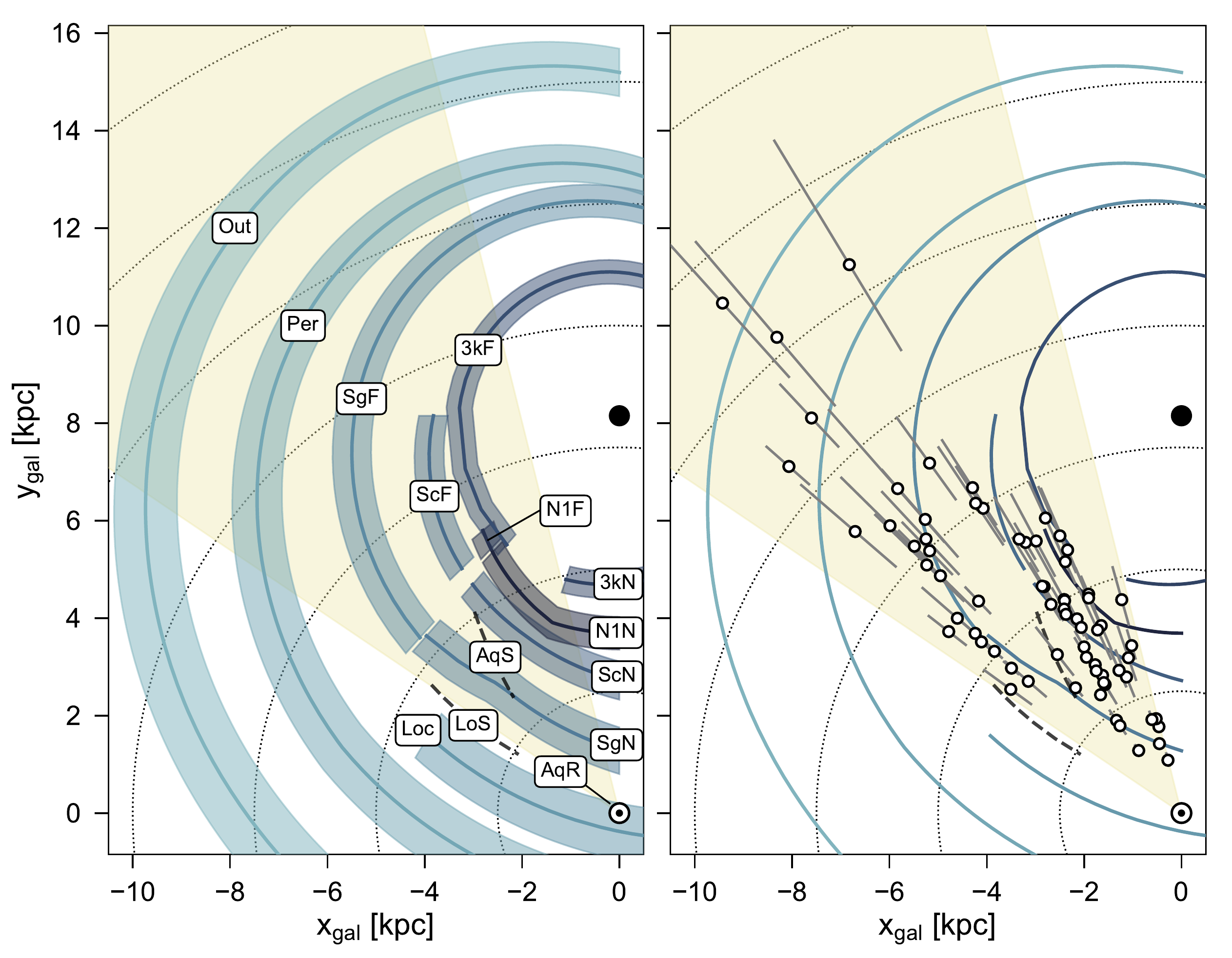}
    \caption[Overview about the spiral arm model and maser parallax sources]{
    Face-on view of the $1^{\text{st}}$ Galactic quadrant showing the GRS coverage (beige shaded area), and the positions of the Sun (Sun symbol) and Galactic centre (black dot).
    Dotted black lines indicate distances to the Sun in $2.5$~kpc intervals.
    \textit{Left panel}: Positions of Galactic features as determined by \citet{Reid2019}.
    Spiral arms are shown with green solid lines and the shaded green areas indicate $\sim\! 3.3\sigma$ widths of the arms.
    Spurs are shown with dashed black lines.
    The features are labelled as follows: $3$ kpc far arm ($3$kF), Aquila Rift (AqR), Aquila Spur (AqS), Local arm (Loc), Local spur (LoS), Norma $1^{\text{st}}$ quadrant near and far portions (N$1$N, N$1$F), Outer (Out), Perseus (Per), Scutum near and far portions (ScN, ScF), Sagittarius near and far portions (SgN, SgF).
    \textit{Right panel}: Position and uncertainties in distance of $71$ maser sources overlapping with the GRS coverage. 
    Spiral arm and spur positions are the same as in the \textit{left panel}.
    }
    \label{fig:schematic_spiral_arms+masers}
\end{figure}

The BDC results are strongly influenced by the choice of the spiral arm model and the included parallax measurements to maser sources.
It is therefore instructive to discuss and illustrate how many spiral arm features and maser sources overlap with the GRS coverage as these will be decisive factors in the distance estimation.
The left panel in \fig\ref{fig:schematic_spiral_arms+masers} shows Galactic features, such as spiral arms and spurs, that were inferred from distance measurements to maser parallax sources and archival CO and \hi\ surveys \citep{Reid2016, Reid2019} and are used as spiral arm model for the SA prior.
The width of the spiral arm features shows the approximate extent for associations of data points with these features.
In the right panel we show the position and distance uncertainties of $71$ maser sources from \citet{Reid2019} that are overlapping with the spatial and spectral coverage of the GRS.
These maser sources all have parallax uncertainties $< 20\%$, which is the BDC default requirement for the inclusion of parallax sources for the PS prior.
The PS prior for GRS sources is determined by association with one or more of these parallax sources (see \citealt{Reid2016} for how this association is performed).
Most of the measured maser sources are associated with the Scutum and Sagittarius spiral arms, thus leading to an additional emphasis of these features in the distance determination.

Since we only have access to the radial velocity component of the gas, we do not use the PM prior that would require knowledge about the proper motion of the gas.
In the following, we motivate and explain the chosen settings for our two BDC runs:

\begin{description}
	\item Run A: For this run we used all priors (KD, GL, PS, SA).
	We used the default weights for $\prob{KD}{}$ and $\prob{PS}{}$.
	In test runs of the BDC, we found that the default weight of $0.85$ for $\prob{SA}{}$ led to a strong domination of the spiral arm model (see Appendix~\ref{sec:runs_comp}) compared to the remaining priors. 
	We thus opted to reduce $\prob{SA}{}$ to $0.5$, which led to a more balanced ratio between the priors in our tests. 
	Since in the default settings of the BDC the priors for the spiral arm model and the Galactic latitude are combined, we also set $\prob{GL}{}$ to $0.5$ to keep the ratio between these priors intact.
	\item Run B: For this run we did not use the priors for the proper motion and the spiral arm model.
By default, the BDC combines the SA and GL priors, which means that setting $\prob{SA}{} = 0$ has the effect of also setting $\prob{GL}{} = 0$.
As the Galactic latitude information contains important prior information for the distances, we slightly modified the BDC source code so that we could use the $\prob{GL}{}$ prior without using the $\prob{SA}{}$ prior. 
However, we found that in this case the default settings of $\prob{GL}{}=0.85$ could yield a strong bias towards the far KD solution.
To reduce this bias, we opted to decrease $\prob{GL}{}$ to a value of $0.5$, which yielded a more balanced ratio between the priors in our tests.
\end{description}

\noindent In addition to these settings, we include priors that incorporate literature KDA resolutions and fold in information from the fitted linewidth in both BDC runs. 
These additional priors are described in more detail in the next two sections. 

\subsection{Prior for the kinematic distance ambiguity}
\label{sec:prior_kda}

\begin{figure*}
    \centering
    \includegraphics[width=\textwidth]{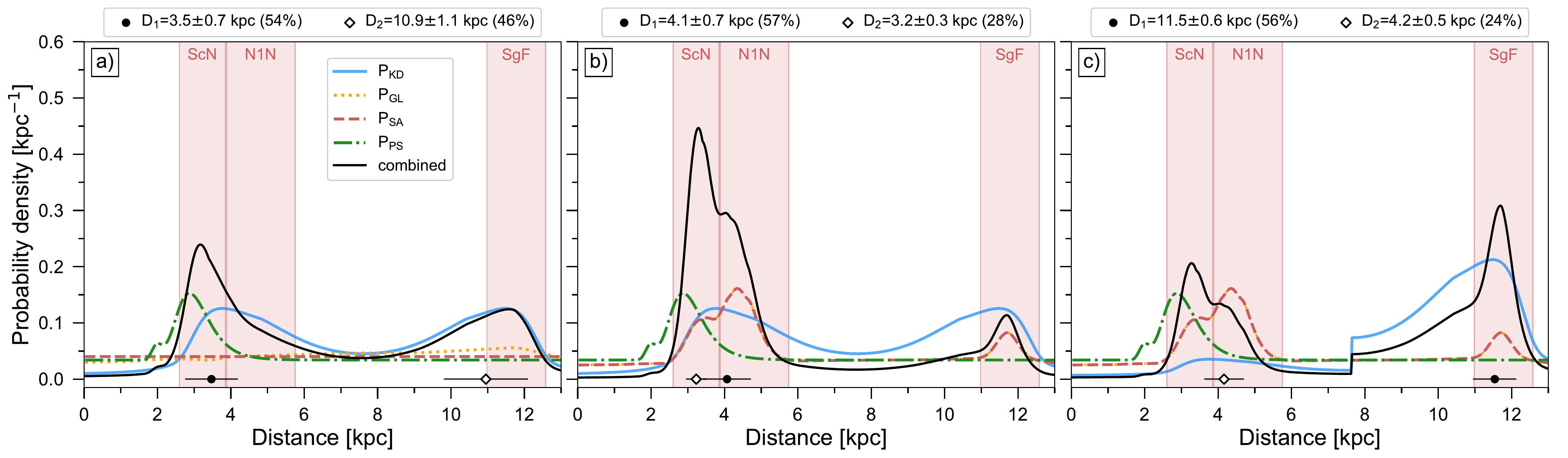}
    \caption[Effect of BDC priors on the distance estimation]{Effect of BDC priors on the distance estimation for a source located at $\ell=20.66\degr, b=-0.14\degr,\ \text{and}\ \vlsr=56.2$~\kms.
    For \textit{panel~a} we set $\prob{SA}{}=0$ and $\prob{far}{}=0.5$, for \textit{panel~b} we set $\prob{SA}{}=0.5$ and $\prob{far}{}=0.5$, and for \textit{panel~c} we set $\prob{SA}{}=0.5$ and $\prob{far}{}=0.875$.
    Coloured lines show the probabilities of the priors for kinematic distances (blue solid line), latitude probability (orange dotted line), spiral arms (red dashed line), and associations with parallax sources (green dash-dotted line).
    The black solid line shows the combined distance PDF.  
    The red-shaded areas show the distance ranges for which the source would be associated with individual Galactic features.
    Black circles and white diamonds indicate the first and second choice for the distance estimates, with the horizontal bars showing their corresponding uncertainties. 
    The boxes above the panels list the resulting distances, their uncertainties, and estimated probabilities.
    }
    \label{fig:bdc_examples}
\end{figure*}

For all sources located within the solar circle the KD prior yields two possible distance solutions (called the \emph{near} and \emph{far} distances).
However, over recent years many works have already solved the KDA for many objects such as molecular clouds and clumps that overlap with the GRS coverage.
Many of these studies even used the GRS data set directly in their distance estimation.
To take advantage of these previous works, we implemented a new scheme that uses these literature KDA solutions to inform the $\prob{far}{}$ prior of the BDC, which results in a preference for  the near or far distance solution.
In Appendix~\ref{sec:kda} we list all literature KDA solutions that we incorporated in our method and describe in detail how we use this information to determine the $\prob{far}{}$ weight for individual sources.
In total, this prior was used in the distance estimation of about $30\%$ of the $\co{13}{}$ fit components (see App.~\ref{app:bdc_prob_pfar_choice} for more details).
In Appendix~\ref{sec:kda} we also discuss the performance of this prior; we found that its inclusion leads to a significant increase in consistency of the BDC distance results with the reported literature distances. 

We illustrate the effect of the KDA prior on the distance estimation with an example in \fig\ref{fig:bdc_examples}, which shows the resulting distance PDFs for the individual priors.
In panel~(a) the spiral arm prior ($\prob{SA}{}$, in red) was switched off and no KDA prior was supplied (i.e. $\prob{far}{}=0.5$), so the only remaining contributions to the combined distance PDF come from the priors for the kinematic distances ($\prob{KD}{}$, in blue), the association with parallax sources ($\prob{PS}{}$, in green), and the Galactic latitude ($\prob{GL}{}$, in orange).
Since the source is located close to the centre of the Galaxy, the two peaks of the kinematic distance PDF are not Gaussian-shaped, but were down-weighted to reflect expected large peculiar motions near the Galactic bar \citep{Reid2019}.
Distances are estimated by fitting Gaussians to the peaks of the combined distance PDF (in black); the most likely distance value corresponds to the Gaussian component with the highest integrated probability density, so the highest peak of the combined distance PDF need not result in the most likely distance estimate.
The distance uncertainty is given by the standard deviation of the Gaussian fit component.

With no associated parallax sources or conclusive latitude information the two distance solutions would have corresponded to the two peaks of the kinematic distance PDF and would have received the same probability ($50\%$).
In our case, the prior incorporating the Galactic latitude position favours the far distance, but associated parallax sources shift the balance towards the near kinematic distance solution, yielding a most likely distance value for the source of $\sim 3.4$~kpc.
We note that even though the SA prior is switched off the BDC still gives the information of whether the distance results do overlap with locations of spiral arm and interarm features; the extent for such associations is indicated with the red-shaded areas in \fig\ref{fig:bdc_examples}. 
For the example depicted in panel~(a), D$_1$ is associated with the near portion of the Scutum spiral arm, whereas D$_2$ corresponds to an interarm position.

If the spiral arm prior is included (panel~b), the most likely distance shifts to a higher value of about $4.2$~kpc.\footnote{In this case, the spiral arm and Galactic latitude probabilities are by default combined.}
Also the distance estimate with the second highest probability corresponds to a near distance solution, which illustrates the strength of the spiral arm prior.

Finally, panel~(c) shows the effect of adding a prior for the KDA, which in our case favours the far kinematic distance solution (for this example we assume $\prob{far}{}=0.875$; see App.~\ref{sec:kda} for how exactly $\prob{far}{}$ is determined from literature KDA solutions).
Setting the KDA prior has the effect of rescaling the kinematic distance PDF, which in this example shifts the most likely distance value to a far distance solution.

This example illustrated that the KDA prior can be a decisive factor for the distance estimation.
However, while the $\prob{far}{}$ prior can give a strong preference for one of the kinematic distance solutions, we note that the combination with the other priors can still result in a different choice for the most likely distance.

\subsection{Prior for the fitted linewidth}
\label{sec:prior_linewidth}

\begin{figure*}
    \centering
    \includegraphics[width=\textwidth]{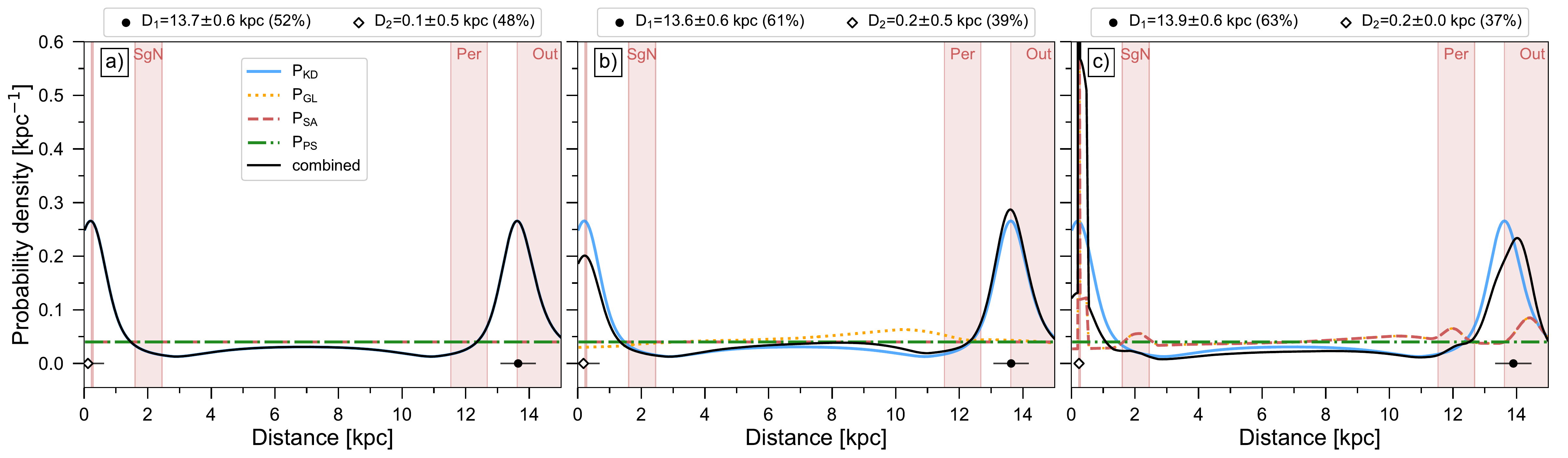}
    \caption[Illustration of BDC bias for sources with low $\vlsr$ velocities]{BDC examples for a source located at $\ell=32\degr, b=0\degr,\ \text{and}\ \vlsr=5$~\kms, illustrating how sources with low $\vlsr$ velocities are biased towards the far distance solution.
    The panels show BDC results using the KD and PS priors (\textit{a}), in addition to the GL (\textit{b}) as well as the SA priors (\textit{c}).
    The meaning of the lines and symbols is the same as in \fig\ref{fig:bdc_examples}.
    }
    \label{fig:effect_lowvlsr}
\end{figure*}

In our tests of the BDC, we noticed that sources with low $\vlsr$ velocities are preferentially placed at larger distances (see \fig\ref{fig:effect_lowvlsr}).
This effect is strongest for sources with $\vlsr \lesssim 5$~\kms; for sources with $\vlsr \lesssim 0$~\kms\ the KD prior permits essentially only the far distance solution. 
This effect can be mitigated by the inclusion of the SA prior as sources can receive a strong association with the nearby Aquila Rift cloud complex.
However, since the association with Aquila Rift is only performed over a very limited distance range, this leads to narrow high peaks in the distance PDF, which in turn yield associated Gaussian fit components with a lower integrated area than for the far distance solution (\fig\ref{fig:effect_lowvlsr}c).
This effect thus has a large impact on our distance results, since we expect strong confusion between local emission from the solar neighbourhood and the far side of the Galactic disk at $-5 < \vlsr \lesssim 20$~\kms\ \citep{Riener2020}.

\begin{figure}
    \centering
    \includegraphics[width=\columnwidth]{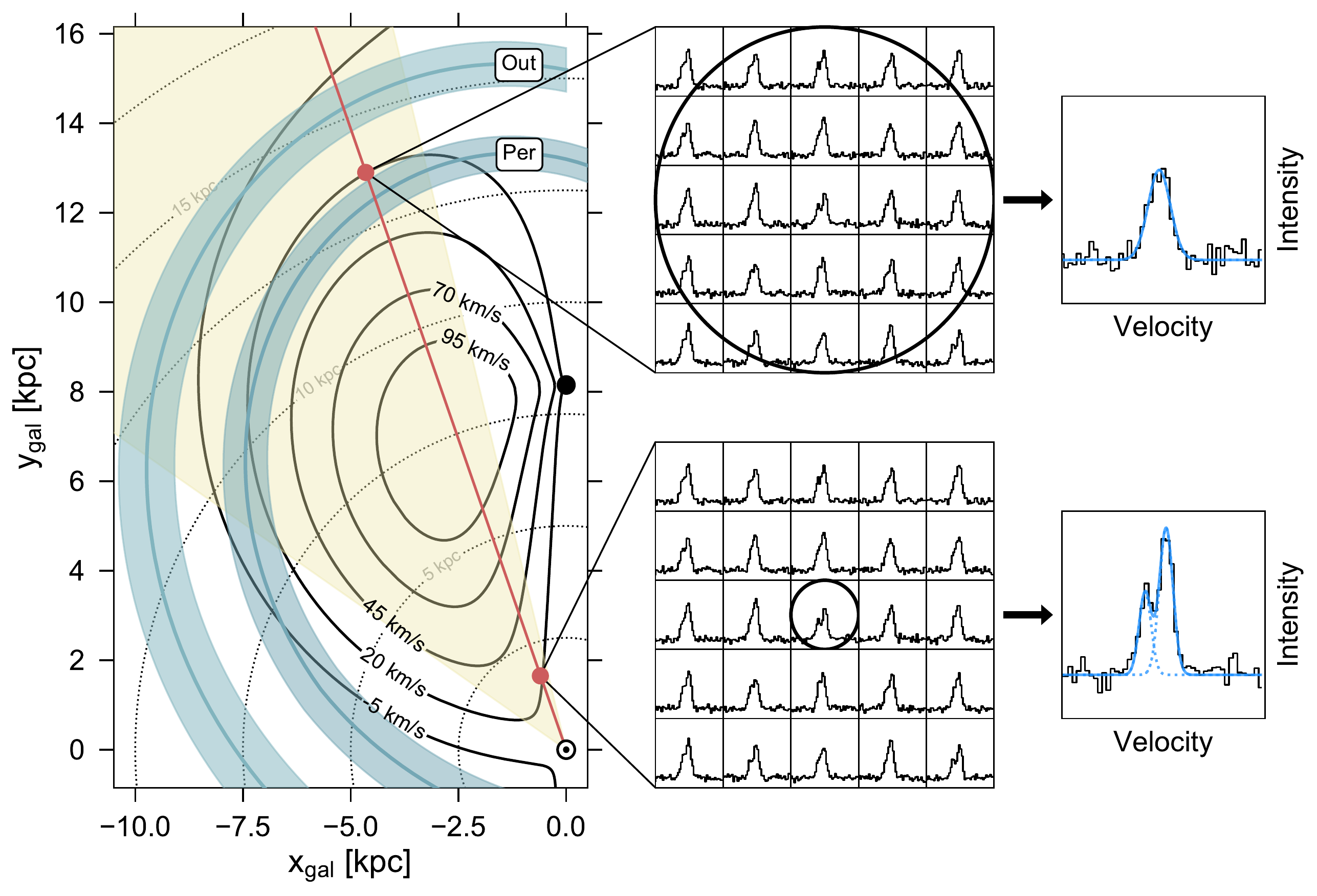}
    \caption[Illustration of linewidth broadening caused by beam averaging]{Illustration of linewidth broadening caused by beam averaging.
    \textit{Left panel}: Same as the \textit{left panel} of \fig\ref{fig:schematic_spiral_arms+masers}, but showing only the positions and estimated widths of the Perseus (Per) and Outer (Out) spiral arms.
    Black solid lines show curves of constant projected $\vlsr$ values. 
    The red line shows a random line of sight with the corresponding intersections with the $\vlsr = 20$~\kms\ curve indicated with red dots.
    The centre panels illustrate the change in spatial extent of the beam (black circle) for a region with two blended velocity components embedded at the near (bottom centre) and far (top centre) distance.
    The right panels illustrate the resulting observed spectra (black line) and Gaussian fit components (blue lines).
    }
    \label{fig:schematic_beam_averaging}
\end{figure}

However, as suggested in \citet{Riener2020}, we can try to use the velocity dispersion values of the fit components as an additional prior information for the distance calculation.
Figure~\ref{fig:schematic_beam_averaging} recaps the argument put forth in \citet{Riener2020}: due to averaging of bigger spatial areas at larger distances, we expect broadened lines due to, for example, sub-beam structure and velocity crowding, velocity gradients of the line centroids (either along the line of sight or in the plane of the sky), or fluctuations in the non-thermal contribution to the linewidth (e.g. due to regions with higher turbulence).
The example shown in \fig\ref{fig:schematic_beam_averaging} highlights the effect of sub-beam structure and velocity crowding.
If a region with two strongly blended velocity components is located at close distances, the individual emission peaks can be well resolved and fitted with two narrow Gaussian components (bottom centre and right panels in \fig\ref{fig:schematic_beam_averaging}).
However, if the same region is located at far distances, the individual velocity components might not be resolved, leading to a decomposition with a single broad Gaussian component (top centre and right panels in \fig\ref{fig:schematic_beam_averaging}).

Given these expected differences due to beam averaging effects, it is unlikely that very narrow fitted linewidths are associated with emission at large distances.
For most of the molecular gas in the GRS, the molecular gas temperatures will be about $10$ to $20$~K, which is the typical temperature of gas at intermediate density ($\sim\! 10^{3}$~cm$^{-3}$) in molecular clouds.
The thermal broadening of the spectral lines for these temperatures is about $0.2$ to $0.3$~\kms, so effectively the spectral resolution of the GRS.
The physical extent of the GRS beam is $\sim\! 0.1$~pc at the distance of the Aquila Rift complex and increases to $\gtrsim 2$~pc at distances beyond the solar radius.
Therefore the physical areas covered by the beam at the nearby distances of the Aquila Rift and the far distances of the Perseus and Outer arm are different by a factor of $> 400$.
Even in the case of no sub-beam structure and velocity crowding and no significant non-thermal contributions to the linewidth, expected variations in the line centroids across the beam-averaged area are enough to broaden the lines significantly (see Appendix~\ref{sec:beam_avg_veldisp}). 

\begin{figure}
    \centering
    \includegraphics[width=\columnwidth]{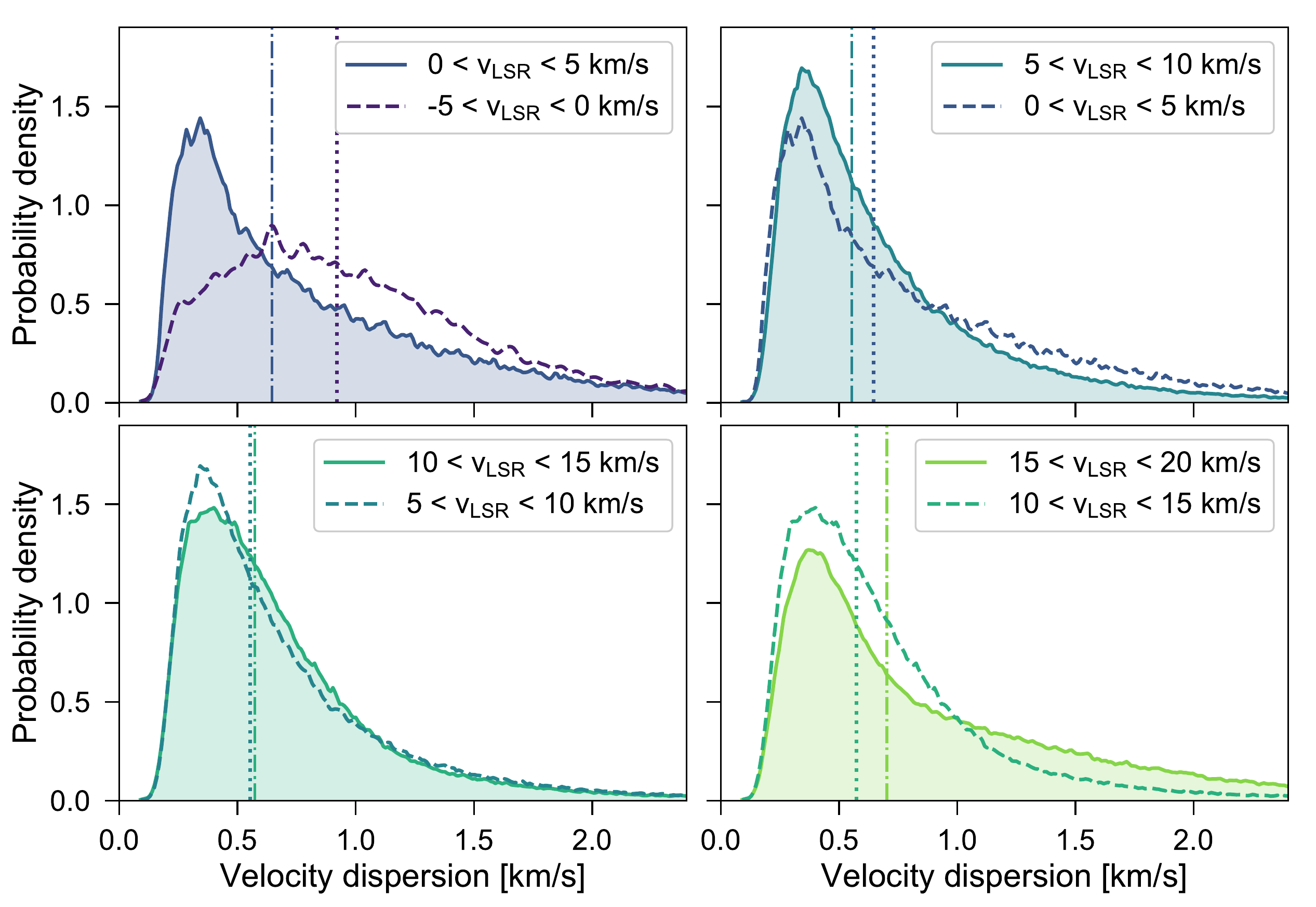}
    \caption[PDFs of fitted velocity dispersion values in different $\vlsr$ ranges]{PDFs of fitted velocity dispersion values in different $\vlsr$ ranges.
    Each panel compares PDFs of two different $\vlsr$ ranges; the median values of the PDFs with the solid and dashed lines are indicated with the dash-dotted and dotted vertical lines, respectively.
    }
    \label{fig:hist_lowvlsr_veldisp}
\end{figure}

The effect of broader linewidths for emission originating at larger heliocentric distances is already noticeable in the fitted linewidths (\fig\ref{fig:hist_lowvlsr_veldisp}).
We would expect fit components in the interval of $-5 < \vlsr < 0$~\kms\ (left upper panel of \fig\ref{fig:hist_lowvlsr_veldisp}) to predominantly originate from large distances and indeed the distribution of $\veldisp$ values is shifted towards larger values compared to similar $\vlsr$ ranges between $0$ and $20$~\kms\ (remaining panels of \fig\ref{fig:hist_lowvlsr_veldisp}).
The distribution of these other ranges has a strong peak at $\veldisp < 0.5$~\kms, consistent with the assumption that this corresponds to emission lines originating from nearby spatially resolved regions. 

\begin{figure}
    \centering
    \includegraphics[width=\columnwidth]{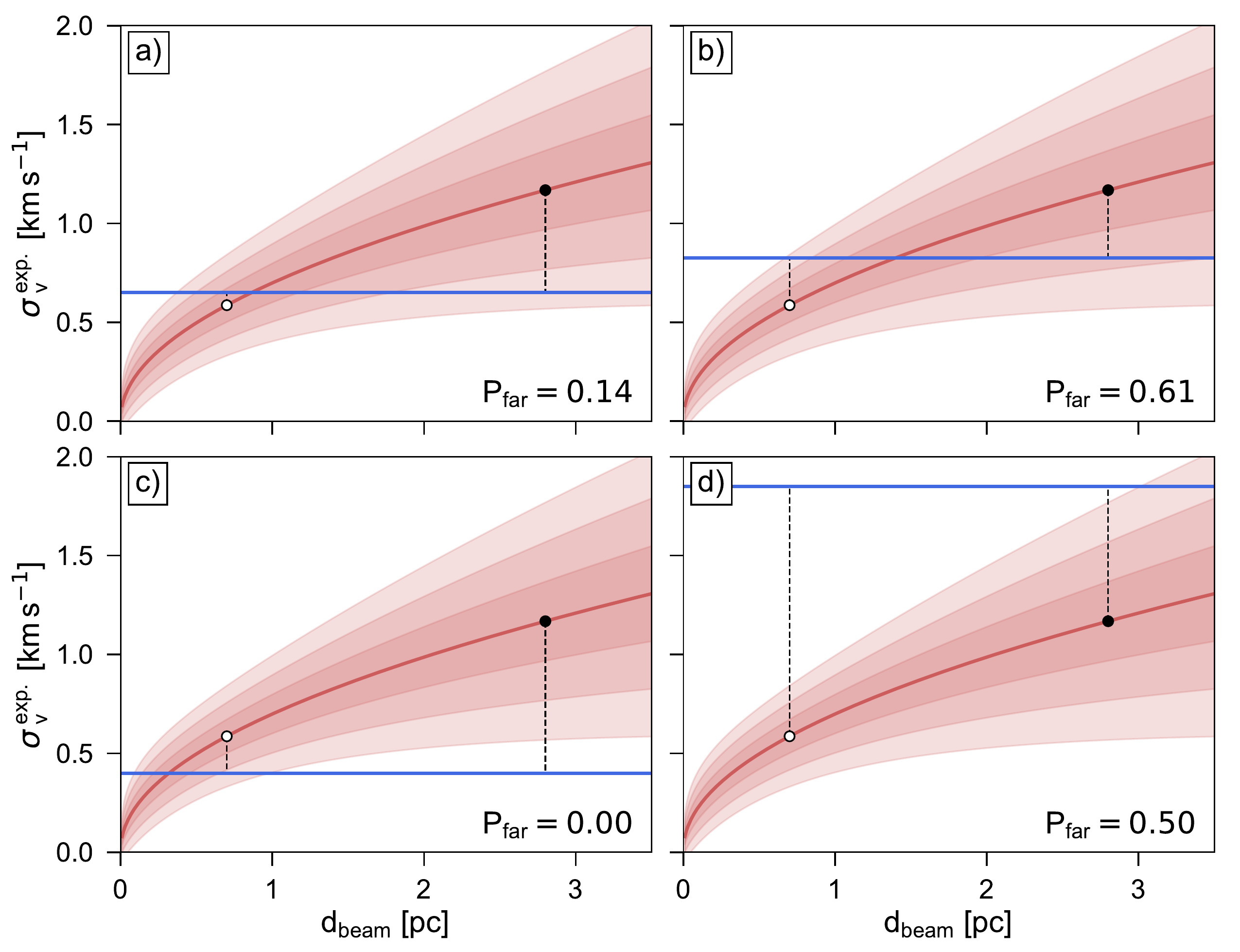}
    \caption[Illustration of the distance prior based on the fitted linewidth]{Illustration of the distance prior based on the fitted linewidth.
    The red line shows the size-linewidth relation from \citet{Solomon1987} corrected for the most recent distance estimates to the Galactic centre.
    Red-shaded areas show the $1\tilde{\sigma}$, $2\tilde{\sigma}$, and $3\tilde{\sigma}$ intervals determined from variations of the parameters of the size-linewidth relation.
    The two dots show expected velocity dispersion values for corresponding physical extents of the beam for the near (white dot) and far (black dot) kinematic distance solutions.
    Blue horizontal lines show velocity dispersion values of fit components and dashed vertical lines indicate the relevant distances to the expected values, from which the $\prob{far}{}$ prior is determined. 
    See \sect\ref{sec:prior_linewidth} for more details.
    }
    \label{fig:schematic_prior_veldisp}
\end{figure}

Having established that the fitted velocity dispersion values can contain information about the distance to the gas emission, we will in the following explain how we implement this as prior information for our distance calculation.
Similar to \citet{Riener2020}, we will use the size-linewidth relationship established by \citet{Solomon1987} for molecular clouds in the Galactic disk to inform our decision about whether a fitted $\veldisp$ value is more likely associated with a region at near or far distances.
This size-linewidth relationship has the form of:

\begin{equation}
	\sigma_{v}^{\text{exp.}} = \sigma_{v,\,0} \cdot \left(\dfrac{L}{1~\text{pc}}\right)^{\gamma},
	\label{eq:size_lw}
\end{equation}

\noindent with $\gamma = 0.5$ and $\sigma_{v,\,0}=0.7$ (corrected for the most recent distance estimates to the Galactic centre; \citealt{Abuter2019, Reid2019}).
In \fig\ref{fig:schematic_prior_veldisp} we show the expected velocity dispersion values based on this relation as a function of physical extents of the beam (d$_{\text{beam}}$) with the solid red line.
The shaded red areas indicate $1\tilde{\sigma}$, $2\tilde{\sigma}$, and $3\tilde{\sigma}$ intervals for the size-linewidth relation assuming variations in $\gamma$ and $\sigma_{v,\,0}$ of $\pm 0.1$.
The magnitude of these variations was motivated for consistency with results obtained from more local molecular clouds \citep{Larson1981, Shetty2012}.

We use this size-linewidth relationship to inform the KDA prior as follows.
We first calculate the physical extent of the beam (d$_{\text{beam}}$) for the two kinematic distance solutions that are always obtained for positive $\vlsr$ values in the inner Galaxy. 
We then use the size-linewidth relationship to calculate the expected velocity dispersions for both d$_{\text{beam}}$ values.
Subsequently, we compare the actual fitted velocity dispersion with these expected velocity dispersion values to decide whether it is more consistent with the near or far distance value.
This decision is driven by how close the fitted $\veldisp$ value is to the expected values from the near and far distances.
We calculate for both distances the difference between the fitted and expected $\veldisp$ values; if the difference is within the $3\tilde{\sigma}$ interval indicated in \fig\ref{fig:schematic_prior_veldisp} we give it the corresponding weight from a normalised Gaussian function:

\begin{equation}
   \weight{\sigma}{} = \text{exp}
    \left(-0.5 \cdot 
    	\left(\dfrac{\veldisp - \veldispexp}{\tilde{\sigma}}
    	\right)^{2}
    \right),
\end{equation}

\noindent where $\tilde{\sigma}$ is the standard deviation for $\veldispexp$.
From this we calculate values for the $\prob{far}{}$ prior as:

\begin{equation}
    \prob{far}{} = \sfrac{1}{2} + \sfrac{1}{2}\left(\weight{\sigma}{far} - \weight{\sigma}{near}\right),
\end{equation}

\noindent with $\weight{\sigma}{near}$ and $\weight{\sigma}{far}$ indicating the weights for the near and far distance value.
If the fitted $\veldisp$ value falls above the $3\tilde{\sigma}$ interval, we set the corresponding weight $\weight{\sigma}{}$ to zero.
If the fitted $\veldisp$ value falls below the $3\tilde{\sigma}$ interval for the far distance but is not above the $3\tilde{\sigma}$ interval for the near distance, we automatically assume $\prob{far}{} = 0$.

We illustrate this procedure for four different cases in \fig\ref{fig:schematic_prior_veldisp}; for all these cases the kinematic distance solution and the corresponding d$_{\text{beam}}$ values are the same but the values of the fitted $\veldisp$ values (horizontal blue lines) vary.
For the first case (panel~a) the $\veldisp$ value is more consistent with the near distance; we obtain $\weight{\sigma}{near} = 0.76$ and $\weight{\sigma}{far} = 0.04$, yielding $\prob{far}{} = 0.14$ and thus strongly favouring the near distance.
In the second case (panel~b) the far distance is favoured, as $\weight{\sigma}{near} = 0.17$ and $\weight{\sigma}{far} = 0.24$.
In the third case, the fitted $\veldisp$ value is much lower than the expected $\veldisp$ value and falls below the $3\tilde{\sigma}$ range ($\weight{\sigma}{far} = 0$); in such cases we always assume $\prob{far}{} = 0$ unless the $\veldisp$ value is above the $3\tilde{\sigma}$ interval for the near distance (in which case $\prob{far}{}$ would be $0.5$).
Finally, the last case (panel~d) yields no $\prob{far}{}$ prior as the fitted $\veldisp$ value is much higher than the expected $\veldisp$ values for both the near ($\weight{\sigma}{near} = 0$) and far ($\weight{\sigma}{far} = 0$) distance.
This ensures that we do not exclude the possibility that a source with high $\veldisp$ value can come from a nearby region with high non-thermal contributions to the linewidth.

Recent studies have found large dispersions of the size-linewidth relation across the Galactic disk \citep[e.g.][]{Heyer2009, Miville-Deschenes2017} and advocate a scaling relation that also takes the surface density into account.
Moreover, especially in the inner part of the Galaxy, linewidths can be systematically higher than predicted by the size-linewidth relation, indicating that $\veldisp$ is at least partly set by Galactic environment \citep[][]{Shetty2012, Henshaw2016, Rice2016, Henshaw2019}.
We want to emphasize here that we do not use the size-linewidth relation to make conclusive decisions about the distance to a gas emission peak, but only use it as additional prior KDA information for sources with $\vlsr < 20$~\kms. For sources with larger $\vlsr$ values the difference between the $\veldispexp$ values for the two KD solutions gets smaller and the size-linewidth prior might bias components with narrower fitted linewidths to be preferentially placed at the near distance solution. 
We also do not use the $\veldisp$ prior in case the literature solutions for the KDA (\sect\ref{sec:prior_kda}) already yielded a $\prob{far}{}$ value $\neq 0.5$.

\subsection{Choice of distance solution}
\label{sec:dist_choice}

\begin{figure}
    \centering
    \includegraphics[width=0.75\columnwidth]{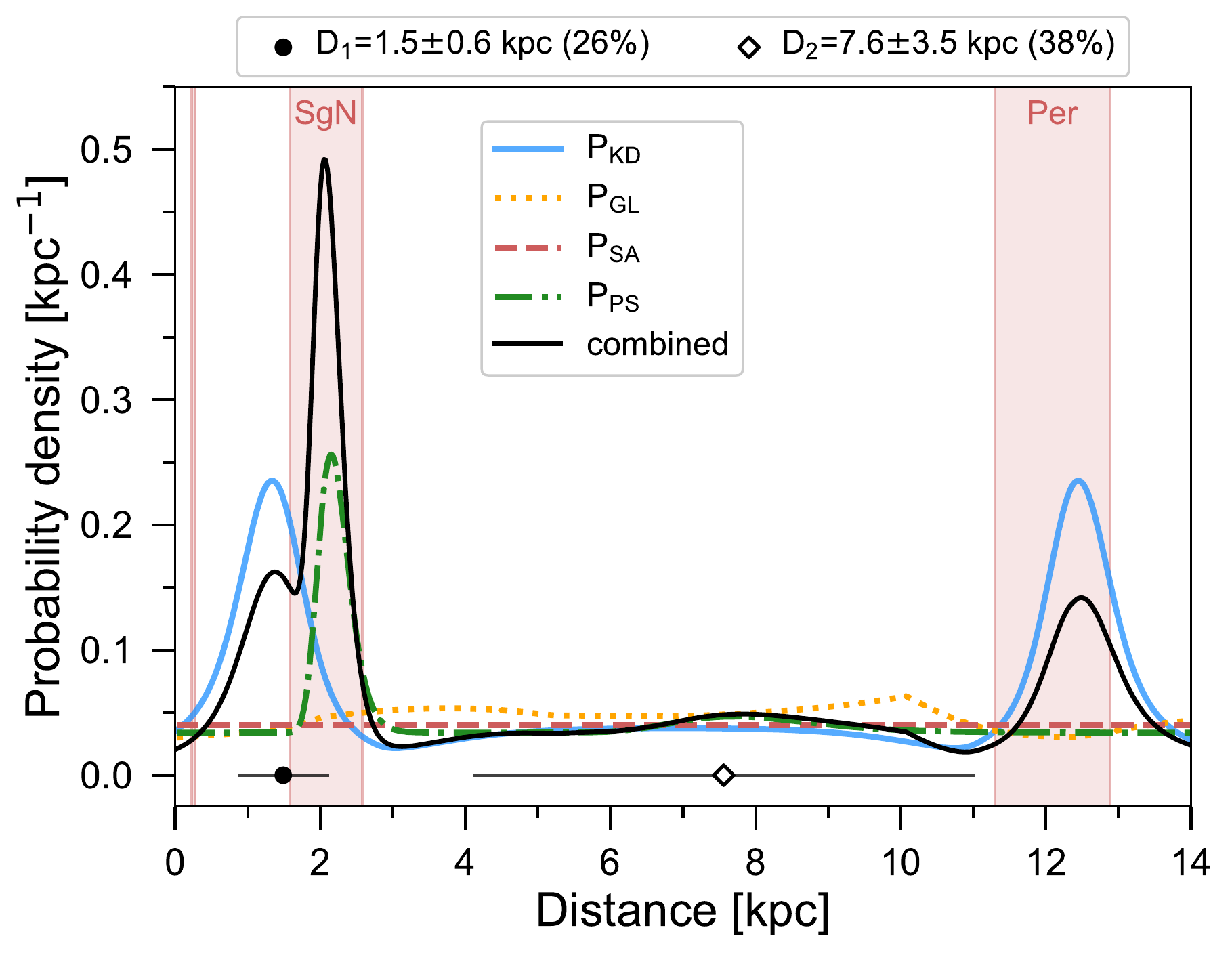}
    \caption[Example of final distance choice for BDC results]{Example of distance choice in case one of the distance components has a high integrated area but low peak amplitude value.
    The meaning of the lines and symbols is the same as in \fig\ref{fig:bdc_examples}.
    }
    \label{fig:bdc_broad_comp}
\end{figure}

The distance calculation with the BDC yields multiple alternative distance solutions with corresponding estimates of their probability.
These probabilities are obtained from Gaussian fits to the combined distance PDF \citep{Reid2016}.
By default, the Gaussian distance component with the highest integrated area is chosen as the most likely distance value.
So even if the distance PDF shows a clear peak, this need not correspond to the selected most likely distance value.
Our tests showed that this could be problematic, as very broad Gaussian components with low peak values can be selected as the most probable distance component, resulting in unlikely distance solutions (\fig\ref{fig:bdc_broad_comp}).
For our BDC runs we found that such broad components with low peak values would be chosen as the preferred distance value in $\sim 2.5\%$ (Run~A) and $\sim 9\%$ (Run~B) of the distance assignments.

To avoid the selection of such broad components with low peak values, we adapted the choice for the most likely distance as follows. 
In case of two reported distance solutions (as is the default in v$2.4$ of the BDC), we first check whether the peaks of the individual Gaussian fit components exceed a pre-defined limit. 
We set this limit to $0.12$, which corresponds to three times the value of a flat  distance PDF\footnote{The distance PDF is evaluated from $0$ to $25$~kpc.
Requiring that the integrated area of a flat distance PDF is equal to unity yields a value of $0.04$ for the PDF at all distances.}.
If one of the distance components does not satisfy this criterion, we choose the remaining distance solution, regardless of whether its integrated area was less (see \fig\ref{fig:bdc_broad_comp}).
If both of the distance components exceed or fail the amplitude limit, we choose the distance component with the highest assigned probability (i.e. the Gaussian fit component having the highest integrated area).
In case both distance components have the same assigned probability, we choose the distance solution with the lower absolute distance error.
If both components are also tied in the distance errors (as can happen if the combined distance PDF is dominated strongly by the KD prior), we choose the distance component with the lower distance value. 
The last two conditions were only used in $\sim 1\%$ of the distance choices for the two BDC runs (see App.~\ref{app:bdc_prob_pfar_choice} for more details).


\section{Galactic distribution of the gas emission}
\label{sec:distribution}

In this section we report the distance results obtained for the BDC runs including (Run~A) and excluding (Run~B) the prior for the spiral arm model (\sect\ref{sec:bdc_settings}).
In the subsections discussing the results, we always show and compare both BDC runs; if not indicated otherwise, the left- and right-hand panels depict the results of Run~A and B, respectively. 
We first present an overview of the results and then discuss the differences in terms of the face-on and vertical distribution of the gas emission and its variation with heliocentric and Galactocentric distance.
Finally, we discuss problems and biases of the two distance runs and compare our results with previous studies. 

\subsection{Catalogue description}
\label{sec:catalogue}

With this work, we also make a catalogue of all our distance results for the GRS available. 
In this section we describe the entries of the catalogue, which includes useful parameters that help gauge the performance of the distance results.

\begin{table*}
    \caption{Distance results.}
    \centering
    \scriptsize
    \renewcommand{\arraystretch}{1.2}
    \setlength{\tabcolsep}{4.5pt}
\begin{tabular}{ccccccccccccccccccc}
\hline\hline
 & & & \multicolumn{8}{c}{Run A: with SA prior} & \multicolumn{8}{c}{Run B: without SA prior} \\\cmidrule(lr){4-11} \cmidrule(lr){12-19}
$\ell$ & $b$ & v$_{\text{LSR}}$ & d$_{\odot,\,\text{A}}$ & $\Delta$d$_{\odot,\,\text{A}}$ & R$_{\text{gal,\,A}}$ & P$_{\text{A}}$ & Arm$_{\text{A}}$ & P$_{\text{far,\,A}}$ &Ref$_{\text{A}}$ & F$_{\text{A}}$ &d$_{\odot,\,\text{B}}$ & $\Delta$d$_{\odot,\,\text{B}}$ & R$_{\text{gal,\,B}}$ & P$_{\text{B}}$ & Arm$_{\text{B}}$ & P$_{\text{far,\,B}}$ &Ref$_{\text{B}}$ & F$_{\text{B}}$\\
{[$^{\circ}$]} & [$^{\circ}$] & [\kms] &[kpc] & [kpc] & [kpc] &  &  &  &  &  & [kpc] & [kpc] & [kpc] &  &  &  &  & \\
(1) & (2) & (3) & (4) & (5) & (6) & (7) & (8) & (9) & (10) & (11) & (12) & (13) & (14) & (15) & (16) & (17) & (18) & (19) \\
\hline

54.241 & -1.088 & 45.187 & 4.27 & 0.86 & 6.63 & 1.0 & LoS & 0.5 & -- & 2 & 4.01 & 0.79 & 6.66 & 1.0 & LoS & 0.5 & -- & 2 \\
54.241 & -1.088 & 24.809 & 1.59 & 0.64 & 7.33 & 0.54 & ... & 0.5 & -- & 2 & 1.68 & 0.58 & 7.30 & 0.78 & ... & 0.5 & -- & 2 \\
54.235 & -1.088 & 44.955 & 4.27 & 0.87 & 6.63 & 1.0 & LoS & 0.5 & -- & 2 & 4.0 & 0.8 & 6.66 & 1.0 & LoS & 0.5 & -- & 2 \\
54.235 & -1.088 & 24.849 & 1.6 & 0.64 & 7.33 & 0.54 & ... & 0.5 & -- & 2 & 1.69 & 0.58 & 7.29 & 0.78 & ... & 0.5 & -- & 2 \\
54.229 & -1.088 & 24.878 & 1.6 & 0.64 & 7.33 & 0.54 & ... & 0.5 & -- & 2 & 1.69 & 0.58 & 7.29 & 0.78 & ... & 0.5 & -- & 2 \\
54.217 & -1.088 & 24.836 & 1.6 & 0.64 & 7.33 & 0.54 & ... & 0.5 & -- & 2 & 1.69 & 0.58 & 7.29 & 0.78 & ... & 0.5 & -- & 2 \\
54.210 & -1.088 & 24.755 & 1.59 & 0.64 & 7.33 & 0.54 & ... & 0.5 & -- & 2 & 1.68 & 0.58 & 7.30 & 0.78 & ... & 0.5 & -- & 2 \\
54.204 & -1.088 & 24.842 & 1.6 & 0.64 & 7.33 & 0.54 & ... & 0.5 & -- & 2 & 1.69 & 0.58 & 7.29 & 0.78 & ... & 0.5 & -- & 2 \\
54.192 & -1.088 & 24.794 & 1.59 & 0.64 & 7.33 & 0.54 & ... & 0.5 & -- & 2 & 1.68 & 0.58 & 7.30 & 0.78 & ... & 0.5 & -- & 2 \\
54.186 & -1.088 & 24.863 & 1.6 & 0.64 & 7.33 & 0.54 & ... & 0.5 & -- & 2 & 1.69 & 0.58 & 7.29 & 0.78 & ... & 0.5 & -- & 2 \\
54.180 & -1.088 & 24.721 & 1.59 & 0.64 & 7.33 & 0.54 & ... & 0.5 & -- & 2 & 1.68 & 0.58 & 7.29 & 0.78 & ... & 0.5 & -- & 2 \\

    \hline
    \end{tabular}
    \label{tbl:table_dist}
    \tablefoot{ 
    \scriptsize 
    This table is available in its entirety in electronic form at the CDS via anonymous ftp to cdsarc.u-strasbg.fr (130.79.128.5) or via \url{http://cdsweb.u-strasbg.fr/cgi-bin/qcat?J/A+A/}. A portion is shown here for guidance regarding its form and content.
    }
\end{table*}

We show a subset of the distance results in Table~\ref{tbl:table_dist}.
Each row corresponds to a single Gaussian fit component; a spectrum fitted with eight Gaussian components thus occupies eight consecutive rows in the table.

Columns~(1) and (2) show the Galactic coordinate values and column~(3) gives the mean position $\vlsr$ of the fit component.
Columns~(4-11) list the parameters of the distance results for Run~A.
Columns~(4) and (5) give the heliocentric distance d$_{\odot}$ and its associated uncertainty $\Delta$d$_{\odot}$, and column~(6) gives the Galactocentric distance $\rgal$.
Columns~(7) and (8) give the estimated probabilities P and the associated Galactic features (\emph{Arm}) for the distance results.
Column~(9) and (10) list the probability $\prob{far}{}$ that was used for the KDA prior and the corresponding reference for an associated literature distance ($\prob{far}{}=0.5$ corresponds to the default value, in case no literature sources could be associated; see \sect\ref{sec:prior_kda} and App.~\ref{sec:kda}) for more details.
Column~(11) gives the flag that indicates which criterion was used for the choice of the final distance solutions (see \sect\ref{sec:dist_choice} and App.~\ref{app:bdc_prob_pfar_choice} for more details).
Columns~(12-19) list the same parameters as columns~(4-11), but for the distance results of Run~B.

\subsection{Face-on view of the \texorpdfstring{\textsuperscript{13}}{13}CO emission}
\label{sec:faceon}

\begin{figure}
    \centering
    \includegraphics[width=\columnwidth]{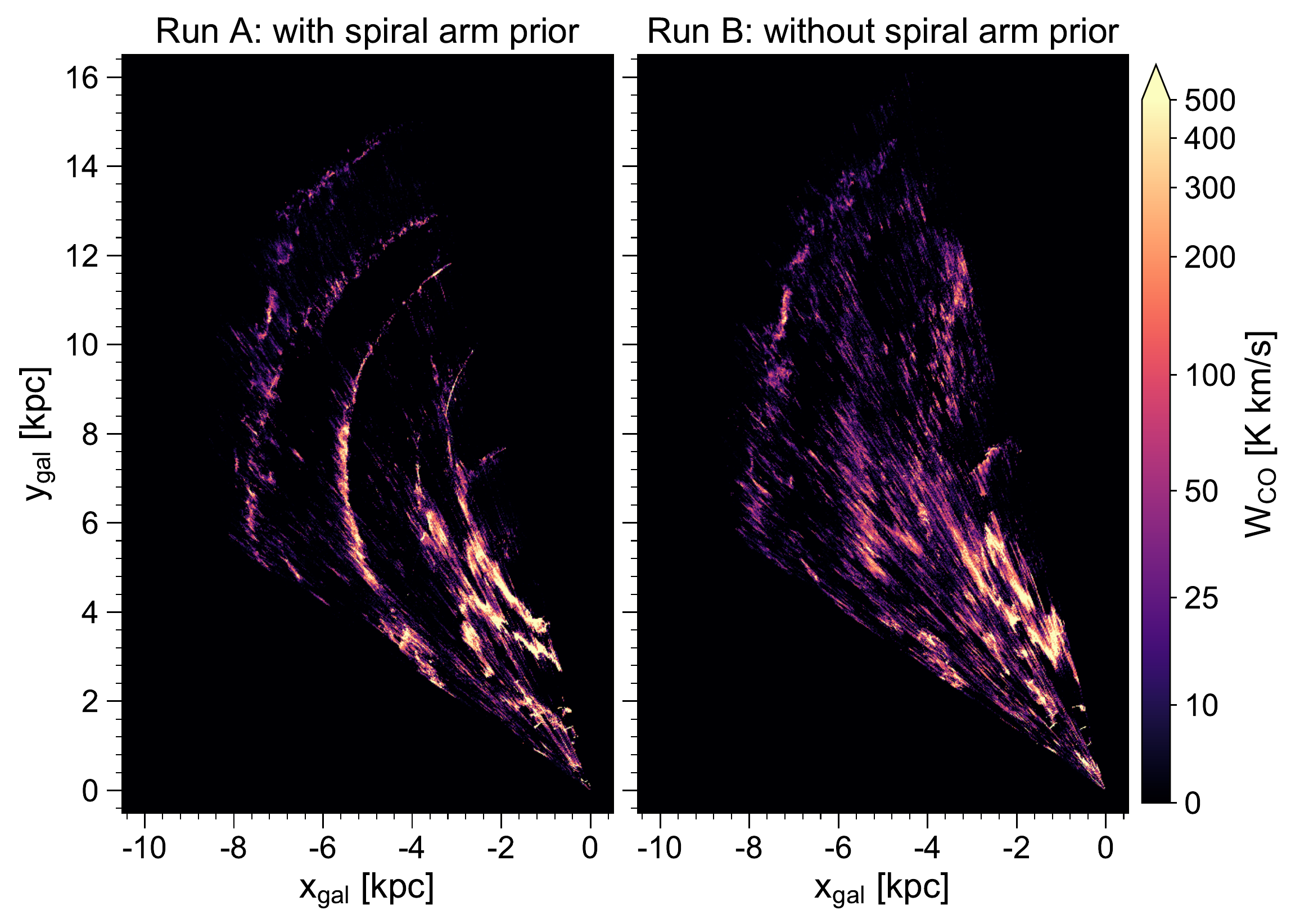}%
    \hspace{-\columnwidth}%
    \begin{ocg}{fig:grid_off_1}{fig:grid_off_1}{0}%
    \end{ocg}%
    \begin{ocg}{fig:grid_on_1}{fig:grid_on_1}{1}%
    \includegraphics[width=\columnwidth]{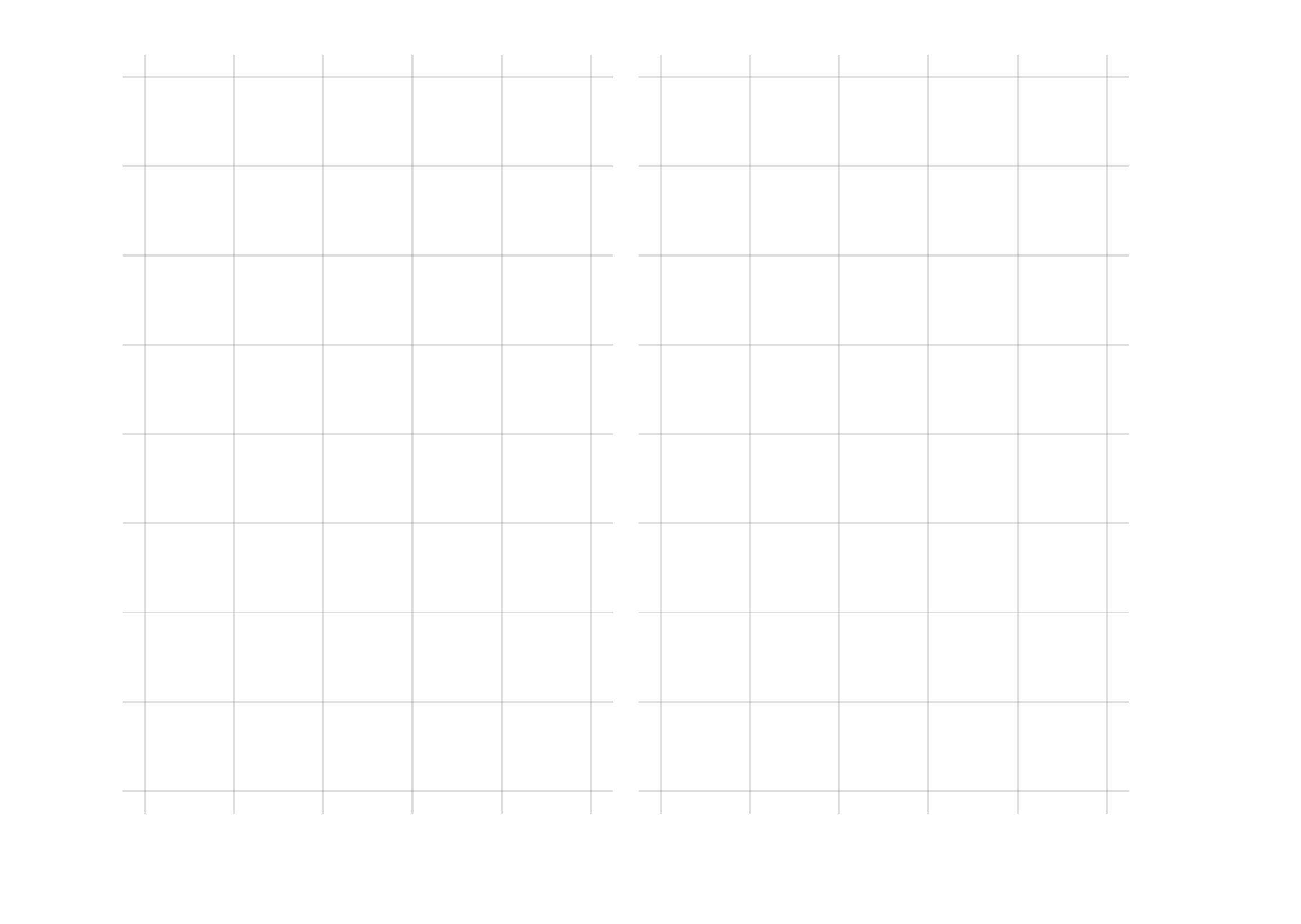}%
    \end{ocg}%
    \hspace{-\columnwidth}%
    \begin{ocg}{fig:arms_off_1}{fig:arms_off_1}{0}%
    \end{ocg}%
    \begin{ocg}{fig:arms_on_1}{fig:arms_on_1}{1}%
    \includegraphics[width=\columnwidth]{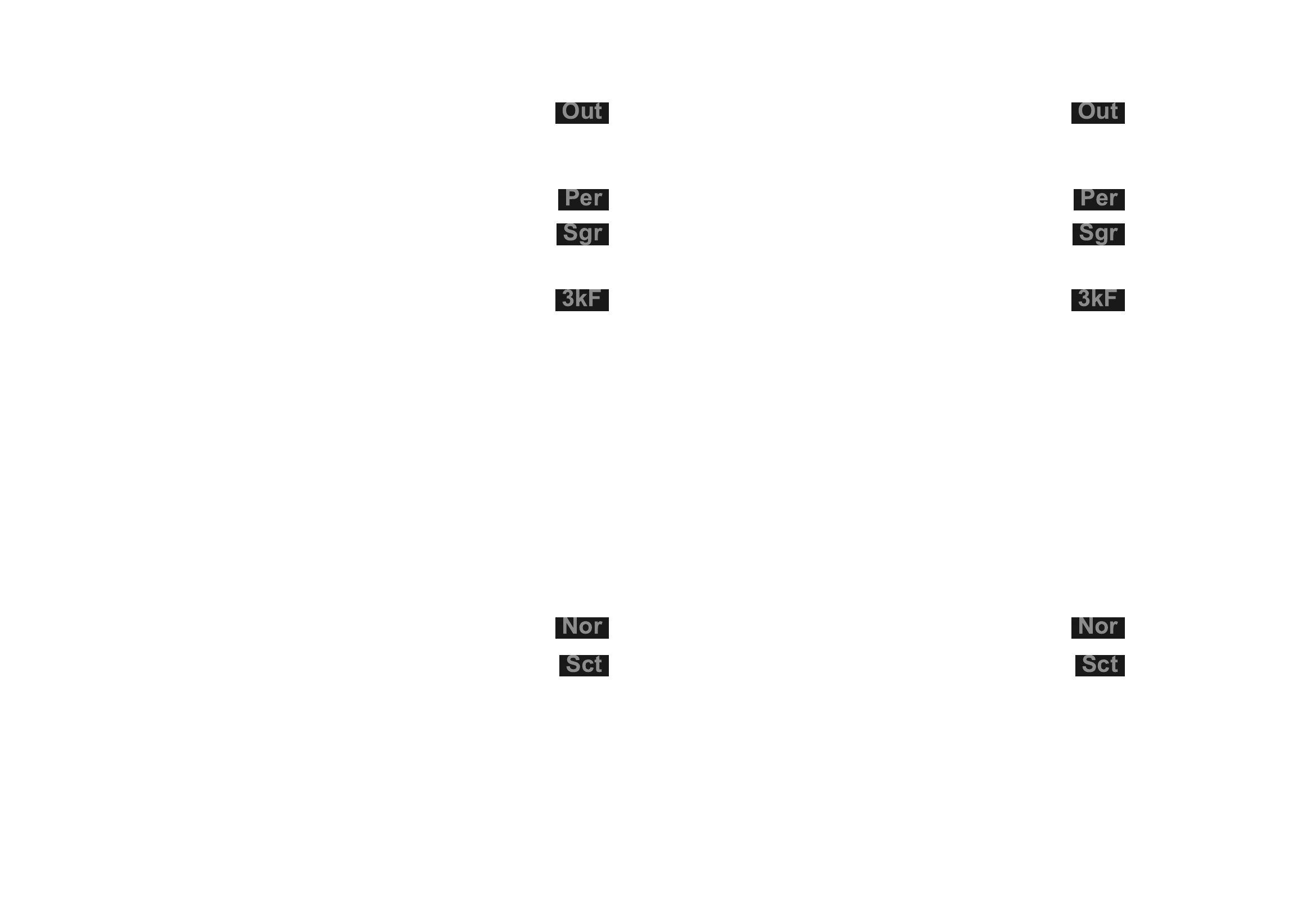}%
    \end{ocg}%
    \hspace{-\columnwidth}%
    \includegraphics[width=\columnwidth]{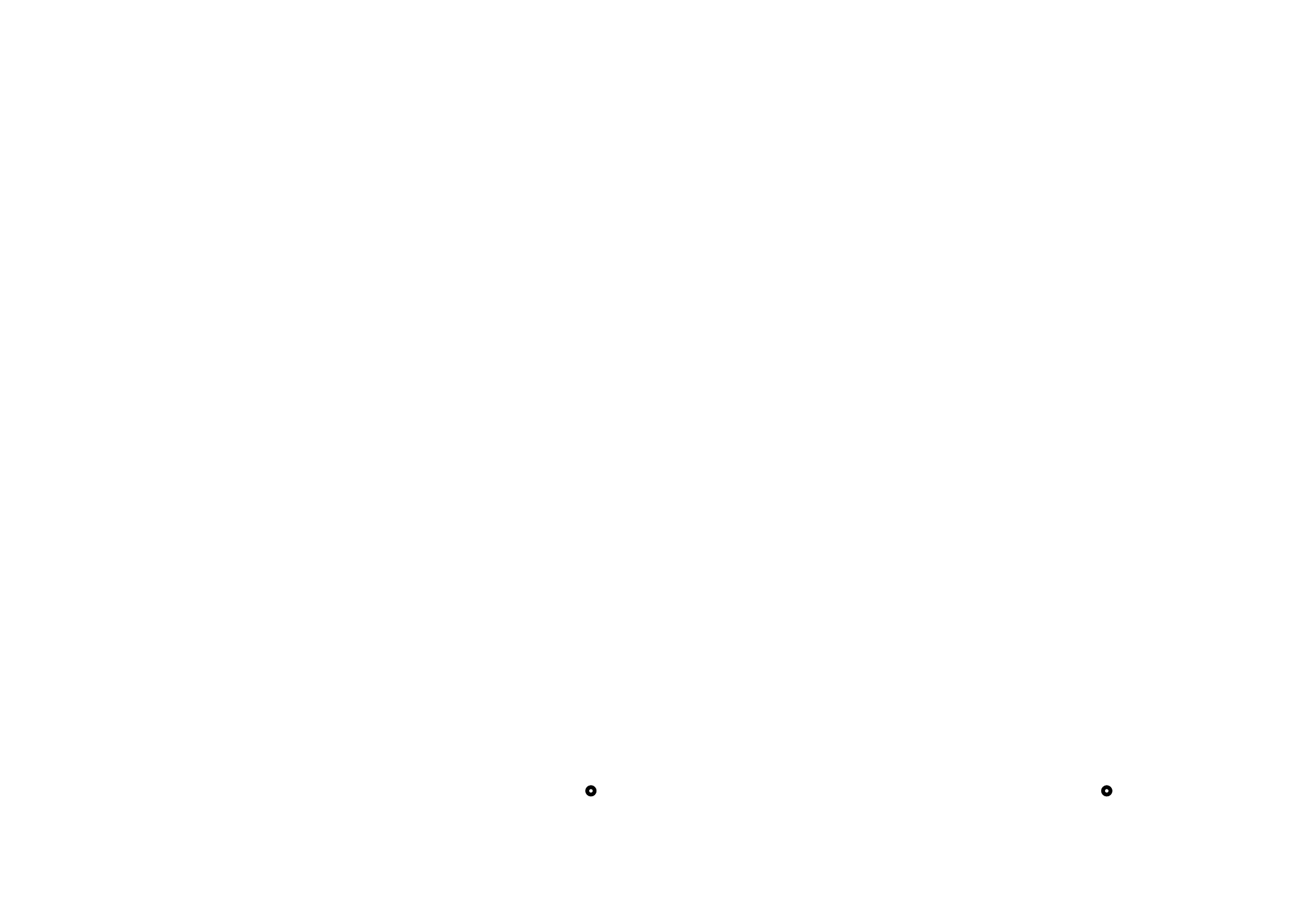}%
    \caption[Face-on view of the integrated $\co{13}{}$ emission]{Face-on view of the integrated $\co{13}{}$ emission for the BDC results obtained with (\textit{left}) and without (\textit{right}) the spiral arm prior.
    The values are binned in $10\times10$~pc cells and are summed up along the $\zgal$ axis.
    The position of the Sun and Galactic centre are indicated by the Sun symbol and black dot, respectively.
    When displayed in Adobe Acrobat, it is possible to hide the \ToggleLayer{fig:arms_on_1,fig:arms_off_1}{\protect\cdbox{spiral arm positions}} and the \ToggleLayer{fig:grid_on_1,fig:grid_off_1}{\protect\cdbox{grid}}.
    }
    \label{fig:fov_inttot}
\end{figure}

\begin{figure}
    \centering
	\includegraphics[width=\columnwidth]{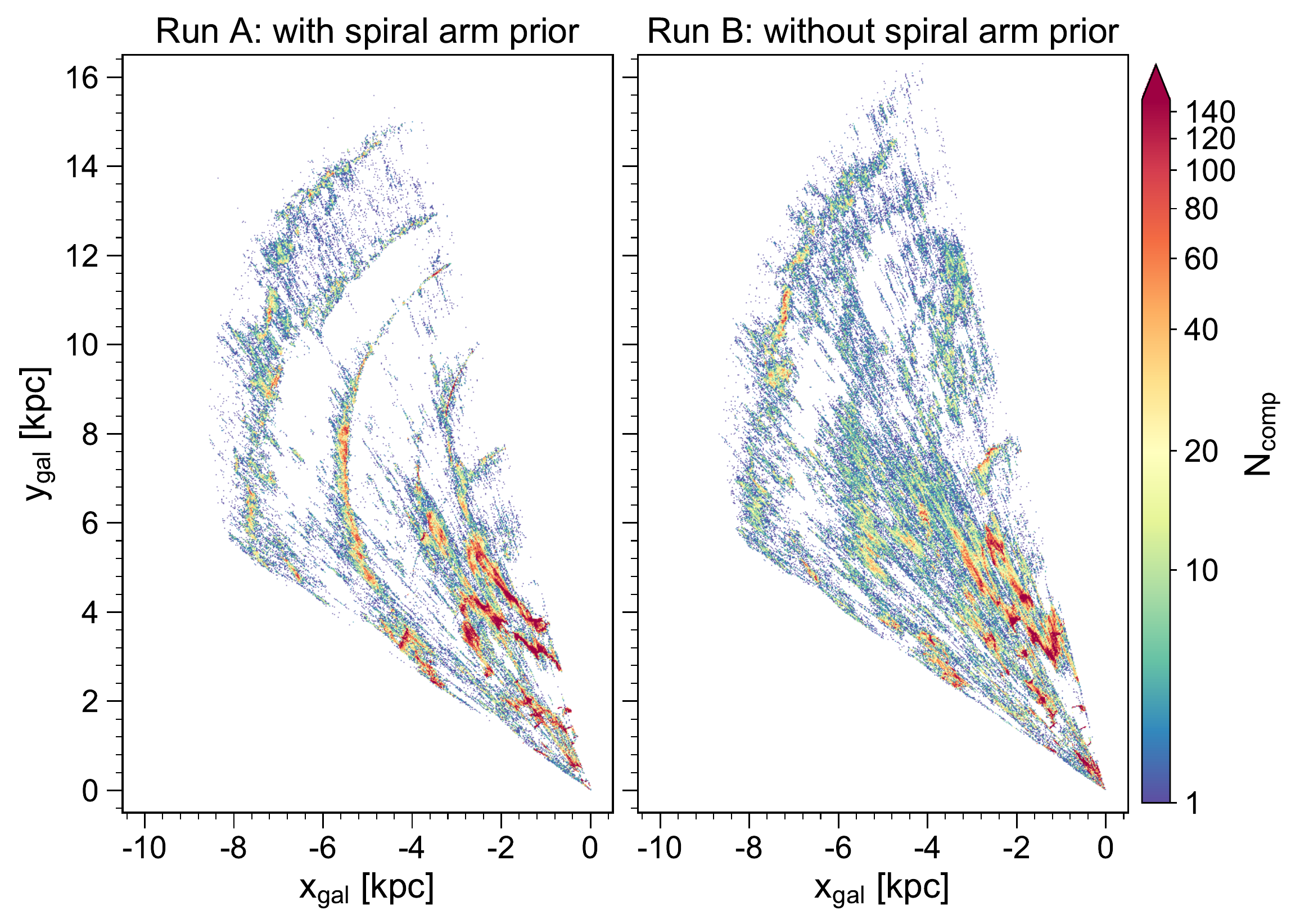}%
    \hspace{-\columnwidth}%
    \begin{ocg}{fig:grid_off_2}{fig:grid_off_2}{0}%
    \end{ocg}%
    \begin{ocg}{fig:grid_on_2}{fig:grid_on_2}{1}%
    \includegraphics[width=\columnwidth]{grs_fov_layer_grid.pdf}%
    \end{ocg}%
    \begin{ocg}{fig:arms_off_2}{fig:arms_off_2}{1}%
    \end{ocg}%
    \begin{ocg}{fig:arms_on_2}{fig:arms_on_2}{0}%
    \hspace{-\columnwidth}%
    \includegraphics[width=\columnwidth]{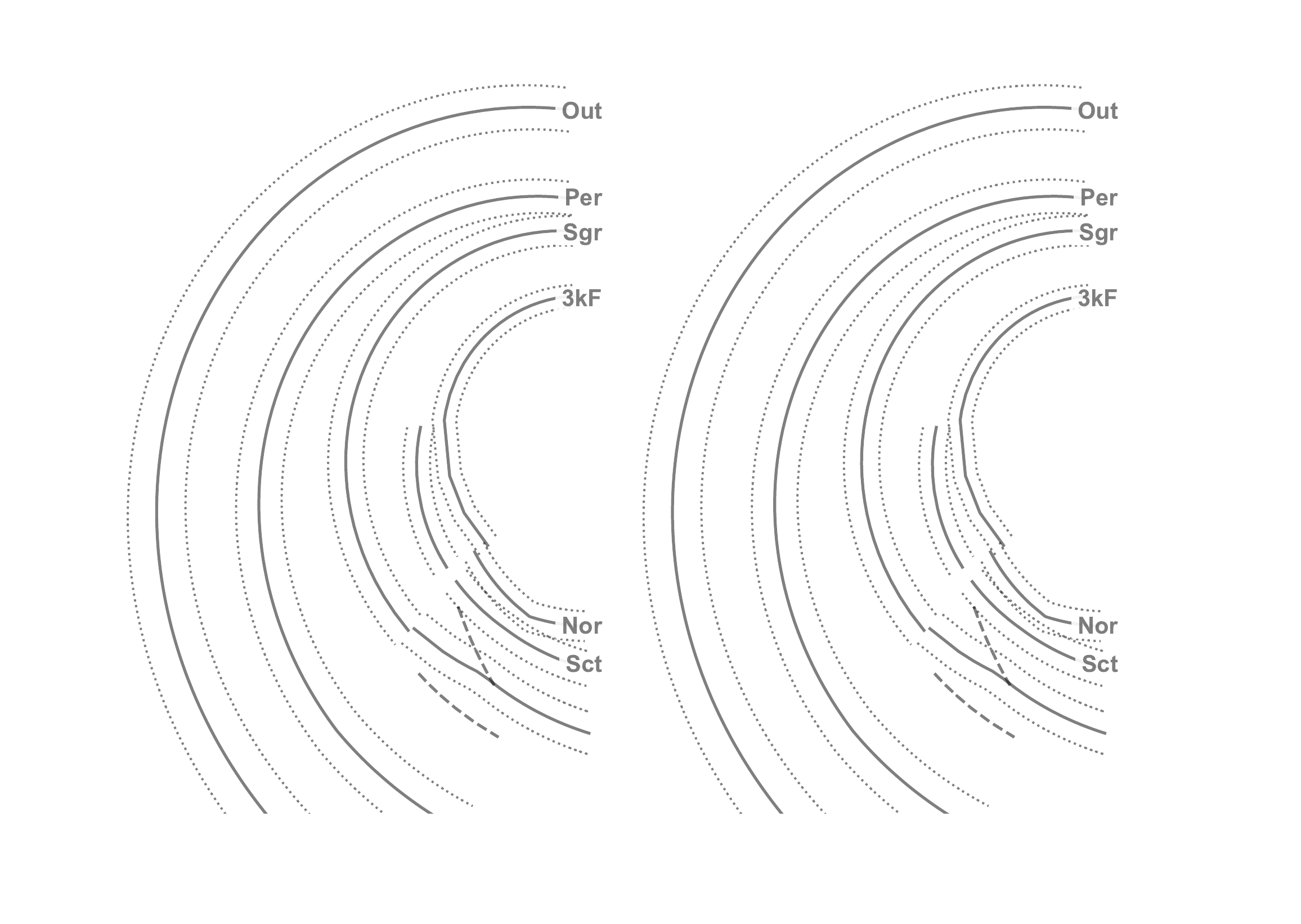}%
    \end{ocg}%
    \hspace{-\columnwidth}%
    \includegraphics[width=\columnwidth]{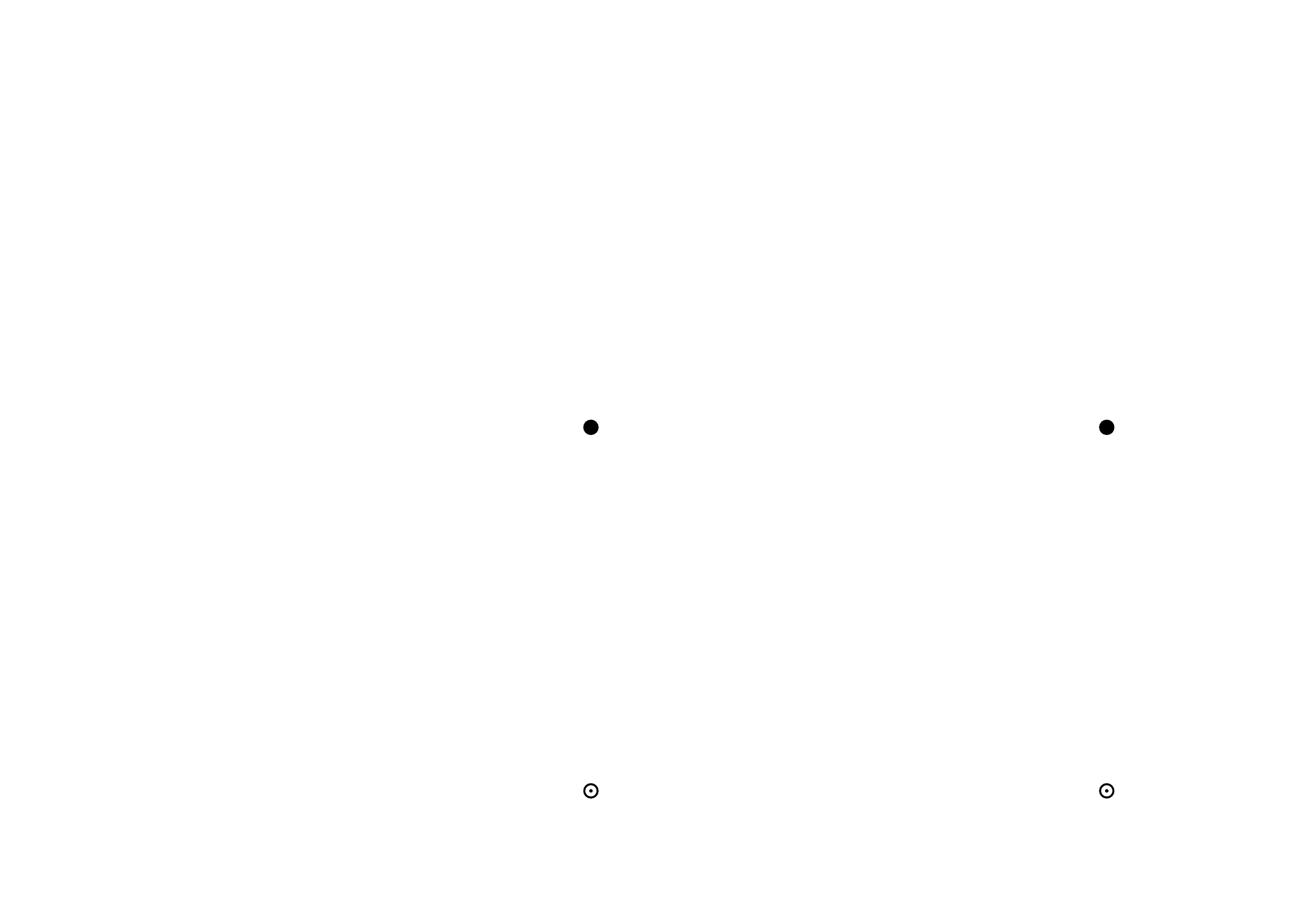}%
    \caption[Face-on view of the number of Gaussian fit components]{Face-on view of the number of Gaussian fit components for the BDC results obtained with (\textit{left}) and without (\textit{right}) the spiral arm prior.
    The values are binned in $10\times10$~pc cells and are summed up along the $\zgal$ axis.
    The position of the Sun and Galactic centre are indicated by the Sun symbol and black dot, respectively.
     When displayed in Adobe Acrobat, it is possible to show the \ToggleLayer{fig:arms_off_2,fig:arms_on_2}{\protect\cdbox{spiral arm positions}} and hide the \ToggleLayer{fig:grid_on_2,fig:grid_off_2}{\protect\cdbox{grid}}.
    }
    \label{fig:fov_ncomps}
\end{figure}

\begin{figure}
    \centering
    \includegraphics[width=\columnwidth]{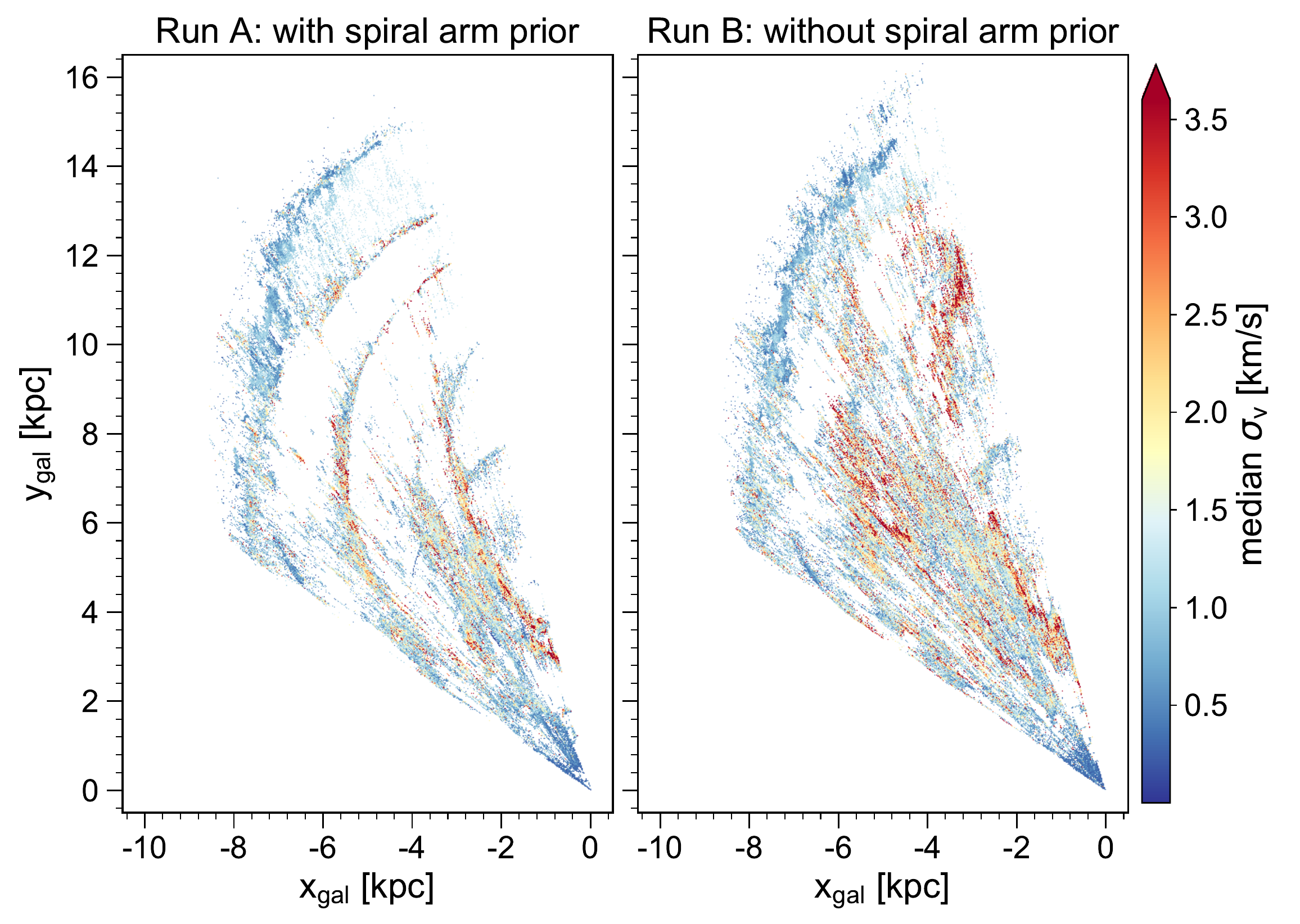}%
    \hspace{-\columnwidth}%
    \begin{ocg}{fig:grid_off_3}{fig:grid_off_3}{0}%
    \end{ocg}%
    \begin{ocg}{fig:grid_on_3}{fig:grid_on_3}{1}%
    \includegraphics[width=\columnwidth]{grs_fov_layer_grid.pdf}%
    \end{ocg}%
    \begin{ocg}{fig:arms_off_3}{fig:arms_off_3}{1}%
    \end{ocg}%
    \begin{ocg}{fig:arms_on_3}{fig:arms_on_3}{0}%
    \hspace{-\columnwidth}%
    \includegraphics[width=\columnwidth]{grs_fov_layer_spiral_arms_black.pdf}%
    \end{ocg}%
    \hspace{-\columnwidth}%
    \includegraphics[width=\columnwidth]{grs_fov_layer_sun+gc_black.pdf}%
    \caption[Face-on view of the median velocity dispersion values]{Face-on view of the median velocity dispersion values of Gaussian fit components for the BDC results obtained with (\textit{left}) and without (\textit{right}) the spiral arm prior.
    The values are binned in $10\times10$~pc cells and the median was calculated along the $\zgal$ axis.
    The position of the Sun and Galactic centre are indicated by the Sun symbol and black dot, respectively.
     When displayed in Adobe Acrobat, it is possible to show the \ToggleLayer{fig:arms_off_3,fig:arms_on_3}{\protect\cdbox{spiral arm positions}} and hide the \ToggleLayer{fig:grid_on_3,fig:grid_off_3}{\protect\cdbox{grid}}.
    }
    \label{fig:fov_veldisp}
\end{figure}

We show face-on view maps of the integrated $\co{13}{}$ emission, the number of Gaussian fit components, and the median $\veldisp$ value in Figs.~\ref{fig:fov_inttot}-\ref{fig:fov_veldisp}.
Comparing the maps of the $\co{13}{}$ emission (\fig\ref{fig:fov_inttot}), we can clearly see the effect of the SA prior in the left panel, which tends to concentrate most of the emission close to the Galactic features as they are defined in the spiral arm model (\fig\ref{fig:schematic_spiral_arms+masers}).
By neglecting the SA prior we get a distribution of the $\co{13}{}$ emission that is much more spread out and extends over a much larger area in between the arms, which can also be clearly observed in \fig\ref{fig:fov_ncomps}. 
This spreading of the emission to interarm locations is to a large part due to our use of archival KDA solutions to inform the $\prob{far}{}$ prior.
We present a comparison of the face-on map of $\co{13}{}$ emission with and without the use of archival KDA solutions in Appendix~\ref{sec:runs_comp}.
While we find only moderate differences in the fraction of emission assigned to interarm locations, the distribution of the gas emission itself changes significantly.

Even though Run~B shows a larger spreading of emission into interarm regions, we can still identify $\co{13}{}$ overdensities at the positions of the Galactic features of the SA model (right panels of Figs.~\ref{fig:fov_inttot} and \ref{fig:fov_ncomps}). 
This is not surprising, as Run~B has still a contribution from the maser parallax sources, which tend to be concentrated at spiral arms and spurs as well (cf. right panel in \fig\ref{fig:schematic_spiral_arms+masers}).
Moreover, the Galactic features for the spiral arm model are also based on overdensities in archival \hi\ and $\co{12}{}$ Galactic plane surveys, so we would expect that the $\co{13}{}$ emission is also present at these same locations.

Looking at the maps of the median $\veldisp$ values (\fig\ref{fig:fov_veldisp}), we qualitatively observe that spiral arm features seem to be associated with $\co{13}{}$ components with larger linewidths.
In general, we can see increased median $\veldisp$ values within Galactocentric distances $\lesssim 6$~kpc; as already speculated in \citet{Riener2020}, these increasing $\veldisp$ values towards the inner Galaxy could be due to the presence of the Galactic bar and the observed overdensity of star-forming regions \citep{Anderson2017, Ragan2018}, but could also partly result from our inability to correctly decompose strongly blended emission lines. 
We can further see an increase of the median $\veldisp$ values with heliocentric distance; for emission lines with $\vlsr < 20$~\kms\ this is partly due to our use of the size-linewidth prior (\sect\ref{sec:prior_linewidth}).
However, this effect is also present if we do not use this prior (see Appendix~\ref{sec:runs_comp} for a comparison between the maps of median $\veldisp$ values obtained with and without the size-linewidth prior).

Figures~\ref{fig:fov_inttot}-\ref{fig:fov_veldisp} also show a persistent feature at a Galactic longitude range of $29\degr \lesssim \ell \lesssim 38\degr$ that seemingly connects the Perseus and Outer arm\footnote{The position of this feature is indicated with a yellow dashed ellipse in \fig\ref{fig:survey_limits}.}. 
This is very likely emission originating close to the Sun that has been erroneously placed at far distances.
We can find evidence for this in \fig\ref{fig:fov_veldisp}, where the median $\veldisp$ value shows significantly lower values ($\lesssim 0.5$ \kms) at these locations than for most of the other parts of the Perseus arm.
This erroneously placed local emission is also clearly identifiable in the offset positions from the Galactic midplane, which we will discuss in \sect\ref{sec:height}.

\begin{figure}
    \centering
    \includegraphics[width=\columnwidth]{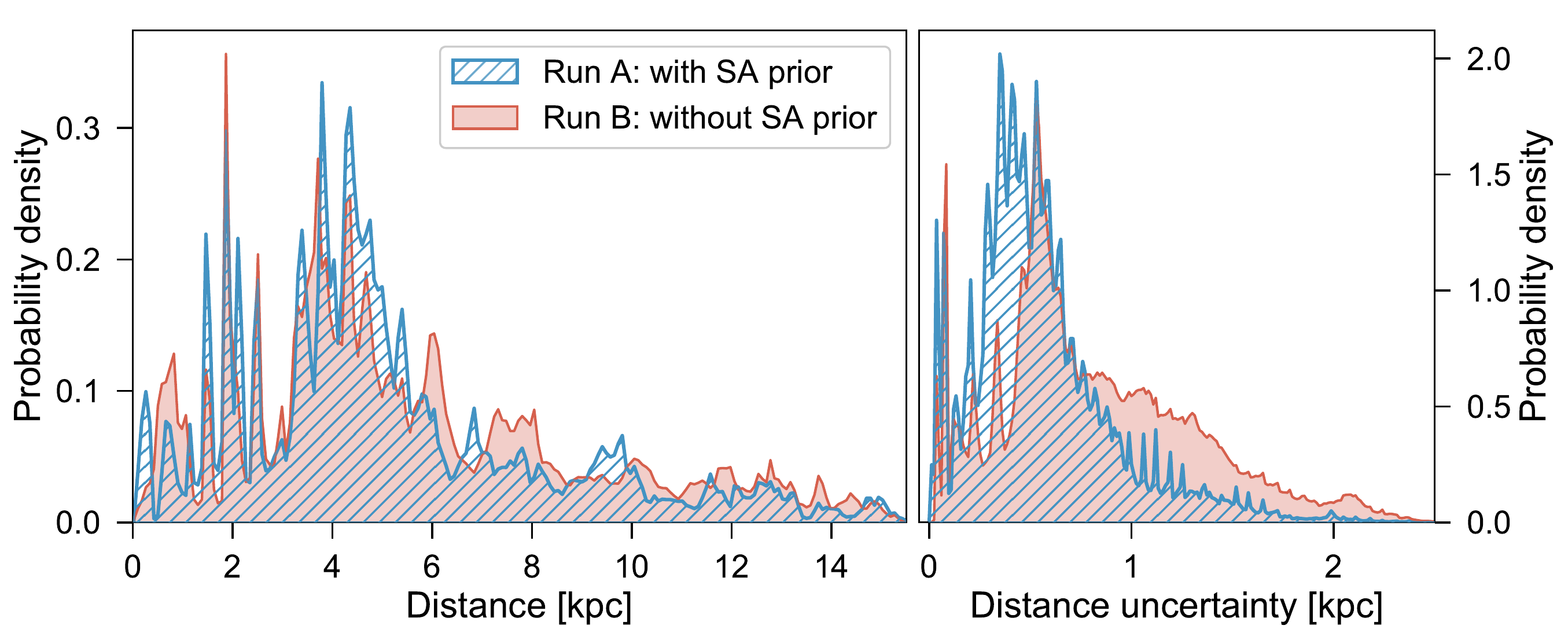}
    \caption[PDFs for estimated heliocentric distances (\textit{left}) and corresponding uncertainties]{PDFs for the estimated heliocentric distances (\textit{left}) and corresponding uncertainties (\textit{right}).
    }
    \label{fig:hist_dist_edist}
\end{figure}

\subsection{Comparison of the distance results}
\label{sec:comp_dist}

\begin{figure*}
    \centering
    \includegraphics[width=\textwidth]{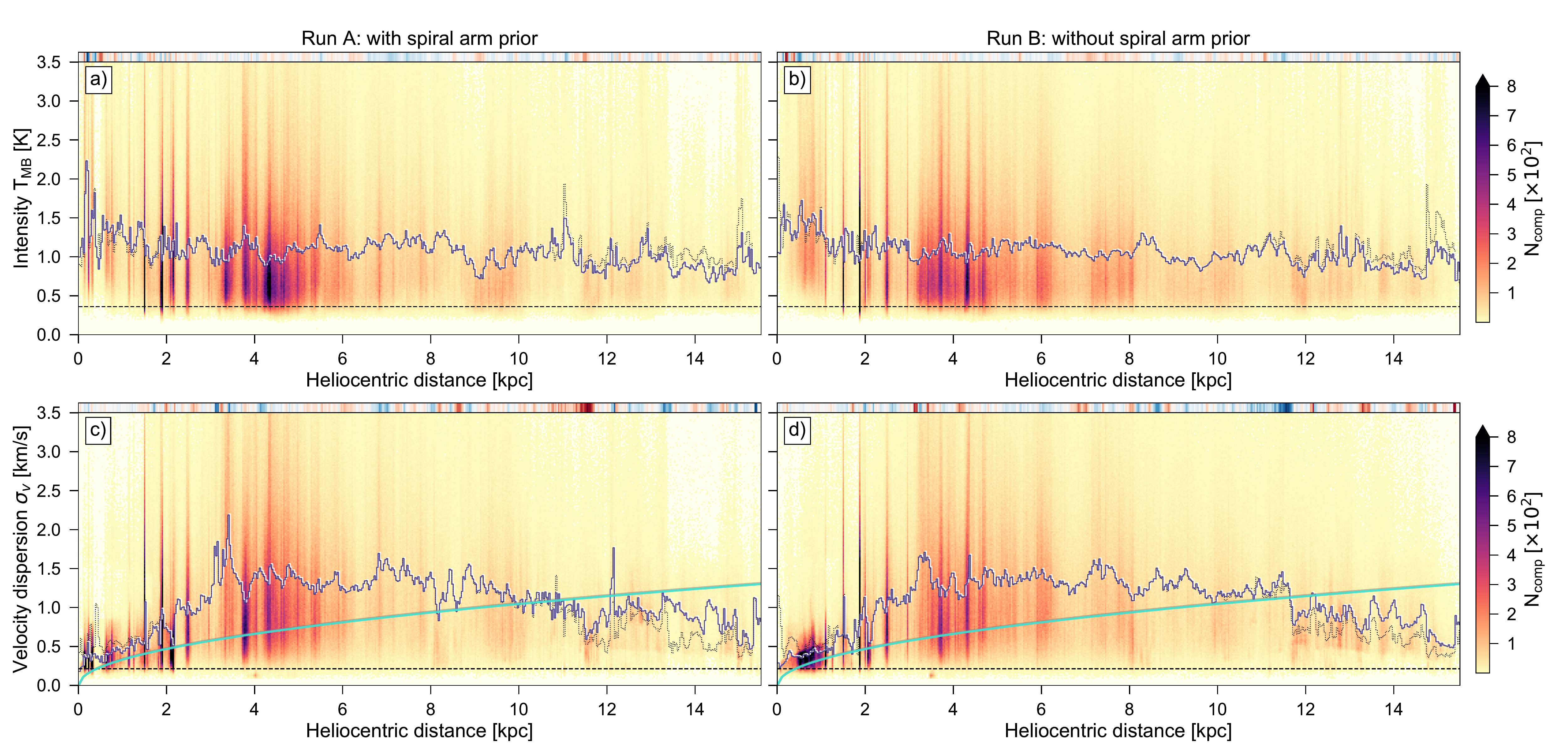}
    \caption[$2$D histograms of estimated distance values against intensity and velocity dispersion values]{$2$D histograms of estimated distance values and intensity (\textit{top}) and velocity dispersion (\textit{bottom}) values for the BDC run with (\textit{left column}) and without (\textit{right column}) the spiral arm prior.
    The blue lines show the respective median values per distance bin and dotted black lines give the corresponding values for distances obtained without the size-linewidth prior.
    The small strips at the top of the individual panels show where the median value is higher (blue) or lower (red) compared to the opposite BDC run, with the strength of the colour corresponding to the magnitude of the difference.
    The turquoise line in the bottom panels indicates the expected values from the size-linewidth relationship (Eq.~\ref{eq:size_lw}).
    The dashed horizontal line in the top panels at $\Tb=0.36$~K corresponds to the $3\times$ S/N limit for the $0.1$st percentile of the GRS noise distribution \citep[see][]{Riener2020}.
    The dashed horizontal line in the bottom panels indicates the velocity resolution of the GRS ($0.21$~\kms).
    }
    \label{fig:hist2d_dhelio}
\end{figure*}

Comparing the resulting distances of the two BDC runs, we find that $68.2\%$ are compatible with each other within their estimated uncertainties.\footnote{We note that each percentage point corresponds to about $46\,500$ independent distance assignments.}
The distance uncertainties are given by the standard deviation of the chosen Gaussian component fit to the combined distance PDF (see \sect\ref{sec:dist_choice}).
In terms of differences in absolute distance uncertainty values, $67.2\%$, $78.8\%$, and $83.8\%$ of the distance results are compatible within $\pm 0.5$, $\pm 1.0$, and $\pm 1.5$~kpc, respectively.
We thus conclude that for the majority of the GRS fit components the two distance runs yielded similar results. 
We can use the PDFs of the estimated heliocentric distances and corresponding distance uncertainties (\fig\ref{fig:hist_dist_edist}) to identify where the distance estimates deviated. 
For example, Run~B yielded more distances above $8$~kpc ($21.2\%$) but produced fewer distance assignments $< 0.5$~kpc ($1.2\%$) compared to Run~A ($17.1\%$ and $2.4\%$, respectively).

The difference between the BDC runs is even more pronounced in the distance uncertainties. 
Half of the distance assignments of Run~A have distance uncertainties $< 0.5$~kpc, but only a quarter of the distance assignments for Run~B are below this distance uncertainty threshold.
This difference is also reflected in the estimated probabilities of the distances: about $44\%$ of the results from Run~A have high-confidence probabilities $> 0.75$; for Run~B only $\sim 34\%$ of the distance results exceed this probability threshold (see App.~\ref{app:bdc_prob_pfar_choice} for more details).
We caution that the estimated uncertainties and probabilities do not allow for a straightforward comparison of the quality of the distance results.
Strongly favouring the distance assignments towards a particular prior may yield small uncertainties and high probabilities but the prior itself may lead to biased distance results.
We discuss these issues further in \sect\ref{sec:problems}. 
    
In the top panels of \fig\ref{fig:hist2d_dhelio} we show how the intensity and velocity dispersion values of the Gaussian fit components vary with heliocentric distance for both BDC runs.
While the intensity values cover a large range, their median values stay flat over all considered distances.

The bottom panels in \fig\ref{fig:hist2d_dhelio} show how the $\veldisp$ values of the fit components vary with their estimated distances.
We can see a clear increase in the median $\veldisp$ values up until heliocentric distances of about $3.5$~kpc, after which it stays at increased values of $> 1$~\kms, until it drops again at distances $\gtrsim 11.5$~kpc.
This drop at the largest distances is due to a bias in the distance calculation that erroneously puts emission from nearby regions at large distances from the Sun (see \sect\ref{sec:faceon}).
We also show the median $\veldisp$ values we would have gotten if we had not used the size-linewidth prior (\sect\ref{sec:prior_linewidth}), which shows an even bigger drop at these large distances. 
However, for d $< 4$~kpc we recover a similar trend of increased linewidths with larger heliocentric distances, indicating that beam averaging effects play a crucial role in producing these increased linewidths.
Another explanation could be a larger non-thermal contribution to the linewidth for emission located in the inner part of the Galaxy. 
The comparison of the median $\veldisp$ curve with the size-linewidth relationship from Eq.~\ref{eq:size_lw} shows that most of the fit components have linewidths that are significantly larger than those expected values.
In \fig\ref{fig:hist2d_dhelio} we also indicate the median intensity and velocity dispersion values without the use of the size-linewidth prior (black dotted lines).
Since we restricted the use of the size-linewidth prior to $\vlsr$ values $< 20$~\kms, the distribution does not change between $2 \lesssim \dsun \lesssim 10$~kpc.
However, we can see that in absence of the size-linewidth prior the distribution of the velocity dispersion values (bottom panels) contain much more gas emission with $\veldisp < 1$~\kms\ at heliocentric distances $\dsun > 11$~kpc, which based on our considerations in \sect\ref{sec:prior_linewidth} is likely not correct.
We thus conclude that while we need to exercise caution in the use of the size-linewidth prior its restricted use for $\vlsr$ values $< 20$~\kms\ led to significant improvements. 

\subsection{Gas fraction in spiral arm and interarm regions}
\label{sec:distresults}

\begin{table}
    \caption[Distance results for the two BDC runs]{Distance results for the two BDC runs.}
    \centering
    \footnotesize
    \renewcommand{\arraystretch}{1.3}
    \setlength{\tabcolsep}{5pt}
\begin{tabular}{ccccccc}
	\hline\hline
 & \multicolumn{3}{c}{Run A: with SA prior} & \multicolumn{3}{c}{Run B: without SA prior} \\\cmidrule(lr){2-4} \cmidrule(lr){5-7}
\multirow{2}{*}{Feature\tablefootmark{a}} & W$_{\text{CO}}$ & N$_{\text{comp}}$ & $\sigma_{\text{v,\,med.}}$\tablefootmark{b} & W$_{\text{CO}}$ & N$_{\text{comp}}$ & $\sigma_{\text{v,\,med.}}$\tablefootmark{b} \\
 & [$\%$] & [$\%$] & [\kms] & [$\%$] & [$\%$] & [\kms] \\
	\hline\\[-2.75ex]
$3$kF & $2.6$ & $2.4$ & $1.4$ $\left[^{0.9}_{2.2}\right]$ & $1.4$ & $1.6$ & $1.3$ $\left[^{0.8}_{2.0}\right]$ \\[0.5ex]
AqR\tablefootmark{c} & $1.1$ & $2.8$ & $0.4$ $\left[^{0.3}_{0.5}\right]$ & $1.0$ & $2.5$ & $0.4$ $\left[^{0.3}_{0.6}\right]$ \\[0.5ex]
AqS & $7.0$ & $7.2$ & $1.4$ $\left[^{0.9}_{2.2}\right]$ & $4.3$ & $4.5$ & $1.3$ $\left[^{0.8}_{2.1}\right]$ \\[0.5ex]
LoS & $1.8$ & $2.5$ & $1.0$ $\left[^{0.7}_{1.6}\right]$ & $1.5$ & $2.1$ & $1.0$ $\left[^{0.7}_{1.5}\right]$ \\[0.5ex]
N$1$F & $1.3$ & $1.0$ & $1.3$ $\left[^{0.9}_{2.0}\right]$ & $3.2$ & $2.3$ & $1.5$ $\left[^{1.0}_{2.3}\right]$ \\[0.5ex]
N$1$N & $20.6$ & $14.9$ & $1.5$ $\left[^{1.0}_{2.5}\right]$ & $16.8$ & $11.7$ & $1.5$ $\left[^{1.0}_{2.5}\right]$ \\[0.5ex]
Out\tablefootmark{c} & $0.6$ & $1.3$ & $0.7$ $\left[^{0.5}_{1.0}\right]$ & $0.6$ & $1.4$ & $0.7$ $\left[^{0.5}_{1.0}\right]$ \\[0.5ex]
Per\tablefootmark{c} & $3.9$ & $6.0$ & $0.9$ $\left[^{0.6}_{1.4}\right]$ & $4.0$ & $6.0$ & $1.0$ $\left[^{0.6}_{1.5}\right]$ \\[0.5ex]
ScF & $5.7$ & $4.3$ & $1.4$ $\left[^{0.9}_{2.2}\right]$ & $6.2$ & $5.1$ & $1.4$ $\left[^{0.9}_{2.2}\right]$ \\[0.5ex]
ScN & $24.6$ & $20.1$ & $1.4$ $\left[^{0.9}_{2.3}\right]$ & $23.6$ & $19.2$ & $1.4$ $\left[^{0.9}_{2.3}\right]$ \\[0.5ex]
SgF & $12.2$ & $10.3$ & $1.4$ $\left[^{0.9}_{2.1}\right]$ & $11.9$ & $9.7$ & $1.4$ $\left[^{0.9}_{2.3}\right]$ \\[0.5ex]
SgN & $12.1$ & $15.8$ & $0.8$ $\left[^{0.5}_{1.4}\right]$ & $8.0$ & $9.4$ & $1.0$ $\left[^{0.6}_{1.5}\right]$ \\[0.5ex]
N/A & $6.5$ & $11.2$ & $0.7$ $\left[^{0.5}_{1.1}\right]$ & $17.5$ & $24.5$ & $0.8$ $\left[^{0.5}_{1.4}\right]$ \\[0.5ex]
\hline\\[-2.75ex]
Spiral arms & $83.6$ & $76.2$ & 1.2 $\left[^{0.8}_{2.0}\right]$ & $75.6$ & $66.3$ & 1.3 $\left[^{0.8}_{2.1}\right]$ \\[0.5ex]
Interarm & $16.4$ & $23.8$ & 0.8 $\left[^{0.5}_{1.4}\right]$ & $24.4$ & $33.7$ & 0.8 $\left[^{0.5}_{1.4}\right]$ \\[0.5ex]
	\hline
\end{tabular}
\tablefoot{
\small
\tablefoottext{a}{
$3$ kpc far arm ($3$kF), Aquila Rift (AqR), Aquila Spur (AqS), Local spur (LoS), Norma $1^{\text{st}}$ quadrant near and far portions (N$1$N, N$1$F), Outer (Out), Perseus (Per), Scutum near and far portions (ScN, ScF), Sagittarius near and far portions (SgN, SgF), unassociated (N/A).}\\
\tablefoottext{b}{
The two values in the brackets give the corresponding IQR.}\\
\tablefoottext{c}{
Values are likely severely impacted by confusion between emission from the solar neighbourhood and far distances; see Sects.~\ref{sec:height} and \ref{sec:problems}.}
}
\label{tbl:stats_arms}
\end{table}

In this section we discuss the fraction of $\co{13}{}$ residing in spiral arm and interarm environments, which also serves to give a more quantitative overview of the distance results.
In Table~\ref{tbl:stats_arms} we split our distance results into different subsamples that correspond to the determined association with Galactic features (\textit{left panel} in \fig\ref{fig:schematic_spiral_arms+masers}) by the BDC. 
This association is based on the ($\ell, b, \vlsr$) coordinates and the position and extents of the spiral arm and interarm features (see Sect.\,2.1 in \citealt{Reid2016} for more details about this association). 
For each subsample we report the fraction of the total integrated $\co{13}{}$ intensity ($\wco$), the fraction of the total number of fit components ($\Ncomp$), and the median velocity dispersion value ($\sigma_{\text{v,\,med.}}$) with the corresponding interquartile range (IQR) in brackets.
We also list the combined values for all spiral arm ($3$kF, N$1$F, N$1$N, Out, Per, ScF, ScN, SgF, and SgN) and interarm (AqR, AqS, LoS, N/A) features as \emph{Spiral arm} and \emph{Interarm}, respectively.

In the two BDC runs, about $76 - 84\%$ of the integrated $\co{13}{}$ emission and $66 - 76\%$ of the $\co{13}{}$ fit components were associated with spiral arm features, mostly with the Norma, Scutum, and Sagittarius arms.
Run~B placed about $1.5$ times more $\co{13}{}$ emission in interarm regions not associated with any of the Galactic features shown in \fig\ref{fig:schematic_spiral_arms+masers}.
To put these numbers into perspective and check whether also the gas distribution in Run~B shows a significant concentration towards spiral arm features, we determined the fraction of $\co{13}{}$ gas in spiral arms based on only kinematic distances.
We calculate the kinematic distances using methods contained in the BDC v$2.4$ and solve for the KDA by using the Monte Carlo approach outlined in \sect\,3.1 of \citet{RomanDuval2016}, assuming a Gaussian vertical density profile of the molecular gas with a FWHM of $110$~pc as was done in that study.
For these pure kinematic distance solutions we find that $\sim 58\%$ of the integrated $\co{13}{}$ emission and $\sim 52\%$ of the fit components overlap with the positions of spiral arms from our assumed model.
These results demonstrate that compared to pure kinematic distances both our BDC runs contain a significant enhancement of $\co{13}{}$ emission at the position of spiral arm features. 

To further check the robustness of our results we also looked at the distance results of only the $\sim\! 75\%$ of fit components that had a signal-to-noise (S/N) ratio $> 3$.
We do not find significant deviations from the trends presented in Table~\ref{tbl:stats_arms}.
In particular, we recover the same difference in $\sigma_{\text{v,\,med.}}$ between the Galactic features, which we discuss in the next section.


\subsection{Velocity dispersion in spiral arm and interarm regions}
\label{sec:var_arms}

\begin{figure}
    \centering
    \includegraphics[width=\columnwidth]{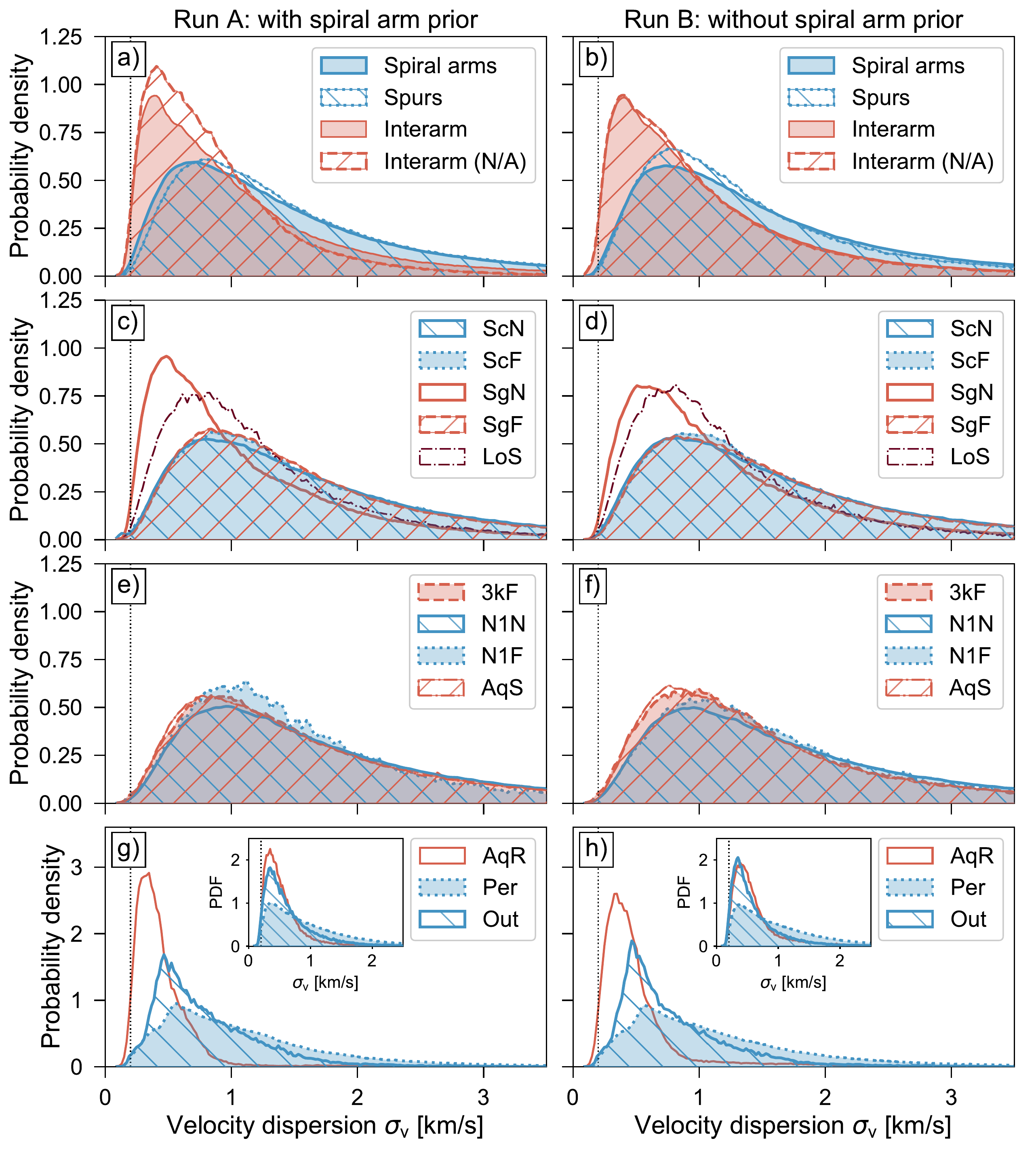}
    \caption[PDFs of velocity dispersion values associated with Galactic features]{PDFs of velocity dispersion values associated with Galactic features for the distance results with (\textit{left columns}) and without (\textit{right columns}) the SA prior.
    The dotted vertical line indicates the GRS velocity resolution ($0.21$~\kms).
    The insets in the bottom panels show the corresponding PDFs without the use of the size-linewidth prior.
    }
    \label{fig:hist_arms_veldisp}
\end{figure}

One interesting exercise is to look for possible variations of the gas velocity dispersion between spiral arm and interarm regions, which has been observed for the nearby spiral galaxy M$51$ \citep{Colombo2014}.
To split our data points into spiral arm and interarm features, we again use the BDC assignment with Galactic features from the previous section.
Figure~\ref{fig:hist_arms_veldisp} shows $\veldisp$-PDFs for these Galactic features and Table~\ref{tbl:stats_arms} gives the corresponding median values and interquartile ranges for these distributions.
Generally speaking, spiral arm structures are associated with larger $\veldisp$ values than interarm structures, with the spiral arm PDF peaking at larger $\veldisp$ values.
We note that the PDF labelled \emph{Interarm} contains also associations with the spur features (AqS, LoS) and the nearby Aquila Rift complex (AqR).
To check how this might skew the results, we also show PDFs for interarm emission not associated with any of the Galactic features from the SA model (labelled \emph{Interarm (N/A)}) and emission only associated with spur features (\emph{Spurs}).
Interestingly, the PDF for the spurs is almost indistinguishable from the PDF of the spiral arms. 

We make a more detailed comparison between emission associated with spiral arm and spur structures in \fig\ref{fig:hist_arms_veldisp}c--f.
The emission associated with the two major spiral structures covered by the GRS, the Scutum and Sagittarius arms, essentially has identical $\veldisp$-PDFs apart from the near portion of the Sagittarius arm (SgN), whose distribution peaks at much lower $\veldisp$ values and is more similar to the PDF of the Local Spur (LoS) and the interarm PDFs in panel~(a) and (b).
Other structures in the inner Galaxy -- the Norma arm (N$1$N, N$1$F), the far portion of the $3$-kpc-arm ($3$kF), and the Aquila spur (AqS) -- all show a very similar $\veldisp$ distribution that is essentially identical to the PDFs of the Scutum arm and the far portion of the Sagittarius arm.
Since the near portion of the Sagittarius arm and the Local Spur are located at the highest longitude ranges covered by the GRS, this might point to real differences in terms of the linewidth distribution in the innermost and more outer parts of the GRS coverage.
However, since parts of the SgN are also located close to the Sun ($\dsun<3$~kpc), its emission lines might simply be better resolved spatially, leading to narrower linewidths (see also discussion in \sect\ref{sec:prior_linewidth}).
The difference in the $\veldisp$-PDFs might also be explained by difficulties in the decomposition of strongly blended emission lines in the inner Galaxy, which could have led to higher fitted $\veldisp$ values.

The bottom panels (g, h) show PDFs for the Aquila Rift (AqR) complex and the Perseus (Per) and Outer (Out) arms.
As already mentioned, we are not able to fully separate the near and far contribution of this emission with low $\vlsr$ values.
This problem is reflected in the shape of the PDFs, which are moreover impacted by our use of the size-linewidth prior.
For comparison, we also show how the PDFs would look like if we did not use the size-linewidth prior (small insets in panels~g and h).
In this case their $\veldisp$-PDFs become more similar, which is in contrast to expectations based on beam averaging effects (see \sect\ref{sec:prior_linewidth} and Appendix~\ref{sec:beam_avg_veldisp}) and the other spiral arm PDFs, which show much higher $\veldisp$ values.
We currently also have no reason to suspect that the Perseus and Outer arms should be peculiar in terms of their linewidth distribution compared to other spiral arms.

\begin{figure}
    \centering
    \includegraphics[width=\columnwidth]{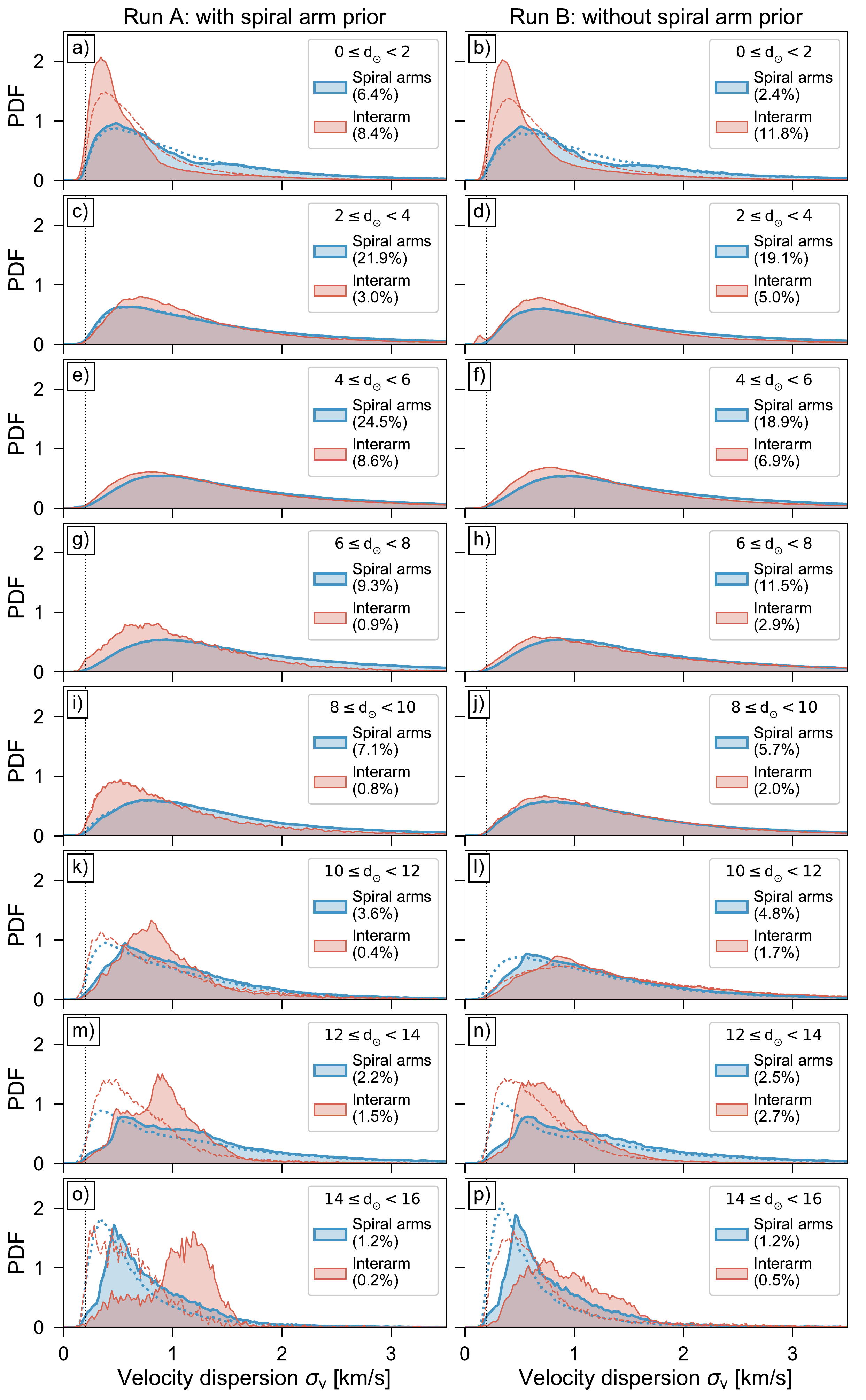}
    \caption[PDFs of velocity dispersion values for $2$~kpc heliocentric distance bins]{PDFs of velocity dispersion values for $2$~kpc heliocentric distance bins for the distance results with (\textit{left columns}) and without (\textit{right columns}) the SA prior.
    The dotted vertical line indicates the GRS velocity resolution ($0.21$~\kms).
    Dotted and dashed histograms show the distribution for distance results obtained without the size-linewidth prior.
    Percentages in the legend indicate the respective fraction of Gaussian fit components associated with spiral arm and interarm structures.
    }
    \label{fig:hist_dsun_veldisp}
\end{figure}

To further check the significance of the difference in the $\veldisp$-PDFs of spiral arm and interarm structures, we looked at the $\veldisp$-PDFs in $2$~kpc heliocentric bins (\fig\ref{fig:hist_dsun_veldisp}).
About one third of the fit components associated with interarm structures have distances $< 2$~kpc (panels~a, b), compared to a much lower fraction of fit components associated with spiral arms in this distance range.
This difference seems to be the major cause for the difference in the total $\veldisp$-PDFs in \fig\ref{fig:hist_arms_veldisp}~(a) and (b).
The remaining interarm distributions in \fig\ref{fig:hist_dsun_veldisp} show much closer resemblance to the spiral arm PDFs, and indicate no consistent or considerable trend towards lower linewidths.

Figure~\ref{fig:hist_dsun_veldisp} once more highlights the problem of confusion between emission originating from the near and far side of the Galactic disk.
For most of the PDFs in \fig\ref{fig:hist_dsun_veldisp}~(a)--(j) we do see a shift towards higher linewidths with increasing distance ranges, which would match our expectations based on beam averaging effects.
The interarm PDFs in panels~(g) and (i) show a deviation from this trend, which could be indicative of an increased confusion between near and far emission at these distance bins. 
The second bump at low $\veldisp$ values ($< 0.2$~\kms) in panel~(d) is due to an instrumental artefact in the GRS data set that led to the fitting of very narrow components (see Appendix~A.4 in \citealt{Riener2020}).

The strong confusion for emission at low $\vlsr$ values ($< 20$~\kms) also becomes apparent again in panels~(a), (b), and (k--p) of \fig\ref{fig:hist_dsun_veldisp}, which also highlight the effect of the size-linewidth prior.
While we do find artefacts induced by the prior in these $\veldisp$-PDFs, the distributions are nonetheless more consistent with the trend of higher $\veldisp$ values with increasing heliocentric distance.
So even though Figs.~\ref{fig:hist_arms_veldisp} and \ref{fig:hist_dsun_veldisp} show that we need to be careful in interpreting the distance results for the emission features with low $\vlsr$ values, we conclude that the use of the size-linewidth prior was justified and successful in disentangling part of the confusion between near and far emission.

\subsection{Galactocentric variation of the gas properties}
\label{sec:var_rgal}

\begin{figure*}
    \centering
    \includegraphics[width=\textwidth]{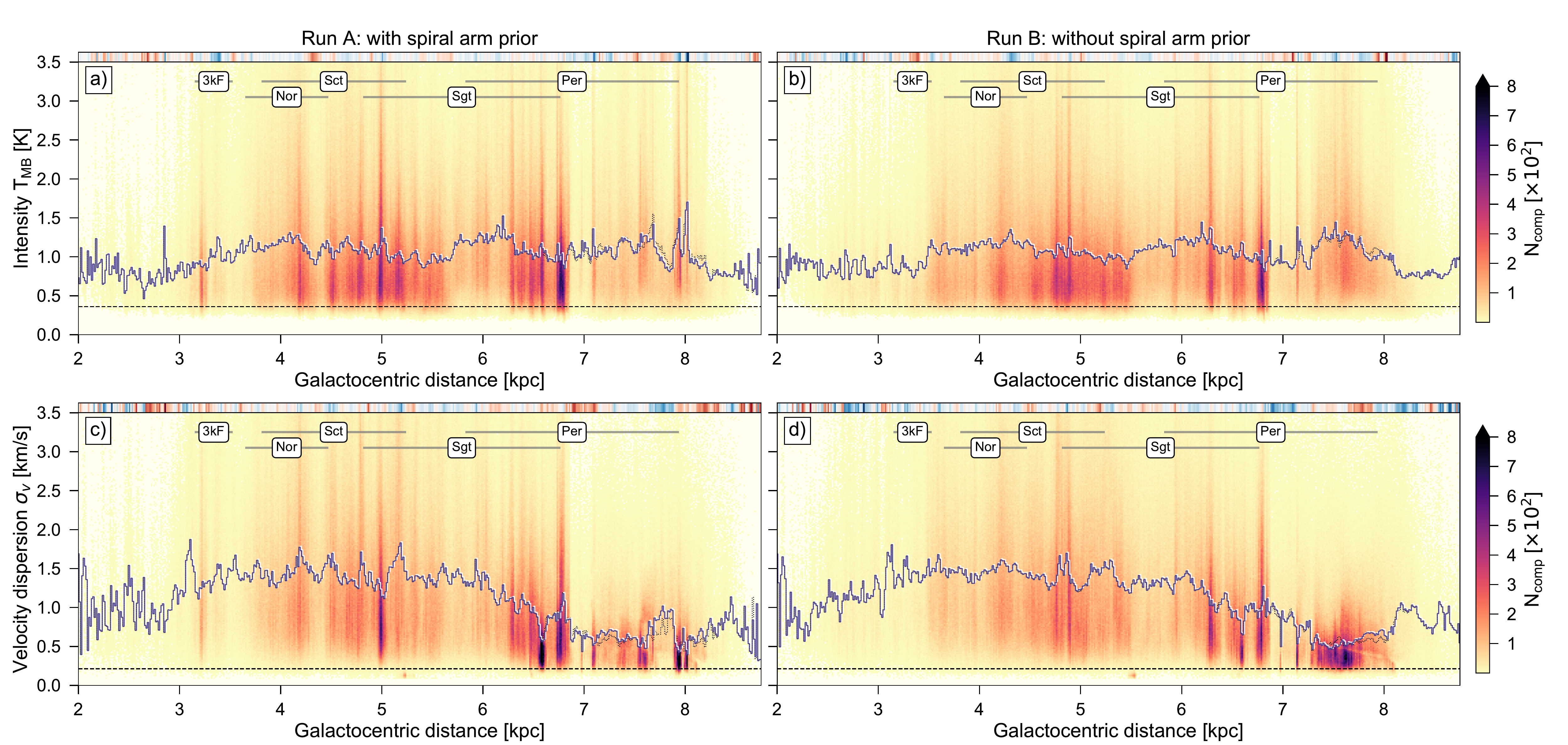}
    \caption[$2$D histograms of estimated Galactocentric distance values against intensity and velocity dispersion values]{$2$D histograms of estimated Galactocentric distance values and intensity (\textit{top}) and velocity dispersion (\textit{bottom}) values for the BDC run with (\textit{left column}) and without (\textit{right column}) the spiral arm prior.
    The blue lines show the respective median values per distance bin and dotted black lines give the corresponding values for distances obtained without the size-linewidth prior.
    The small strips at the top of the individual panels show where the median value is higher (blue) or lower (red) compared to the opposite BDC run, with the strength of the colour corresponding to the magnitude of the difference.
    The grey horizontal lines in all panels show the approximate $\rgal$ extent of five spiral arms overlapping with the GRS coverage.
    The dashed horizontal line in the top panels at $\Tb=0.36$~K corresponds to the $3\times$ S/N limit for the $0.1$st percentile of the GRS noise distribution \citep[see][]{Riener2020}.
    The dashed horizontal line in the bottom panels indicates the velocity resolution of the GRS ($0.21$~\kms).
    }
    \label{fig:hist2d_dgal}
\end{figure*}

We now focus on the distribution of intensity and velocity dispersion values of the $\co{13}{}$ fit components with Galactocentric distance (\fig\ref{fig:hist2d_dgal}).
These distributions also reveal some intriguing differences between the BDC runs. 
For example, for Run~A we can identify an accumulation of data points at the approximate $\rgal$ extent of the far portion of the $3$-kpc arm ($3$kF), which however is almost entirely missing in Run~B.
Indeed, a comparison with \fig\ref{fig:fov_inttot} confirms that Run~B puts significantly less emission at the location of the $3$kF arm than Run~A.
This is most likely due to very large non-circular motions near the Galactic bar that introduces errors and large uncertainties for Run~B, which depends mostly on the KD assumption of circular motions (see \sect\ref{sec:problems}). 
In addition, there is large uncertainty in the rotation velocity at small Galactocentric radii, which also contributes to increased uncertainty for KD estimates. 
Another striking difference occurs at an $\rgal$ value of $\sim 8$~kpc, where Run~A shows large peaks that are missing in Run~B.
This emission corresponds to the position of the nearby Aquila Rift complex, but in Run~B most of its emission is allocated to $\rgal$ distances of $\sim\! 7.5$~kpc.
We can confirm this in the top panels, where the accumulation of data points $<0.5$~kpc for Run~A is shifted to higher distances (between $0.5$ and $1$~kpc) in Run~B.

The intensity distribution (top panels in \fig\ref{fig:hist2d_dgal}) shows large variation but an almost constant median value with no significant trends, similar to \fig\ref{fig:hist2d_dhelio}.
The $\veldisp$ distributions (bottom panels \fig\ref{fig:hist2d_dgal}) show a more interesting behaviour; the median $\veldisp$ value stays at a large value of $\sim 1.5$~\kms\ from $3 \lesssim \rgal \lesssim 6$~kpc, after which it drops significantly to a value of $\sim 0.5$~\kms.
As mentioned before, this could indicate that in the inner Galaxy the $\co{13}{}$ components have higher non-thermal contributions or that there are increased problems in the decomposition of strongly blended emission in the inner parts of the GRS.
We can however also interpret this trend as yet another indication that most of the emission at $\rgal \gtrsim 6.5$~kpc is associated with regions close to the Sun and thus has better resolved emission lines (\sect\ref{sec:prior_linewidth}).

\subsection{Vertical distribution of the \texorpdfstring{\textsuperscript{13}}{13}CO emission}
\label{sec:height}

\begin{figure}
    \centering
    \includegraphics[width=\columnwidth]{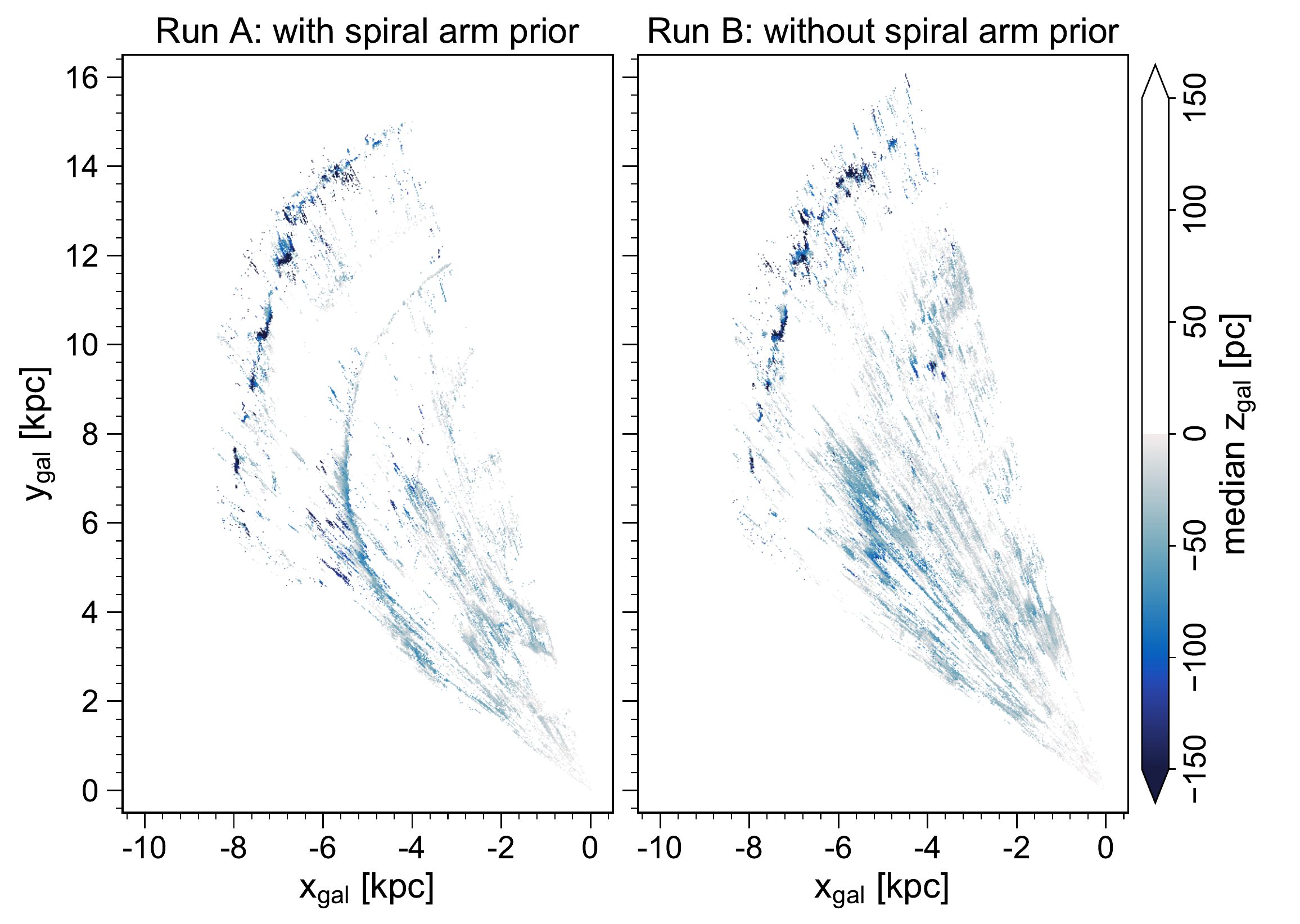}%
    \hspace{-\columnwidth}%
    \begin{ocg}{fig:orig_off_4}{fig:orig_off_4}{0}%
    \end{ocg}%
    \begin{ocg}{fig:orig_on_4}{fig:orig_on_4}{1}%
    \includegraphics[width=\columnwidth]{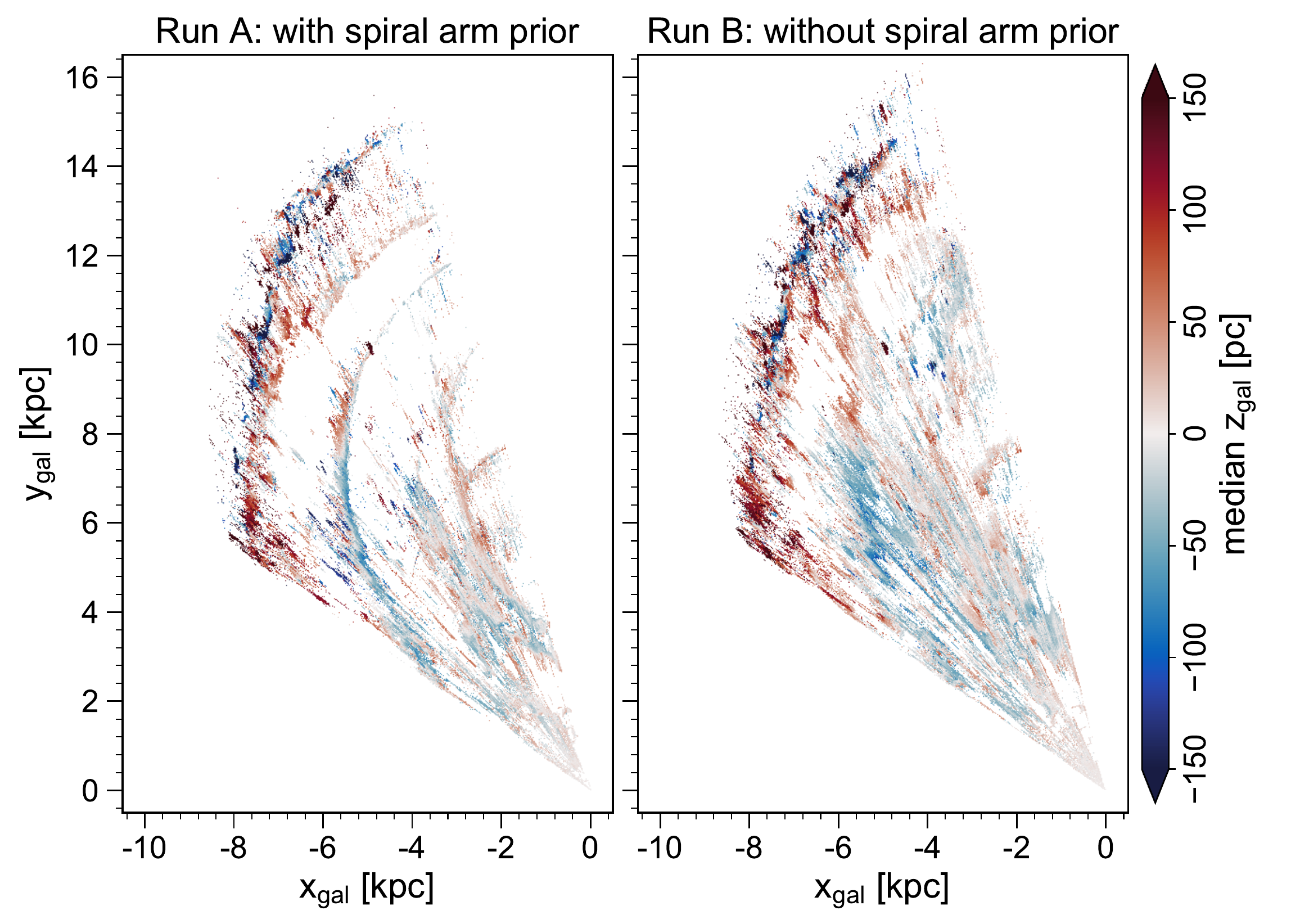}%
    \end{ocg}%
    \hspace{-\columnwidth}%
    \begin{ocg}{fig:grid_off_4}{fig:grid_off_4}{0}%
    \end{ocg}%
    \begin{ocg}{fig:grid_on_4}{fig:grid_on_4}{1}%
    \includegraphics[width=\columnwidth]{grs_fov_layer_grid.pdf}%
    \end{ocg}%
    \begin{ocg}{fig:arms_off_4}{fig:arms_off_4}{1}%
    \end{ocg}%
    \begin{ocg}{fig:arms_on_4}{fig:arms_on_4}{0}%
    \hspace{-\columnwidth}%
    \includegraphics[width=\columnwidth]{grs_fov_layer_spiral_arms_black.pdf}%
    \end{ocg}%
    \hspace{-\columnwidth}%
    \includegraphics[width=\columnwidth]{grs_fov_layer_sun+gc_black.pdf}%
    \caption[Face-on view of the median $\zgal$ values]{Face-on view of the median $\zgal$ values from the BDC results obtained with (\textit{left}) and without (\textit{right}) the spiral arm prior.
    The values are binned in $10\times10$~pc cells and the median was calculated along the $\zgal$ axis.
    The position of the Sun and Galactic centre are indicated by the Sun symbol and black dot, respectively.
     When displayed in Adobe Acrobat, it is possible to show \ToggleLayer{fig:orig_off_4,fig:orig_on_4}{\protect\cdbox{only the negative median $\zgal$ positions}}, show the \ToggleLayer{fig:arms_off_4,fig:arms_on_4}{\protect\cdbox{spiral arm positions}} and hide the \ToggleLayer{fig:grid_on_4,fig:grid_off_4}{\protect\cdbox{grid}}.
    }
    \label{fig:fov_zgal}
\end{figure}

The Galactic plane has long been known to show a warp towards positive $\zgal$ values in the first quadrant at Galactocentric distances $\rgal\gtrsim 7$~kpc \citep{Gum1960}\footnote{The BDC takes into account the effects of this warping in its calculation for the GL prior.}. %
In \fig\ref{fig:fov_zgal} we show a face-on view of the median $\zgal$ values of our estimated distances, which clearly shows this warp of the molecular gas disk at $\rgal\gtrsim 7$~kpc.
However, we can also see patches of negative $\zgal$ values ($< -100$~pc) for regions that coincide with the Perseus and Outer arms. 
A comparison with \fig\ref{fig:fov_veldisp} shows that these patches also correspond to the anomalously low $\veldisp$ values we already pointed out in \sect\ref{sec:faceon}.
This confirms our suspicion that these patches most likely correspond to gas emission that originates from very nearby regions that were erroneously assigned to large distances\footnote{We note that the presence of these incorrect distance assignments do not change our general conclusion about the warp of the Galactic disk towards positive $\zgal$ values.}.

Another conspicuous feature is the presence of substantial negative $\zgal$ values at the location of the Sagittarius arm at Galactic longitude values of $35\degr \lesssim \ell \lesssim 50\degr$ and $5~\text{kpc} \lesssim \rgal \lesssim 7$~kpc.
More quantitatively, the estimated vertical heights for gas emission for these $\ell$ and $\rgal$ ranges associated with the far portion of the Sagittarius arm have a median value of $\zgal = -34$~pc and span an IQR of $-58$ to $-8$~pc for both distance runs.
This bend towards negative $\zgal$ values at this longitude range is already clearly visible in the zeroth moment maps of the GRS data set (cf. Fig.~2 in \citealt{Riener2020}) and has also been observed in the \textit{Herschel} Hi-GAL survey \citep{Molinari2016}. 
Since this distortion seems to be mainly present in the diffuse ISM component of the Milky Way, \citet{Molinari2016} speculated that it might be due to interaction with gas flows that originate from the Galactic halo or the Galactic fountain. 
However, instead of a global phenomenon these negative $\zgal$ values could also simply indicate substructure of the Sagittarius arm.

\begin{figure}
    \centering
    \includegraphics[width=\columnwidth]{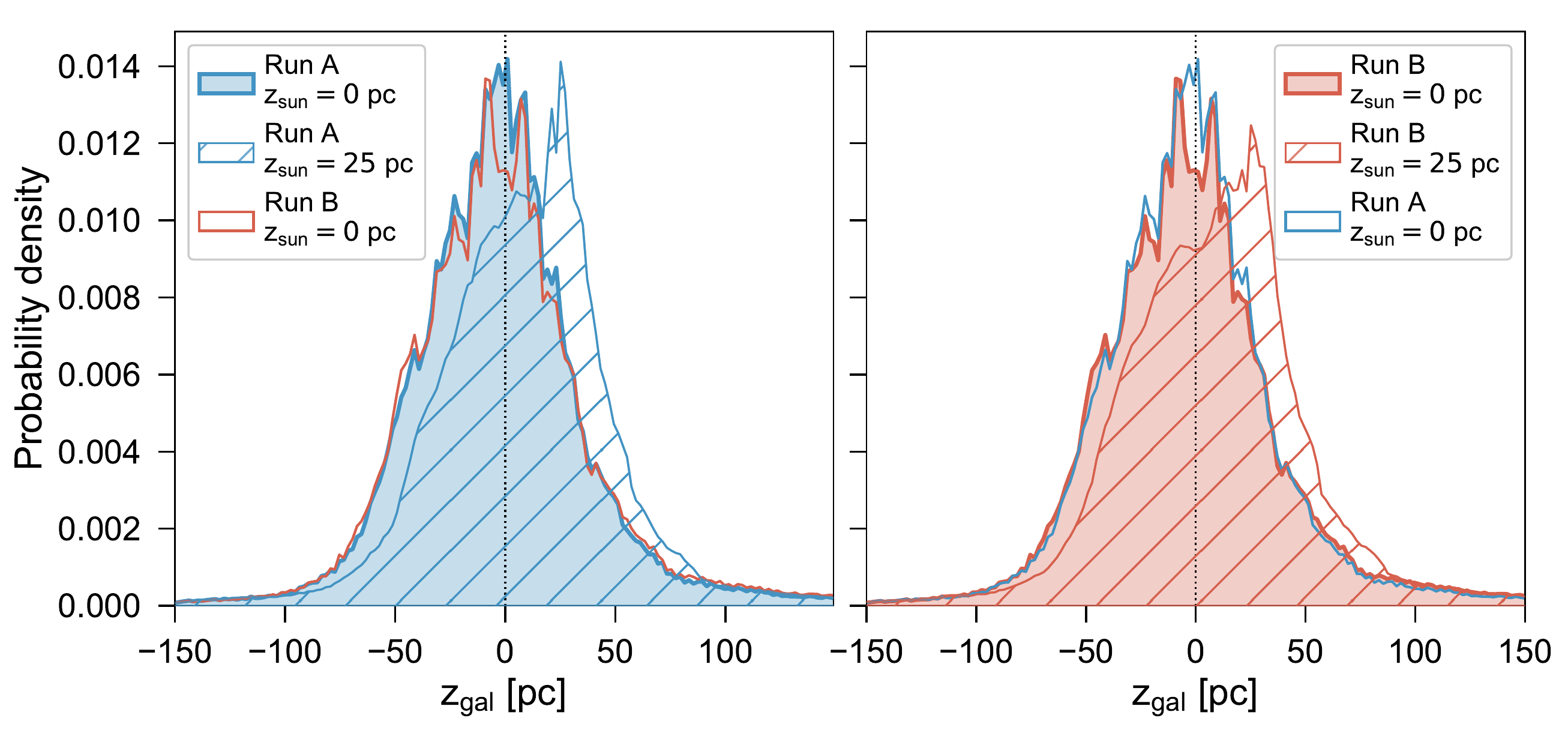}
    \caption[PDFs of the vertical distribution of the $\co{13}{}$ emission]{PDFs of the vertical distribution of the $\co{13}{}$ emission for the entire GRS data set.
    Shaded PDFs are for the BDC runs with (\textit{left}) and without (\textit{right}) the SA prior, and unfilled PDFs show the distribution of the opposite panel for reference. 
    Hatched PDFs show the $\zgal$ distribution assuming an offset of the Sun above the midplane of $\zoff = 25$~pc.
    }
    \label{fig:hist_zgal}
\end{figure}

In \fig\ref{fig:hist_zgal} we present PDFs for the estimated $\zgal$ values, which have very similar shapes in both BDC runs. 
The most notable difference is that Run~A shows a higher concentration at $\zgal = 0$, whereas Run~B shows a dip at this position. 
This difference is mostly due to the association of sources with the Aquila Rift complex in Run~A.

In our calculations we assumed that the Sun is located in the Galactic midplane, which is consistent with results from the most recent studies \citep{Anderson2019, Reid2019}. 
However, previous studies and observations found that the Sun has a vertical offset of $\zoff \sim 25$~pc from the IAU definition of the Galactic midplane \citep{Goodman2014nessie, BlandHawthorn2016}.
Figure~\ref{fig:hist_zgal} shows how the PDFs would change if we correct for this assumed vertical offset of the Sun using Eq.~C3 from \citet{EllsworthBowers2013}.
Accounting for such an offset leads to a shift of the distribution towards positive $\zgal$ values, with an asymmetric peak at $\zgal \sim 25$~pc introduced by emission originating close to the Sun ($\lesssim 1$~kpc).

A Gaussian fit to the PDFs in \fig\ref{fig:hist_zgal} yields full-width at half-maximum (FWHM) values of $72$ and $77$~pc with corresponding mean or peak positions at $\zgal = -5$ and $-6$~pc for Run~A and B, respectively. 
If a $\zoff$ value of $25$~pc is factored in, the FWHM values increase slightly to values of $78$ and $83$~pc and the centroid position changes to $\zgal = 10$ and $8$~pc, respectively.
Our FWHM estimate is lower by about one third than the value of $110$~pc \citet{RomanDuval2016} found for the dense gas (corresponding to H$_2$ surface densities $\gtrsim 25$~\msun\,pc$^{-2}$) in the inner Milky Way.
However, our results correspond very well with scale heights of $\sim 30$ to $40$~pc and peak values of $-4$ to $-10$~pc that have been determined from high mass star forming regions, \hii\ regions, and dust emission surveys in the far-infrared (see Table~$1$ in \citealt{Anderson2019} for a compilation of literature results).
A peak position at $\zgal = -5$~pc also agrees well with $\zoff \sim 5$~pc as found by \citet{Anderson2019} and \citet{Reid2019}.
We also note that the scale height of the \hi\ cold neutral medium ($\sim 150$~pc; \citealt{Kalberla2003}) significantly exceeds our determined scale height for the $\co{13}{}$ gas by about a factor of five.
To check whether our results are impacted by the inclusion of both near and far emission, we also estimated the FWHM estimates for individual $1$~kpc bins in the $\rgal$ range of $3 - 6$~kpc. 
We find a maximum FWHM extent of $\sim 90$~pc for $5 < \rgal < 6$~kpc and FWHM values of $70 - 75$~pc at lower $\rgal$ bins. 

\begin{figure}
    \centering
    \includegraphics[width=\columnwidth]{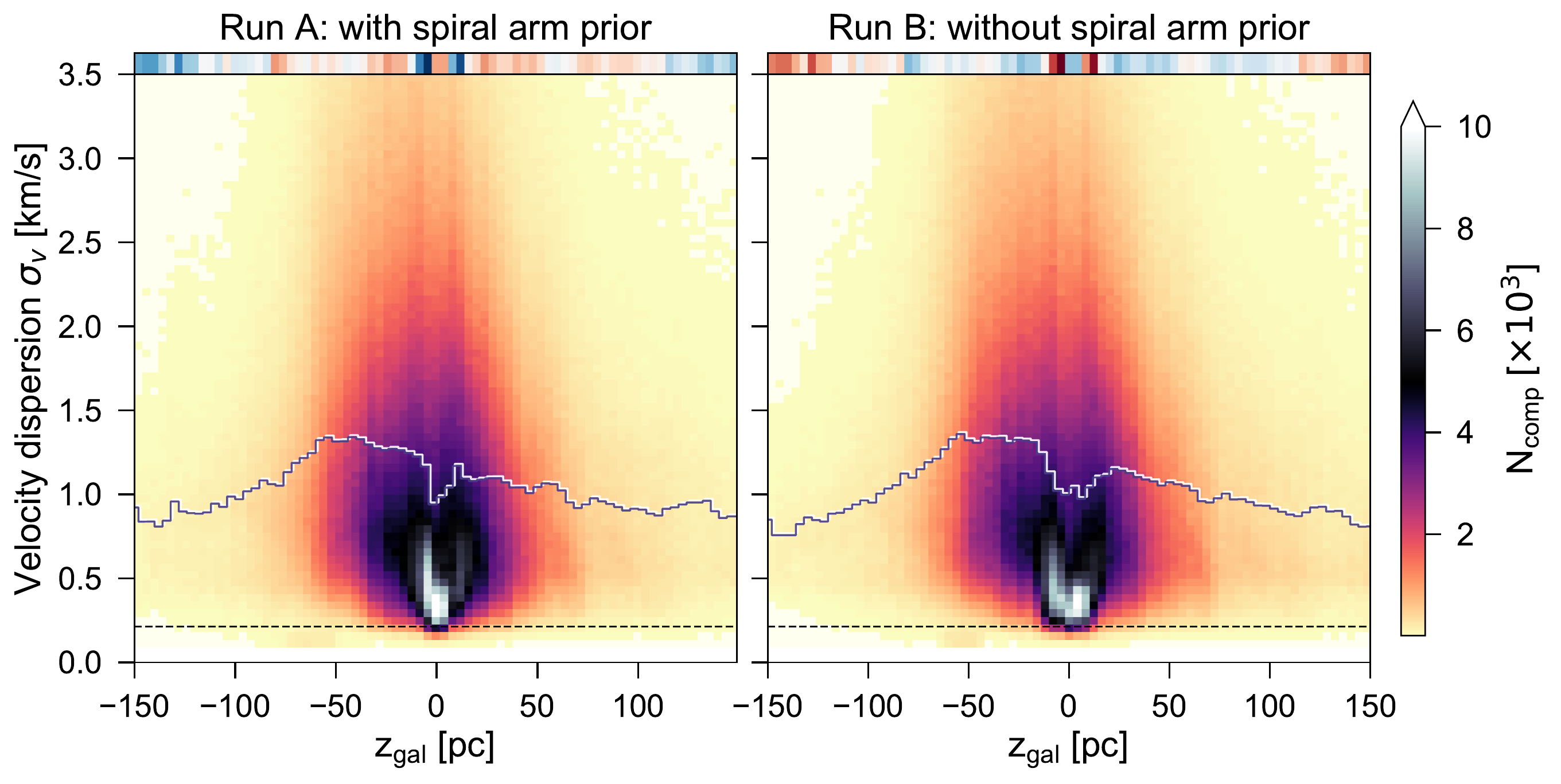}
    \caption[$2$D histograms of velocity dispersion and estimated vertical distances $\zgal$]{$2$D histograms of velocity dispersion and estimated vertical distances $\zgal$ for the BDC run with (\textit{left}) and without (\textit{right}) the spiral arm prior.
    The blue line shows the median $\veldisp$ value per $\zgal$ bin.
    The small strips at the top of the individual panels show where the median value is higher (blue) or lower (red) compared to the opposite BDC run, with the strength of the colour corresponding to the magnitude of the difference.
    The dashed horizontal line indicates the GRS velocity resolution ($0.21$~\kms).
    }
    \label{fig:hist2d_zgal_veldisp}
\end{figure}

Figure~\ref{fig:hist2d_zgal_veldisp} shows the distribution of $\veldisp$ values with vertical height $\zgal$.
For both distance results we can see a clear concentration of data points towards the midplane.
The decrease in the median $\veldisp$ value around a $\zgal$ value of $0$ is due to very nearby emission located $< 1$~kpc from the Sun, which has very narrow linewidths.
We also note the presence of an asymmetry, especially striking in the curve of median values, with a larger fraction of components with broader linewidths located at negative $\zgal$ values. 
\citet{Riener2020} already found a similar asymmetry in the distribution of $\veldisp$ values with Galactic latitude.
As argued in \citet{Riener2020}, such an asymmetry could be explained by an offset position of the Sun above the Galactic midplane.
However, as mentioned, recent results have found that the vertical position of the Sun agrees well with the location of the Galactic midplane \citep{Anderson2019, Reid2019}.

\subsection{Potential problems, artefacts, and biases}
\label{sec:problems}

The BDC tool was designed to estimate distances for spiral arm sources, which means that its default settings have an inherent bias of associating sources with Galactic features from its spiral arm model.
Since we use the BDC in assigning distances to the gas emission of an entire Galactic plane survey, we need to be careful in interpreting its results and should be aware of the biases present in the distance calculation. 

\begin{figure}
    \centering
    \includegraphics[width=\columnwidth]{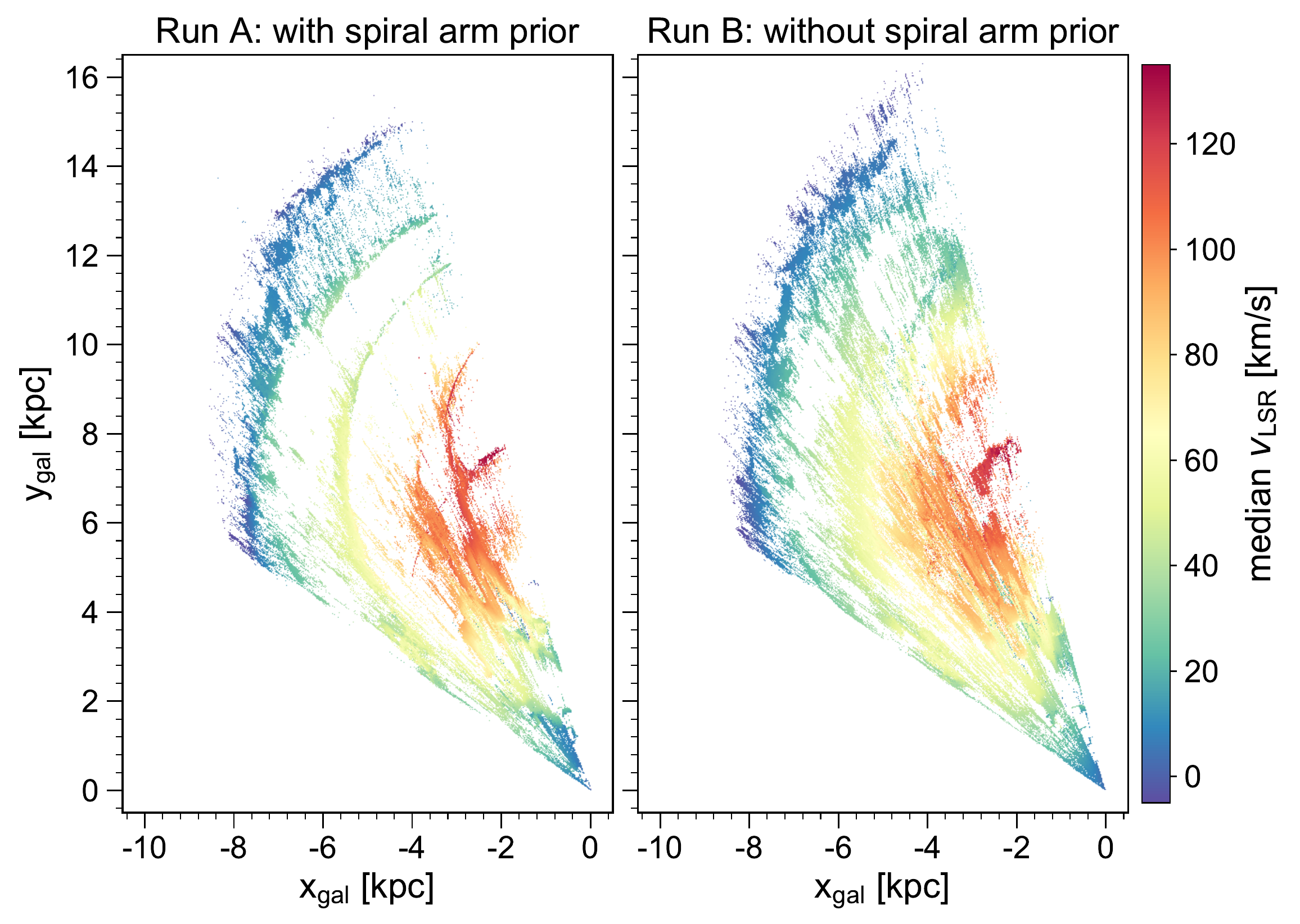}%
    \hspace{-\columnwidth}%
    \begin{ocg}{fig:grid_off_7}{fig:grid_off_7}{0}%
    \end{ocg}%
    \begin{ocg}{fig:grid_on_7}{fig:grid_on_7}{1}%
    \includegraphics[width=\columnwidth]{grs_fov_layer_grid.pdf}%
    \end{ocg}%
    \begin{ocg}{fig:arms_off_7}{fig:arms_off_7}{1}%
    \end{ocg}%
    \begin{ocg}{fig:arms_on_7}{fig:arms_on_7}{0}%
    \hspace{-\columnwidth}%
    \includegraphics[width=\columnwidth]{grs_fov_layer_spiral_arms_black.pdf}%
    \end{ocg}%
    \hspace{-\columnwidth}%
    \begin{ocg}{fig:vlsr_off_7}{fig:vlsr_off_7}{0}%
    \end{ocg}%
    \begin{ocg}{fig:vlsr_on_7}{fig:vlsr_on_7}{1}%
    \includegraphics[width=\columnwidth]{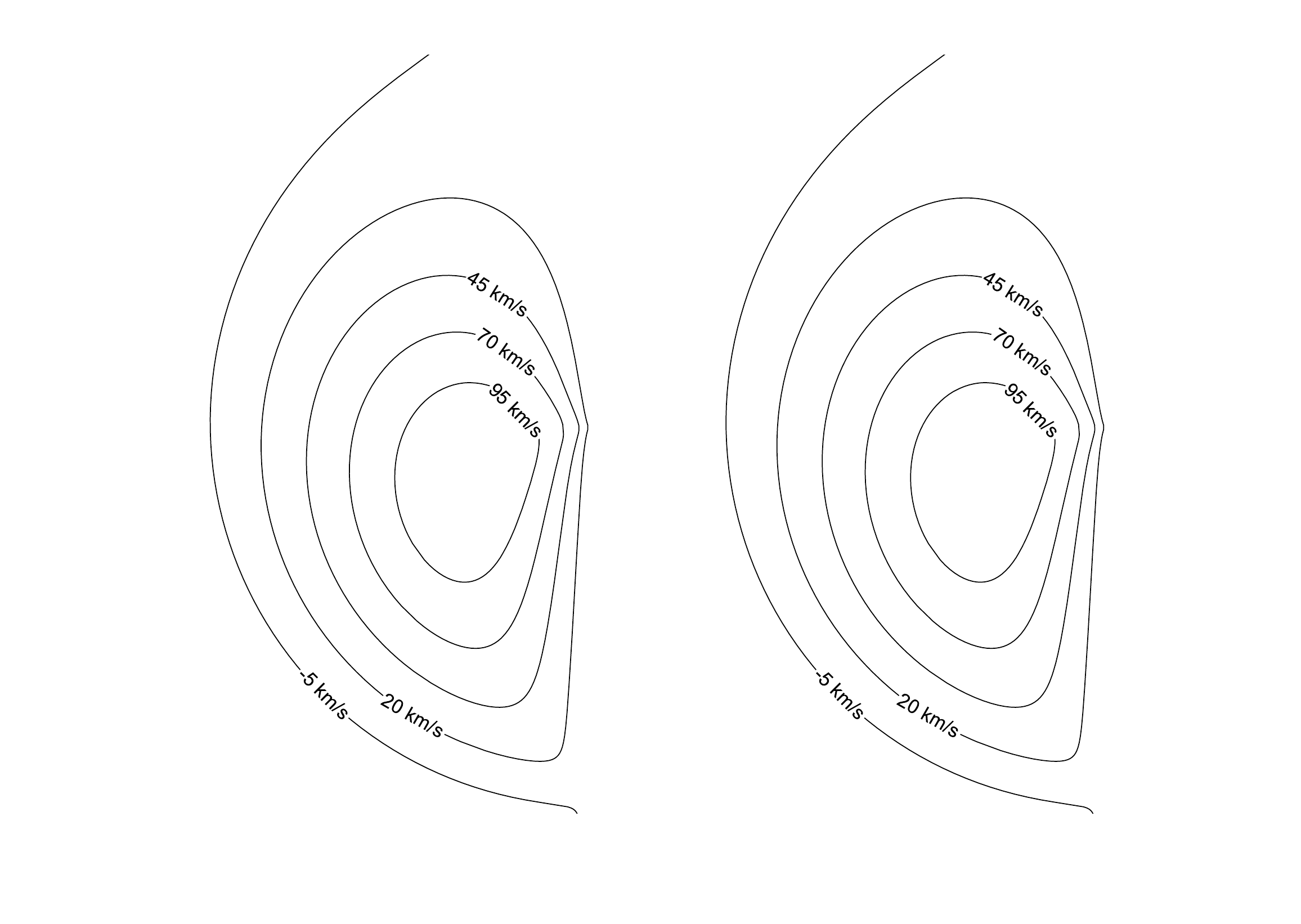}%
    \end{ocg}%
    \hspace{-\columnwidth}%
    \includegraphics[width=\columnwidth]{grs_fov_layer_sun+gc_black.pdf}%
    \caption[Face-on view of the median $\vlsr$ values]{Face-on view of the median $\vlsr$ values from the BDC results obtained with (\textit{left}) and without (\textit{right}) the spiral arm prior.
    The values are binned in $10\times10$~pc cells and the median was calculated along the $\zgal$ axis.
    The position of the Sun and Galactic centre are indicated by the Sun symbol and black dot, respectively.
     When displayed in Adobe Acrobat, it is possible to show the \ToggleLayer{fig:arms_off_7,fig:arms_on_7}{\protect\cdbox{spiral arm positions}}, hide the \ToggleLayer{fig:vlsr_on_7,fig:vlsr_off_7}{\protect\cdbox{curves of constant projected $\vlsr$}}, and hide the \ToggleLayer{fig:grid_on_7,fig:grid_off_7}{\protect\cdbox{grid}}.
    }
    \label{fig:fov_vlsr}
\end{figure}

\begin{figure}
    \centering
	\includegraphics[width=\columnwidth]{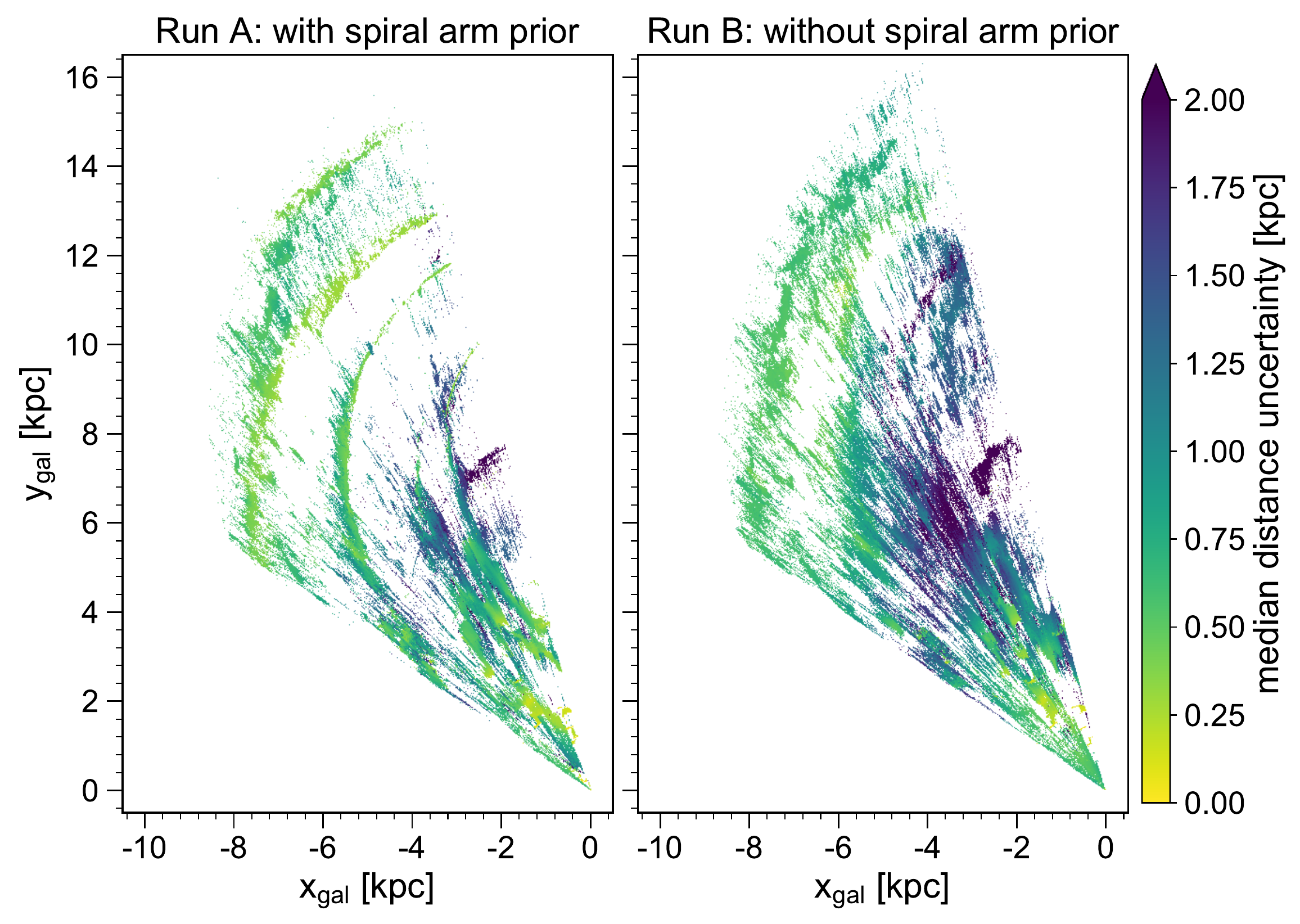}%
    \hspace{-\columnwidth}%
    \begin{ocg}{fig:grid_off_5}{fig:grid_off_5}{0}%
    \end{ocg}%
    \begin{ocg}{fig:grid_on_5}{fig:grid_on_5}{1}%
    \includegraphics[width=\columnwidth]{grs_fov_layer_grid.pdf}%
    \end{ocg}%
    \begin{ocg}{fig:arms_off_5}{fig:arms_off_5}{1}%
    \end{ocg}%
    \begin{ocg}{fig:arms_on_5}{fig:arms_on_5}{0}%
    \hspace{-\columnwidth}%
    \includegraphics[width=\columnwidth]{grs_fov_layer_spiral_arms_black.pdf}%
    \end{ocg}%
    \hspace{-\columnwidth}%
    \includegraphics[width=\columnwidth]{grs_fov_layer_sun+gc_black.pdf}%
    \caption[Face-on view of the median distance uncertainty]{Face-on view of the median distance uncertainty for the BDC results obtained with (\textit{left}) and without (\textit{right}) the spiral arm prior.
    The values are binned in $10\times10$~pc cells and are summed up along the $\zgal$ axis.
    The position of the Sun and Galactic centre are indicated by the Sun symbol and black dot, respectively.
     When displayed in Adobe Acrobat, it is possible to show the \ToggleLayer{fig:arms_off_5,fig:arms_on_5}{\protect\cdbox{spiral arm positions}} and hide the \ToggleLayer{fig:grid_on_5,fig:grid_off_5}{\protect\cdbox{grid}}.
    }
    \label{fig:fov_edist}
\end{figure}

It is a priori not clear which of our BDC runs yields more trustworthy or better distance solutions.
Run~A has the obvious problem that the gas emission will be preferentially located closer to the Galactic features included in the spiral arm model.
For this run we expect biased results in terms of the distribution of emission in spiral arm and interarm regions, with the latter likely severely underestimated.
Run~B gives more unbiased results with regards to the allocation of the gas to arm and interarm regions. 
However, we note that for the distance results from Run~B an association with maser parallax sources can be a decisive factor for the choice of the most likely distance (cf. left panel of \fig\ref{fig:bdc_examples}).
Since these maser sources do mostly overlap with the Galactic features of the spiral arm model (\fig\ref{fig:schematic_spiral_arms+masers}), the distance results thus still contain an implicit, albeit moderate, association with these Galactic features.
Moreover, since Run~B is dominated by the KD prior, it is also more strongly affected by the ambiguities and uncertainties of the KD method. 

\begin{figure}
    \centering
    \includegraphics[width=\columnwidth]{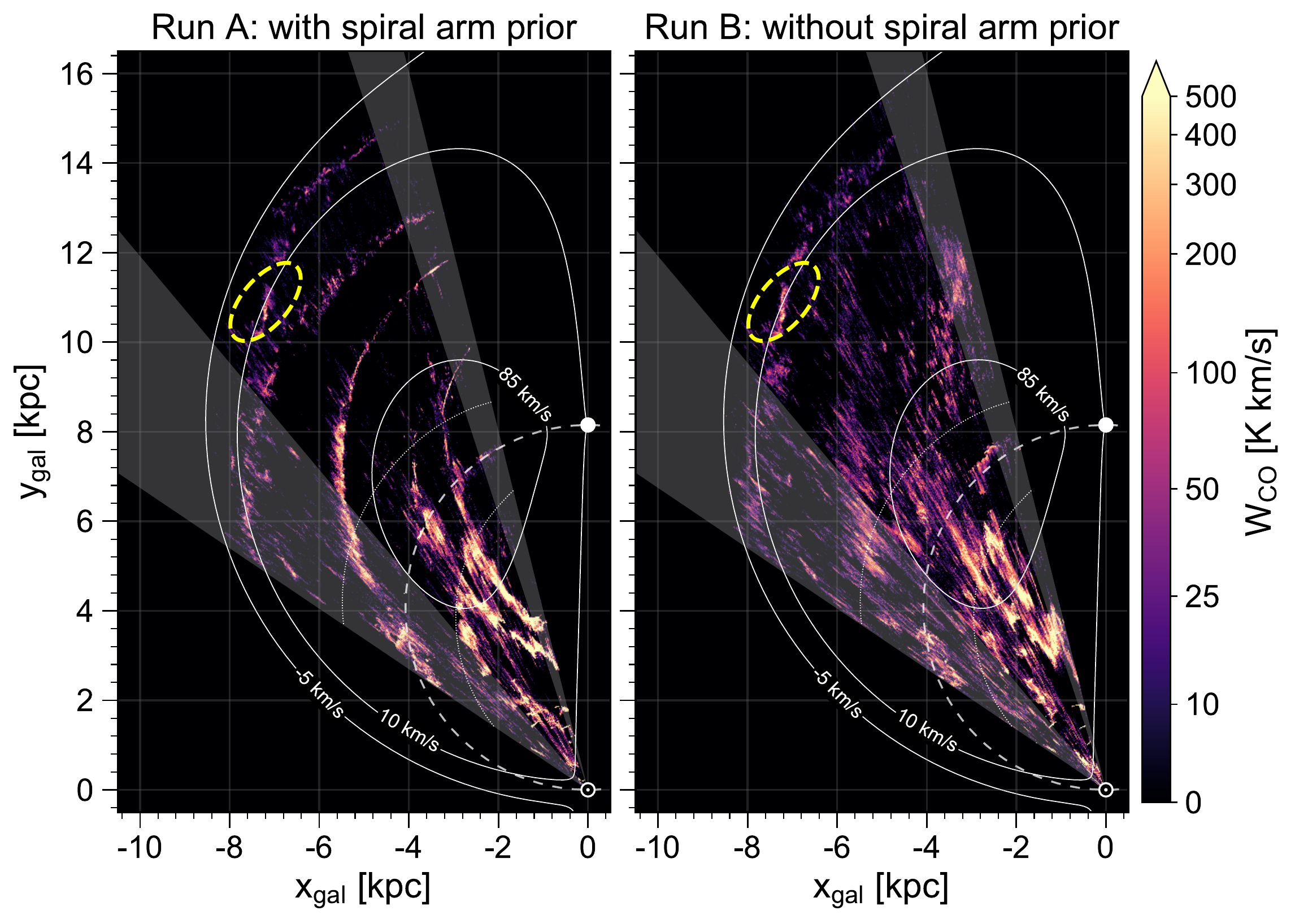}
    \caption[Face-on view with diagnostics to illustrate potential issues in the distance assignment]{Same as \fig\ref{fig:fov_inttot}, but overplotted with diagnostics to illustrate potential issues in the distance assignment.
    The position of the Sun and the Galactic centre are indicated with a Sun symbol and a white dot, respectively.
    Solid white curves show constant projected $\vlsr$ velocities.
  	The dashed white curve marks the locus of tangent points and dotted lines indicate the area where $\vlsr$ values are within $10$~\kms\ of the tangent point velocity.
  	Shaded areas show GRS regions that either had limited latitude or velocity coverage.
  	The yellow dashed ellipse indicates a likely artefact of nearby emission that was erroneously placed at far distances.
  	See \sect\ref{sec:problems} for more details.
    }
    \label{fig:survey_limits}
\end{figure}

We can identify an accumulation of emission features around the locus of tangent points for both distance estimates.
The problem with determining kinematic distances near tangent points is that small changes in the $\vlsr$ value result in large changes in the estimated KD value.
Thus often a threshold is used for the tangent point distance allocation, where for example all sources with $\vlsr$ values within $10$~\kms\ of the tangent point velocity are assigned the tangent point distance \citep[e.g.][]{Urquhart2018}.
We indicate the corresponding region where $\vlsr$ values are within $10$~\kms\ of the tangent point velocity in \fig\ref{fig:survey_limits}.
This threshold of $10$~\kms\ corresponds to expected velocity deviations introduced by streaming motions \citep[e.g.][]{Burton1971, Ramon-Fox2018}.
We thus speculate that some of the empty voids within this region might be at least partly due to this confusion around the tangent point velocity, at least for Run~B that is dominated by the KD prior.
However, a comparison with the default BDC runs (\fig\ref{fig:bdc_default_inttot}) shows that our use of literature distance solutions helped to substantially decrease artefacts around the locus of tangent points.

Figure~\ref{fig:fov_vlsr} shows a comparison of the median associated $\vlsr$ values for the distance results.
We overplot this figure with curves of constant projected $\vlsr$ values that were calculated with the methods included in the BDC.
In general, the BDC runs produced distance results that are well in agreement with the assumed Galactic rotation curve model.
This good correspondence is not surprising, given that the KD solutions are calculated using the rotation curve model.
However, an anticipated problem is that peculiar gas motions, for example introduced by streaming motions within spiral arms, might cause a significant deviation from the expected $\vlsr$ velocities of the assumed rotation curve model \citep[e.g.][]{Ramon-Fox2018}.
This strongly affects regions with $\rgal$ values $\lesssim 5$~kpc, for which we expect large peculiar motions due to the influence of the Galactic bar \citep{Reid2019}.
The BDC takes this into account by down-weighting the KD prior for regions closer to the Galactic centre (see \fig\ref{fig:bdc_examples}); this however has a significant impact on the estimated distance uncertainties, which show a substantial increase with decreasing $\rgal$ values (\fig\ref{fig:fov_edist}).
As already noticed in \fig\ref{fig:hist_dist_edist}, this effect is much stronger for Run~B, since in this case the combined PDF shows broader peaks and corresponding Gaussian fits to these peaks result in higher estimated distance uncertainties. 
We give a more detailed discussion about the deviations from the rotation curve velocities and the regions where they occur in Appendix~\ref{sec:vlsr_deviation}.
We also note that the $\vlsr$ uncertainties of the fit components can have a large impact on the distance calculation routine, as larger uncertainty values can lead to an association with more parallax sources or Galactic features.
We illustrate and discuss this effect with an example in Appendix~\ref{sec:effect_evel}.
Moreover, the BDC is expected to have problems at lower Galactic longitudes for both Run~A and Run~B, since for low longitude ranges emission in the various arm segments along the lines of sight will have similar Galactic latitude and $\vlsr$ values, thus complicating the distance allocation.

Since in this work we do not explicitly correlate the distance results of neighbouring lines of sight, it is possible that assigned distance values can show strong variation between neighbouring lines of sight.
We can see this effect as emission features that are spread out along the line of sight (reminiscent of the `Fingers of God effect'; see e.g. the right panel of \fig\ref{fig:survey_limits}).

The limited spatial and spectral coverage of the GRS also introduces some artefacts in the distance estimation (\fig\ref{fig:survey_limits}).
The limited latitude coverage from $14\degr \lesssim \ell \lesssim 18\degr$  results in missing patches of CO emission at this longitude range.
Moreover, the face-on view is restricted by the lower limit of the velocity coverage ($\vlsr = -5$~\kms) and thus contains no emission past d$\sim\! 13$ kpc at $\ell > 40\degr$.
However, the reduced velocity coverage of $-5 \lesssim \vlsr \lesssim 85$~\kms\ for $40\degr \lesssim \ell \lesssim 56\degr$ should not impact our distance results as we would not expect emission peaks with $\vlsr > 85$~\kms\ at these larger longitude values, as demonstrated in \fig\ref{fig:survey_limits}. 

We note that the GRS decomposition also has uncertainties that might cause problems for the distance estimation.
Especially in the inner part of the Galaxy emission lines can be strongly blended, which could have led to difficulties in the decomposition (see the discussion about flagged components in Sect.~3.1 of \citealt{Riener2020}).
We tried to fold these considerations into the uncertainties of the $\vlsr$ position supplied to the BDC (see discussion in Appendix~\ref{sec:effect_evel}).
Moreover, we tested the effects of a quality cut based on the S/N ratio of the fit components and found that this does not change our overall conclusions (see \sect\ref{sec:distresults}).

As already mentioned in previous sections, the feature that seemingly bridges the Perseus and Outer arm and corresponds to emission with $\vlsr$ values around $10$~\kms\ (yellow dashed ellipse in \fig\ref{fig:survey_limits}) is most likely an artefact introduced by the KD prior.
As discussed in \citet{Riener2020} and \sect\ref{sec:prior_linewidth}, for this $\vlsr$ regime there is strong confusion between emission from the solar neighbourhood and the far disk.
Moreover, the BDC biases emission lines with $\vlsr \lesssim 5$~\kms\ towards far distances (cf. \fig\ref{fig:effect_lowvlsr}).
We can estimate the magnitude of this error by counting all components with an unlikely combination of distance, $\zgal$, and $\veldisp$ values.
Choosing d $> 8$~kpc, $\zgal < -50$~pc, and $\veldisp < 0.5$~\kms\ as unlikely combination of parameters, we find that for both distance results about $0.3\%$ of the fit components satisfy these parameters%
\footnote{If we had not used the size-linewidth prior this fraction would increase to about $0.6$ and $0.9\%$ of the fit components for the runs with and without the SA prior, respectively.}. 
Since the $\wco$ values of these components account for only about $0.1\%$ of the total emission, we conclude that this problem has a very minor impact on our overall conclusions concerning the fraction of emission located in arm and interarm regions.
However, this issue has strong impacts on our conclusions about the $\veldisp$ distribution in the Perseus and Outer arm (see \sect\ref{sec:var_arms}).

\subsection{Comparison with previous results}
\label{sec:comp_prev}

The BDC has already been used in the distance estimation to clouds and clumps extracted from other Galactic plane surveys \citep[][]{Urquhart2018, Rigby2019}.
However, since these works used an older version of the BDC (v$1.0$), it is not straightforward to compare their distance results with the distances obtained in this work.
Moreover, our distance estimations are not independent from these previous results, since we use them as input for our KDA prior.
Notwithstanding these issues, in Appendix~\ref{sec:test_kda_priors} we discuss how well we are able to match these previous distance results and conclude that we recover the vast majority of literature distance results within the degree of expected uncertainties introduced by the updated rotation curve parameters from BDC v$1$ to v$2.4$.

Unfortunately, it is also extremely challenging to compare our results in terms of the spiral arm to interarm fraction and variation of physical properties with previous studies, given that other works used a combination of different tracers, different spiral arm models, or different assumptions about the Galactic rotation curve and the distance to the Galactic centre. 
It would be necessary to homogenise all data sets before any attempted comparison, which potentially requires recalculating and updating the literature distance results with our assumed Galactic parameters.
Since such a homogenisation exceeds the scope of this work, we decided on a strictly qualitative comparison with some of the previous results and do not attempt to account for any of these systematic differences.

Previous works analysing the GRS found a similar overdensity of $\co{13}{}$ with spiral arm features.
\citet{RomanDuval2009grs} found for their sample of GRS molecular clouds that the $\co{13}{}$ surface brightness is strongly enhanced at the location of spiral arms from the model of \citet{Vallee1995}.
\citet{Sawada2012} also found that the GRS emission shows bright and compact concentrations along spiral arm features, whereas more diffuse and extended emission dominates the interarm regions.

However, recent results from other Galactic plane surveys in the first quadrant found a weaker correspondence of molecular clouds with spiral arms.
\citet{Colombo2019} analysed a large $\co{12}{}\transition{3}{2}$ survey overlapping with the GRS and could only attribute about $35\%$ of the flux to molecular clouds associated with spiral arms\footnote{\citet{Colombo2019} used the spiral arm model by \citet{Vallee2017} as a comparison, in which positions for the Scutum and Sagittarius arms deviate by up to $\sim 1$~kpc compared to the corresponding arms defined by \citet{Reid2019}}.
\citet{Colombo2019} attribute this low fraction of flux in spiral arm clouds to difficulties in the distance assignments and optical depth effects of the $\co{12}{}\transition{3}{2}$ emission.

Recently, \citet{Rigby2019} found that clumps from a distance limited ($6 < d_{\odot} < 9$~kpc) sample associated with spiral arms have significantly higher $\veldisp$ values than clumps at the same distances that are located in interarm regions.
They further note that this difference in linewidth is comparable to what has been found in extragalactic work \citep{Colombo2014} and smoothed particle-hydrodynamics simulations \citep{Duarte-Cabral2016}.
We do find a trend for lower $\veldisp$ values at around the same distances for interarm regions for Run~A, but not for Run~B (\fig\ref{fig:hist2d_dhelio}).
This is somewhat surprising, given that \citet{Rigby2019} used the BDC without the SA prior for their distance calculation, which should have yielded a better agreement with our Run~B.
However, we note that \citet{Rigby2019} used a higher-density tracer ($\co{13}{}\transition{3}{2}$) and v$1$ of the BDC\footnote{BDC v$2.4$ includes new maser parallax sources, updated models for the Galactic rotation curve and spiral arm features, and contains significant changes in the distance estimation, such as a down-weighting of the KD prior in the inner Galaxy to accommodate expected large streaming motions introduced by the Galactic bar. 
See \citet{Reid2019} for more information.}, which could both account for any differences compared to our results.


\section{Conclusions}

In this work we present distance estimates for the Gaussian decomposition results of the Galactic Ring Survey presented in \citet{Riener2020}.
Using the most recent version of the Bayesian Distance Calculator tool \citet{Reid2016, Reid2019}, we perform two separate distance calculations for the $\sim\! 4.6$~million individual Gaussian fit components, for which we vary the settings so as to either incorporate or neglect a prior for an association with spiral arm structure (labelled Run~A and Run~B, respectively). 
In addition, we include literature distance information of objects overlapping with the GRS coverage as prior information for solving the kinematic distance ambiguity.
We also incorporate a size-linewidth prior to solve for the confusion between emission from the solar neighbourhood and the far Galactic disk for emission peaks with line centroids of $\vlsr < 20$~\kms.

We find that most of the distance results of the two BDC runs are consistent with each other within their uncertainties, with most of the differences either due to the strong influence of the spiral arm prior for Run~A or larger uncertainties introduced by the stronger effect of the kinematic distance prior for Run~B.
The two distance runs complement each other and show opposing strengths and weaknesses, thus suggesting that the true distribution of the gas emission is closer to a combination of the two results than to each of the individual distance runs. 

In the following we present our main findings based on these two distance results:

\begin{enumerate}[i)]
	\item The majority of the $\co{13}{}$ emission is associated with spiral arm features as defined in the model by \citet{Reid2019}.
	The fraction of $\co{13}{}$ emission located in interarm regions varies from $16\%$ to $24\%$ in terms of the total $\co{13}{}$ integrated emission and $24\%$ to $34\%$ in terms of the total number of $\co{13}{}$ velocity components.
	
	\item The vertical distribution of the gas emission has a FWHM extent of $\sim 75$~pc. 
	We recover a significant warp of the molecular disk towards positive $\zgal$ values of more than $100$~pc for the far side of the disk at $\rgal > 7$~kpc and the entire covered longitude range of $14\degr < \ell < 56\degr$.
	The gas disk shows a significant bend towards negative $\zgal$ values at the position of the Sagittarius arm at Galactic longitude values of $35\degr \lesssim \ell \lesssim 50\degr$ and $5~\text{kpc} \lesssim \rgal \lesssim 7$~kpc.
	
	\item We find a trend of higher velocity dispersion values with increasing heliocentric distance, which we attribute mostly to beam averaging effects. 
	Most of the velocity dispersion values also significantly exceed expected values based on an assumed size-linewidth relationship.  
	
	\item The $\co{13}{}$ emission associated with spiral arms and spur features has a similar distribution of velocity dispersion values, which is shifted to higher values compared to the distribution of velocity dispersion values in interarm structures.
	However, we find that most of this difference is due to the location of a significant fraction of interarm gas at close distances to the Sun, which resulted in spatially better resolved lines and narrower linewidths.
	While we cannot exclude variations in the linewidth between spiral arm and interarm gas, we conclude that our present results do not support strong differences in $\veldisp$ between these environments.
	
	\item There is strong confusion between $\co{13}{}$ emission coming from the local solar neighbourhood and regions associated with the Perseus and Outer arm.
	By using the velocity dispersion values of the fit components as an additional prior we could significantly reduce the confusion between near and far emission for low $\vlsr$ velocities ($-5 < \vlsr < 20$~\kms).
	
\end{enumerate}

\noindent While we use the currently best knowledge about the structure of our Galaxy for our distance results, we anticipate that these will be subject to change, in particular due to updates on the BDC method, the Galactic rotation model, and the position of Galactic features, with additional and more precise maser parallax measurements, and new KDA solutions for sources overlapping with the GRS coverage.
The BDC tool and its enhancements discussed in this work are designed to be versatile enough to incorporate these changes.
We thus conclude that that the approach presented herein should be a helpful contribution to the problem of estimating distances to gas emission features from Galactic plane surveys.

\begin{acknowledgements}
    This project received funding from the European Union’s Horizon 2020 research and innovation program under grant agreement No 639459 (PROMISE). 
    HB acknowledge support from the EuropeanResearch Council under the Horizon 2020 Framework Program via theERC Consolidator Grant CSF-648505.
    HB also acknowledges support from the Deutsche Forschungsgemeinschaft via SFB 881, “The Milky Way System” (sub-projects B1).
    This publication makes use of molecular line data from the Boston University-FCRAO Galactic Ring Survey (GRS). 
    The GRS is a joint project of Boston University and Five College Radio Astronomy Observatory, funded by the National Science Foundation under grants AST-9800334, AST-0098562, \& AST-0100793.   
    We would like to thank the referee, Mark Reid, for constructive and useful suggestions that helped to improve the paper.
    We would moreover like to thank Mark Reid for sharing an advance copy of the BDC v2.4 with us.
    We would also like to thank James Urquhart for sharing the SEDIGISM linewidth measurements for the ATLASGAL sample with us.
      \\\textbf{Code bibliography}:
      This research made use of \textsc{matplotlib} \citep{Hunter2007}, a suite of open-source python modules that provides a framework for creating scientific plots; \textsc{astropy}, a community-developed core Python package for Astronomy \citep{astropy}; \textsc{aplpy}, an open-source plotting package for Python \citep{APLpy}; and \textsc{seaborn} \citep{seaborn}.
\end{acknowledgements}

\bibliographystyle{aa} 
\bibliography{bibliography} 

\begin{appendix} 
\twocolumn

\section{Further details about the KDA prior}
\label{sec:kda}

Here we give more details about the KDA prior (\sect\ref{sec:prior_kda}). 
We describe the method we used to calculate the prior, present the literature results we considered for the prior, and discuss its effect on the distance calculation. 

\subsection{Calculation of priors}
\label{app:calc_priors}

\begin{figure}
    \centering
    \includegraphics[width=\columnwidth]{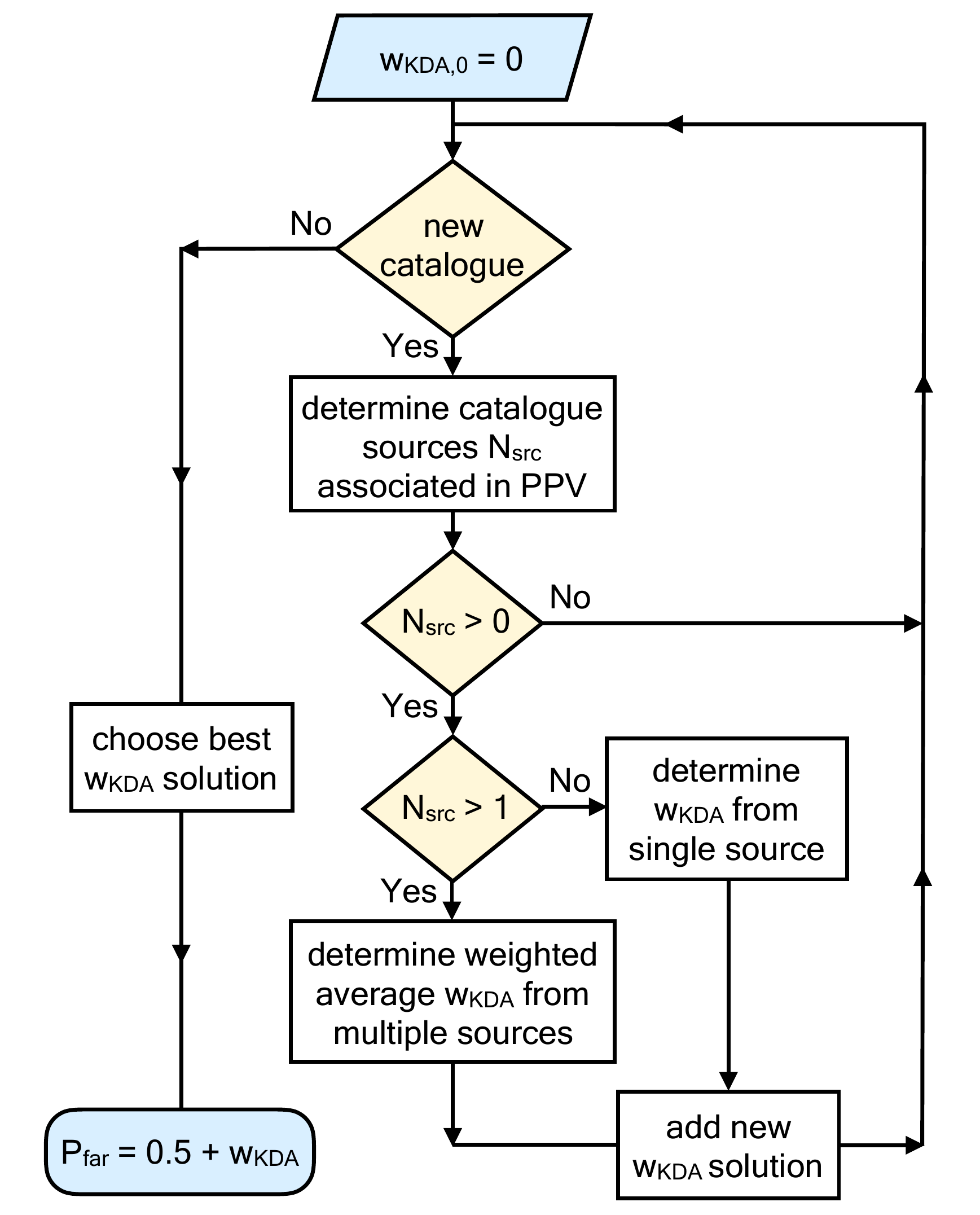}
    \caption[Flowchart outlining the use of literature distances for the KDA prior]{Flowchart outlining how literature distance estimates are used to determine a prior for the KDA solution.}
    \label{fig:flowchart_kda_solution}
\end{figure}

In the following we describe the iterative loop that is used to determine priors for the KDA solution (see also \fig\ref{fig:flowchart_kda_solution}).
For a given ($\ell$, $b$, $\vlsr$) coordinate, we first determine how many sources $\Nsrc$ from a catalogue are associated with this coordinate.
For this we define a Gaussian weight using the definition of a Gaussian function:
\begin{equation}
    f(x) = a \cdot \text{exp}
    \left(-4 \cdot \text{ln}(2) \cdot \dfrac{(x - \mu)^2}{\Theta^2}\right),
\end{equation}

\noindent where $a$, $\mu$, and $\Theta$ denote the amplitude, mean position, and FWHM values, respectively.
From this definition we construct a Gaussian weight $\weight{g}{}$ that evaluates to unity at $x = \Theta / 2$:

\begin{equation}
    \weight{g}{}(x) = \tilde{a} \cdot \text{exp}
    \left(-4 \cdot \text{ln}(2) \cdot x^2\right),
\end{equation}

\noindent with a normalisation factor of $\tilde{a} = \text{exp}\left(\text{ln}(2)\right)$ and $x$ being the distance to the mean position (in our case $\mu = 0$) in fractions of the FWHM ($f_{\Theta}$).
We can apply this Gaussian weighting straightforwardly along the spectral axis, where $x \equiv f_{\Theta} = |\vlsr^{\text{src}} - \vlsr| / \Theta_{\text{src}}$, with $\vlsr^{\text{src}}$ and $\Theta_{\text{src}}$ being the measured centroid velocity and linewidth of the catalogue sources, respectively.
We define the  weight along the velocity axis $\weight{V}{}$ as:

\begin{equation}
    \weight{V}{}=
        \begin{cases}
          1, & \text{for}\ \weight{g}{}(f_{\Theta}) \geq 1 \\
          \weight{g}{}(f_{\Theta}), & \text{for}\ 1 > \weight{g}{}(f_{\Theta}) \geq \weight{V}{min} \\
          0, & \text{otherwise,}
        \end{cases}
\end{equation}

\noindent where $\weight{V}{min}$ is a user-defined threshold. For example, with $\weight{V}{min} = 0.125$ all points along the spectral axis for which $|\vlsr^{\text{src}} - \vlsr| > \Theta_{\text{src}}$ receive a weight of $\weight{V}{} = 0$.

For the spatial association, we use the elliptical or circular FWHM extent of the sources as defined in the respective catalogue.
With the following equation we can check whether a point is located within a rotated ellipse:

\begin{equation}
    \begin{aligned}
    \epsilon = 
    & \dfrac{\left[
        \text{cos}(\alpha)(x_p - x_0) +
        \text{sin}(\alpha)(y_p - y_0)\right]^2}{a^2} + \\
    & \text{~~~~~~~~}\dfrac{\left[
        \text{sin}(\alpha)(x_p - x_0) +
        \text{cos}(\alpha)(y_p - y_0)\right]^2}{b^2},
    \end{aligned}
\end{equation}

\noindent where in our case $x_p$ and $y_p$ refer to the $\ell$ and $b$ coordinates of our PPV data point and $x_0$, $y_0$, $a$, $b$, and $\alpha$ are the central $\ell$ and $b$ coordinates, the semi-major and semi-minor axis, and the rotation of the ellipse, respectively.
If $\epsilon \leq 1$, the point is located within or on the ellipse.
Given that $\sqrt{\epsilon}$ corresponds to the distance to the centre of the source expressed in fractions of its spatial FWHM extent, we can define a weight $\weight{PP}{}$ for the association of Galactic or position-position coordinates with a catalogue source as follows:  

\begin{equation}
    \weight{PP}{}=
        \begin{cases}
          1, & \text{for}\ \weight{g}{}(\sqrt{\epsilon}) \geq 1 \\
          \weight{g}{}(\sqrt{\epsilon}), & \text{for}\ 1 > \weight{g}{}(\sqrt{\epsilon}) \geq \weight{PP}{min} \\
          0, & \text{otherwise,}
        \end{cases}
\end{equation}

\noindent where $\weight{PP}{min}$ is a user-defined threshold.
For example, with $\weight{PP}{min} = 0.125$ all points located beyond twice the extent of the source receive a weight of $\weight{PP}{} = 0$. 

We then combine the spectral and spatial weights to a total PPV weight of $\weight{PPV}{} = \weight{PP}{} \cdot \weight{V}{}$.
We only retain catalogue sources that have a weight $\weight{PPV}{} > 0$.
If there are no sources from the catalogue that have $\weight{PPV}{} > 0$ we proceed to the next catalogue and repeat the source association.

If there was at least one catalogue source associated with the coordinate, we calculate $\weight{KDA}{}$, which gives the weight that the source is located on the near or far distance.
The $\weight{KDA}{}$ weight can range from $-0.5$ (which puts all the weight on the near distance) to $0.5$ (which puts all the weight on the far distance).  
In case of only a single associated catalogue source $\weight{KDA}{} = \weight{PPV}{}\cdot f_{\text{N/F}}\cdot \weight{CAT}{}$, where $f_{\text{N/F}}$ is $-0.5$ or $0.5$ if the catalogue source is associated with the near or far KD solution, respectively, and $\weight{CAT}{}$ is a user-defined weight for the catalogue (see App.~\ref{sec:kda_tables}).
If multiple sources are associated, we calculate $\weight{KDA}{}$ as a weighted average:

\begin{equation}
    \weight{KDA}{} = \dfrac{\sum^{\Nsrc}_{i=1} \weight{PPV, i}{}\cdot f_{\text{N/F, i}}}{\sum^{\Nsrc}_{i=1} \weight{PPV, i}{}}\cdot \weight{CAT}{}.
\end{equation}

\noindent After we searched all catalogues, we use the $\weight{KDA}{}$ with the highest absolute value as our final value to determine our prior for the kinematic distance solution with $\prob{far}{} = 0.5 + \weight{KDA}{}$.
We did not make the final calculation of the weights cumulative or additive as many of the catalogues we use are not fully independent from each other.
For example, the CHIMPS catalogue used many of the other catalogues in resolving the KDA for their clumps.
In case our routine yields multiple $\weight{KDA}{}$ values with the same absolute value, we calculate the average $\weight{KDA}{}$ value from these solutions. 
Thus, if for the same clump two catalogues determined different KDA solutions, the resulting $\prob{far}{}$ value is $0.5$, which means that no prior for the KDA will be supplied.

\subsection{Inferring literature KDA solutions}
\label{sec:kda_det}

We use the literature distance information only for prior information on whether the source is located on the near or far side of the Galactic disk.
In case one of the catalogues already supplied an information about the KDA resolution (that means if the near or far distance was chosen) we adopt these KDA results.
If the KDA resolution was not stated explicitly, we use one of the following two methods to infer it.

\paragraph{Method $1$:} If the corresponding catalogue only gives the information about the heliocentric distance d$_{\odot}$ without any further information on the KDA, we use the relation

\begin{equation}
   d_{\odot} = R_{0}\,\text{cos}\,\ell \pm \sqrt{R_{\text{gal}}^{2} - (R_{0}\,\text{sin}\,\ell)^{2}}
    \label{eq:kindist}
\end{equation}

\noindent to obtain the Galactocentric radius $\rgal$. 
With $\rgal$, we then establish the near and far kinematic distances, from which we determine the chosen KDA solution. 
We only attempt to resolve the KDA if the near and far kinematic distances differ by more than $1$~kpc; otherwise we remove the source from the catalogue.

\paragraph{Method $2$:} In case the KDA cannot be inferred via Method~1, we use functions contained in the BDC tool to calculate kinematic distance solutions and expected tangent point velocities $\vlsr^{\text{TP}}$ for the catalogue sources.
If the $\vlsr$ velocity of a catalogue source is within $10$~\kms\ or is higher than $\vlsr^{\text{TP}}$, we assume that the source is too close to the tangent point to resolve the KDA properly.
These sources are subsequently removed from the catalogue.

Next we check for clear KDA solutions. 
If the $\vlsr$ velocity of the source is lower than the near KD solution or higher than the far KD solution we resolve the KDA as 'N' (near) or 'F' (far), respectively.

For the remaining sources we compare the given catalogue distance to the kinematic distance solutions and the tangent point distance.
In case the difference of the literature distance to the tangent point is lower than the differences to both of the kinematic distance solutions, we do not attempt to resolve the KDA and remove the source from the catalogue. 
Otherwise, we choose the kinematic distance solution that has the smallest difference to the literature distance.

\subsection{Literature results used in the KDA prior}
\label{sec:kda_tables}

We now discuss the catalogues that were used in this work to infer priors for the KDA.
We required that the catalogues contain information about molecular gas $\vlsr$ velocities, either reported directly or reported indirectly via a given kinematic distance.
We only retain the catalogue sources for which we could infer whether the near or far distance solution was chosen; we exclude all catalogue entries which were assigned tangent point distances or which had uncertain KDA resolutions.
For this work we did not attempt to incorporate distances obtained from alternative distance estimation methods, such as dust extinction mapping, if there was no information about the association with molecular gas.
We also chose not to include catalogues based on $\co{12}{}\transition{1}{0}$ observations.

Where available, we incorporated measured $\co{13}{}\transition{1}{0}$ linewidths for the catalogue sources. 
If these did not exist, we either settled on linewidth measurements of a higher-density tracer or used median linewidth measurements obtained for comparable sources (i.e. clumps or clouds).
We use a $\weight{V}{min}$ threshold of $0.125$ for all catalogues, which means that we make a spectral association with a catalogue source if $|\vlsr^{\text{src}} - \vlsr| < \Theta_{\text{src}}$.
We make a spatial association with sources from the clump catalogues if the $(\ell, b)$ coordinate of the PPV point is contained within twice the extent of the clump (i.e. $\weight{PP}{min} = 0.125$).
For the more extended sources of the remaining catalogues, we only make a spatial association if the $(\ell, b)$ coordinate of the PPV point is located within or close to the elliptical or circular extent of the source (with $\weight{PP}{min} = 0.9$)

We give the clump catalogues a higher weight ($\weight{CAT}{} = 0.75$) than the catalogues of more extended objects, such as molecular clouds and IRDCs ($\weight{CAT}{} = 0.5$).
This was done to favour the KDA information from the clumps on small scales, as the clumps are usually embedded within these more extended objects.

In Table~\ref{tbl:overview_kda} we give an overview about the catalogues we used as KDA solutions in this work. 
N$_{\text{src}}^{\text{GRS}}$ gives the number of sources overlapping with the GRS coverage, and N$_{\Theta}$, N$_{\text{N}}$, and N$_{\text{F}}$ give the corresponding numbers of sources with measured linewidths, near, and far KD solutions, respectively.
The columns $\weight{PP}{min}$, $\weight{V}{min}$, and $\weight{CAT}{}$ list the weights we used for the association of catalogue sources (\sect\ref{app:calc_priors}); the KDAR column specifies if the KDA resolution was given in the respective catalogue or how we calculated it otherwise.
In the last two columns we give the abbreviation for the catalogues we refer to further on in the text and list the main references used to obtain information about the location, size, velocity, and distance information of the catalogue sources.
In the following we discuss the individual catalogues in more detail.

\begin{table*}
    \caption[Overview of the catalogues used as KDA solutions]{Overview of the catalogues used as KDA solutions in this work.}
    \centering
    \small
    \renewcommand{\arraystretch}{1.3}
\begin{tabular}{ccccccccccc}
	\hline\hline
Info & N$_{\text{src}}^{\text{GRS}}$ & N$_{\Theta}$ & N$_{\text{N}}$ & N$_{\text{F}}$ & $\weight{PP}{min}$ & $\weight{V}{min}$ & $\weight{CAT}{}$ & KDAR & Abb. & Ref. \\
	\hline
ATLASGAL clumps & $1745$ & $426$ & $1221$ & $524$ & $0.125$ & $0.125$ & $0.75$ & see App.\,\ref{app:atlasgal} & U+18 & $1$, $2$, $3$ \\
BGPS v1 clumps & $146$ & $0$ & $105$ & $41$ & $0.125$ & $0.125$ & $0.75$ & Method $1$ & E+12 & $4$, $5$, $6$ \\
BGPS v$2.1$ clumps & $1046$ & $455$ & $754$ & $292$ & $0.125$ & $0.125$ & $0.75$ & given & EB+15 & $7$, $8$, $9$ \\
CHIMPS clumps & $3294$ & $3294$ & $2318$ & $976$ & $0.125$ & $0.125$ & $0.75$ & given & R+19 & $10$ \\
Hi-GAL clumps & $4021$ & $0$ & $3536$ & $485$ & $0.125$ & $0.125$ & $0.75$ & given & E+17 & $11$ \\
COHRS clouds & $396$ & $396$ & $262$ & $134$ & $0.9$ & $0.125$ & $0.5$ & Method $2$ & C+19 & $12$ \\
GRS clouds & $652$ & $652$ & $453$ & $199$ & $0.9$ & $0.125$ & $0.5$ & Method $1$ & RD+09 & $13$, $14$ \\
GRS clouds (BGPS) & $381$ & $381$ & $203$ & $178$ & $0.9$ & $0.125$ & $0.5$ & Method $1$ & BH14 & $15$ \\
MSX IRDCs & $263$ & $263$ & $261$ & $2$ & $0.9$ & $0.125$ & $0.5$ & Method $2$ & S+06 & $16$, $17$, $18$ \\
GRS \hii\ regions & $169$ & $169$ & $49$ & $120$ & $0.9$ & $0.125$ & $0.5$ & given & A+09 & $19$, $20$ \\
WISE \hii\ regions & $351$ & $0$ & $72$ & $279$ & $0.9$ & $0.125$ & $0.5$ & given & A+14 & $21$ \\
SNRs & $23$ & $0$ & $17$ & $6$ & $0.9$ & $0.125$ & $0.5$ & given & R+18 & $22$, $23$, $24$, $25$, $26$ \\
	\hline
\end{tabular}
\tablefoot{
\small
\tablefoottext{a}{
($1$) \citet{Wienen2012};
($2$) \citet{Urquhart2014};
($3$) \citet{Urquhart2018};
($4$) \citet{Rosolowsky2010};
($5$) \citet{Eden2012};
($6$) \citet{Eden2013};
($7$) \citet{Ginsburg2013bgps};
($8$) \citet{Ellsworth-Bowers2015};
($9$) \citet{Svoboda2016};
($10$) \citet{Rigby2019};
($11$) \citet{Elia2017higal};
($12$) \citet{Colombo2019};
($13$) \citet{Rathborne2009grs};
($14$) \citet{RomanDuval2009grs};
($15$) \citet{Battisti2014};
($16$) \citet{Simon2006a};
($17$) \citet{Simon2006b};
($18$) \citet{Marshall2009};
($19$) \citet{Anderson2009};
($20$) \citet{AndersonBania2009};
($21$) \citet{Anderson2014};
($22$) \citet{Leahy2018};
($23$) \citet{Ranasinghe2018a};
($24$) \citet{Ranasinghe2018b};
($25$) \citet{Ranasinghe2018c};
($26$) \citet{Green2019}.
}}
\label{tbl:overview_kda}
\end{table*}

\paragraph{ATLASGAL clumps:}
\label{app:atlasgal}
 
\citet{Urquhart2018} presented distance results for clumps from the ATLASGAL survey in the inner Galactic plane ($|l| < 60^{\circ}$,  $|b| < 1.5^{\circ}$).
The catalogue does not contain explicit information on how the KDA was resolved but lists the kinematic distance solutions, the distance estimated with the BDC (v1), and the chosen distance. 
With that information we could infer the KDA information ('N', 'F') for $6317$ clumps. 
Of these, $4457$ clumps have linewidth measurements, of which $3139$ were obtained from the SEDIGISM survey (Urquhart et al., in prep.), $668$ measurements were taken from \citet{Urquhart2018}, and $292$ measurements were taken from \citet{Wienen2012}.
For the remaining clumps with resolved KDAs but missing linewidth information we assume a FWHM linewidth of $3.367$~\kms, which corresponds to the median linewidth computed from the clumps with measurements.
We took the size information for the ATLASGAL clumps from \citet{Urquhart2014}.
In total, $1745$ ATLASGAL clumps with resolved KDAs overlap with the GRS coverage.

\paragraph{BGPS clumps (v$1$):}

\citet{Eden2012, Eden2013} presented KD determinations for clumps of the Bolocam Galactic Plane Survey \citep[BGPS;][]{Rosolowsky2010}.
We only use the sample of $165$ BGPS sources whose distances were not inferred via an association with molecular clouds from \citet{RomanDuval2009grs}. 
We established the chosen KDA solution via Method~$1$, using R$_{0} = 8.5$~kpc as assumed by \citet{Eden2012, Eden2013}.
We take the corresponding position and size information of these clumps from v$1$ of the BGPS catalogue \citep{Rosolowsky2010}.
For the linewidth of the clumps we assume a value of $\Theta = 3.316$~\kms, which corresponds to the median linewidth of the BGPS v$2.1$ sample (see next paragraph). 

\paragraph{BGPS clumps (v$2.1$):}

\citet{Ellsworth-Bowers2015} resolved the KDA for $1320$ BGPS clumps.
We took the corresponding position and size information from v$2.1$ of the BGPS catalogue \citep{Ginsburg2013bgps}.
\citet{Svoboda2016} determined linewidths for $610$ of these clumps from NH$_{3}$ observations. 
For clumps the associated $\co{13}{}$ emission has in general broader linewidths than the NH$_{3}$ emission \citep{Wienen2012}.
We thus decided to multiply the measured NH$_{3}$ linewidths of \citet{Svoboda2016} by a factor of two, which is based on the difference found by \citet{Wienen2012}.
For the remaining clumps without measured linewidths we assume the median NH$_{3}$ linewidth value from the \citet{Svoboda2016} sample corrected by a factor two, which corresponds to $3.316$~\kms\ and compares very well to the median value from the ATLASGAL sample.

\paragraph{CHIMPS clumps:}

\citet{Rigby2019} used literature information to resolve the KDA for their sample of clumps compiled from the $^{13}$CO/C$^{18}$O (J=$3$-$2$) Heterodyne Inner Milky Way Plane Survey \citep[CHIMPS;][]{Rigby2016}.
We use $3294$ clumps that have the highest reliability flag and a resolved KDA.
We take position, size, and spectral information of the clumps from \citet{Rigby2019}.

\paragraph{Hi-GAL clumps:}

\begin{figure}
    \centering
    \includegraphics[width=\columnwidth]{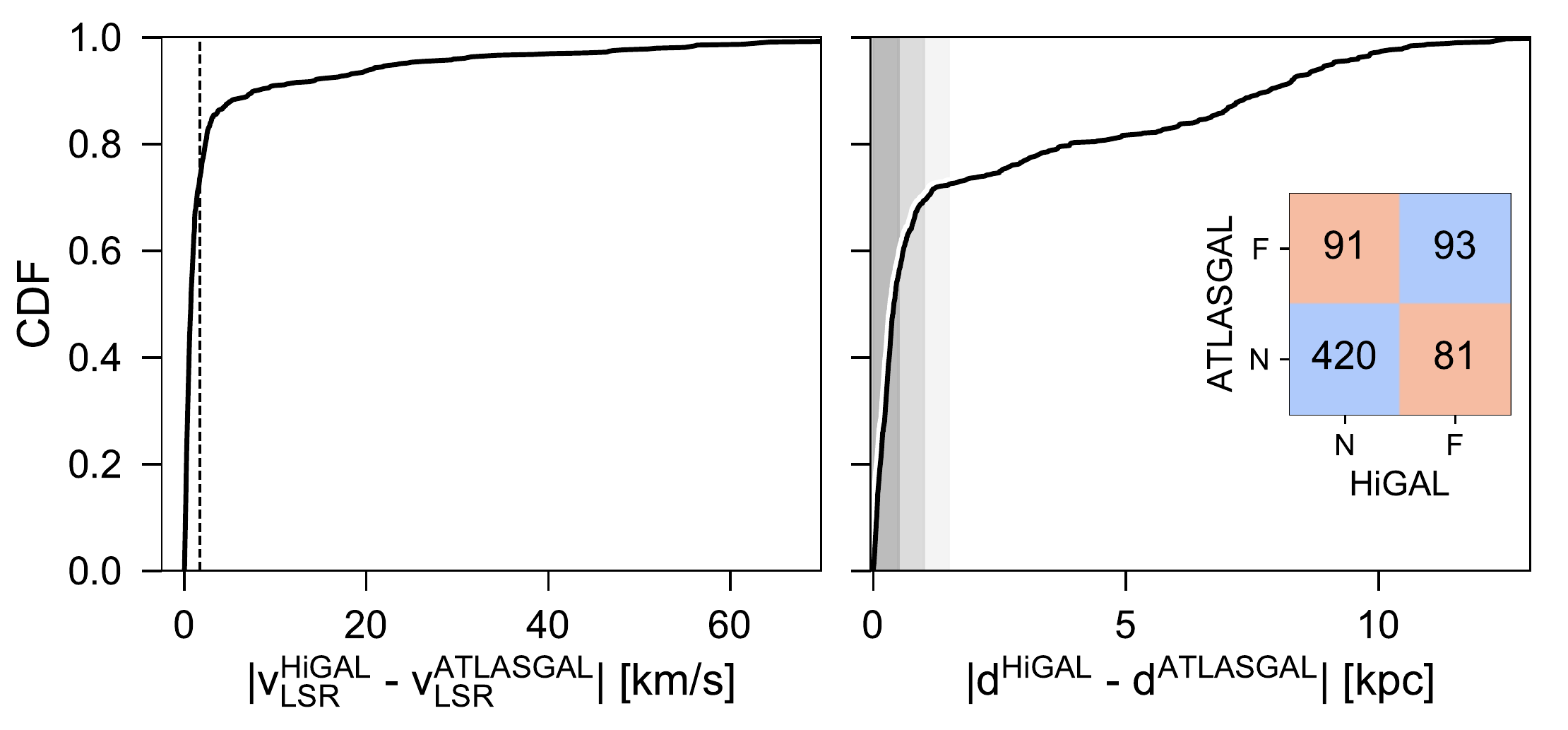}
    \caption[Comparison between associated Hi-GAL and ATLASGAL sources]{Comparison between $685$ associated Hi-GAL and ATLASGAL sources \citep{Urquhart2018}.
    \textit{Left:} CDF of the absolute difference between the inferred and measured $\vlsr$ velocities for the Hi-GAL and ATLASGAL catalogues, respectively.
    The dashed vertical line indicates the assumed linewidth for the Hi-GAL sources. 
    \textit{Right:} CDF of the absolute difference of reported distances for Hi-GAL and associated ATLASGAL sources. 
    The vertical grey areas indicate distance bins of $0.5$~kpc. 
    The inset shows the corresponding confusion matrix for the KDA resolution. 
    }
    \label{fig:higal_atlasgal}
\end{figure}

\citet{Elia2017higal} presented a compact source catalogue for the  Herschel InfraRed Galactic Plane Survey (Hi-GAL), for which KDs were determined via the \citet{Brand1993} rotation curve.
We only retain the sources for which an external indicator was used to solve the KD (flag 'G').
The \citet{Elia2017higal} catalogue does not contain information about the associated $\vlsr$ velocities of the sources. 
We used the function \texttt{brand\_rotcurve.calc\_vlsr} from the kinematic distance package presented in \citet{Wenger2018}\footnote{\url{https://ascl.net/1712.001}} to infer the $\vlsr$ velocities for the longitude and distance values of the sources in the \citet{Elia2017higal} catalogue.
In \fig\ref{fig:higal_atlasgal} we benchmark these estimated Hi-GAL $\vlsr$ velocities for $685$ sources that we could associate with ATLASGAL clumps from \citet{Urquhart2018}.
We associate a Hi-GAL source with an ATLASGAL clump if their central positions are less than $19.2"$ apart, which corresponds to the ATLASGAL beam size.
The left panel in \fig\ref{fig:higal_atlasgal} shows that the majority of estimated Hi-GAL $\vlsr$ velocities match very well with the associated ATLASGAL clump velocities. 
The fraction of sources for which the $\vlsr$ values are not consistent is likely due to wrong associations of spectral lines with the dust features, as there can be multiple molecular gas emission features along the line of sight that the dust feature could be associated with.
For this work, we make no attempt to resolve these inconsistent $\vlsr$ values.
The right panel in \fig\ref{fig:higal_atlasgal} shows the differences in the estimated distances for the $685$ associated sources.
Again, for the majority of associated sources the distance values are similar.
More important for our context, for $75\%$ of the associated sources the KDA resolution is identical.
We take the location and size information for the Hi-GAL clumps from \citet{Elia2017higal}. 
For the linewidth, we assume for each clump the median linewidth estimated from the measured values from the ATLASGAL survey, which corresponds to $\Theta = 3.367$~ \kms.

\paragraph{COHRS molecular clouds:}

\citet{Colombo2019} presented a cloud catalogue for the JCMT CO($3$-$2$) High Resolution Survey (COHRS; \citealt{Dempsey2013cohrs}).
We use their fiducial sample of $540$ molecular clouds with well-defined distance estimations.
We infer best-matching KDA solutions via Method~$2$, which yielded $396$ KDA ('N', 'F') solutions.

\paragraph{GRS molecular clouds:}

\citet{RomanDuval2009grs} determined distances to $750$ clouds from the catalogue of \citet{Rathborne2009grs} but do not explicitly specify the KDA resolution.
Using Method~$1$ (with R$_{0}=8.5$~kpc), we were able to infer the KDA solutions for $652$ clouds.
We took the position, size, and spectral information of the clouds from \citet{Rathborne2009grs}.

\paragraph{GRS molecular clouds crossmatched with BGPS clumps:}

\citet{Battisti2014} compiled a catalogue of $437$ molecular clouds from the GRS survey, which they associated with BGPS sources from the v$1$ catalogue \citep{Rosolowsky2010}.
They resolve the KDA for their sources but do not explicitly list whether they chose the near or far solution. 
Using Method~$1$ (with R$_{0}=8.3$~kpc), we could infer the chosen KDA solution for $389$ clouds.

\paragraph{MSX IRDCs with velocities from GRS:}

\citet{Simon2006b} determined KDs to $313$ IRDCs identified from observations of the \emph{Midcourse Space Experiment} \citep{Simon2006a}, by morphologically matching the IRDCs to $\co{13}{}\transition{1}{0}$ emission from the GRS.
\citet{Simon2006b} argued that since the IRDCs are seen as extinction features they probably will be located in the foreground and thus always resolve the KDA in favour of the near solution.
\citet{Marshall2009} used an alternative approach based on modelling the three-dimensional distribution of interstellar extinction towards $115$ of these IRDCs, which allowed them to obtain distances that do not suffer from near/far ambiguities.
Whenever these were available, we chose the \citet{Marshall2009} distance determinations over the ones obtained by \citet{Simon2006b}.
Using Method~$2$ (\sect\ref{sec:kda_det}), we could resolve the KDA for $272$ IRDCs.
We take the $\vlsr$ and FWHM linewidth information for this sample of IRDCs from \citet{Simon2006b} and take their position and size information from \citet{Simon2006a}.

\paragraph{\hii\ Regions associated with GRS emission:}
\label{sec:grs_hii}

\citet{Anderson2009} associated $301$ Galactic \hii\ regions located within the GRS coverage with the corresponding $\co{13}{}\transition{1}{0}$ properties. 
\citet{AndersonBania2009} resolved the KDA for $266$ of these \hii\ regions using the \hi\ emission/absorption and \hi\ self-absorption methods.
We include \hii\ region sources as prior information if one of the following two conditions was fulfilled: the two methods yielded the same KDA resolution; or one of the methods received a high confidence label, in which case we use its KDA solution.
We thus retained $169$ sources, which had measured associated $\co{13}{}\transition{1}{0}$ properties and a resolved KDA.
We take the position, size, and spectral information of the sources from \citet{Anderson2009}.

\paragraph{WISE \hii\ Regions:}

\citet{Anderson2014} compiled an \hii\ region catalogue from observations of the \textit{Wide-Field Infrared Survey Explorer} (WISE).
Using v$2.0$ of the catalogue, we select all \hii\ regions with resolved KDAs; we also take the position, size, and $\vlsr$ information from this version of the catalogue.
For the spectral extent, we use the median linewidth value from the sample of \hii\ regions of \citet{Anderson2009} (see previous paragraph), which corresponds to a value of $\Theta = 4.05$~\kms.

\paragraph{Supernova remnants:}

We include KDA solutions for $23$ Supernova remnants (SNRs) that have been obtained from \hi\ $21$~cm and GRS $\co{13}{}$ line emission \citep{Ranasinghe2018a, Ranasinghe2018b, Ranasinghe2018c}.
In case no information on the spatial extent was given in these works, we adopted the information given in \citet{Leahy2018} and \citet{Green2019}. 
For the spectral extent, we assume the average linewidth of $3.6$~\kms\ that \citet{Rathborne2009grs} find for their catalogue of molecular clouds of the GRS.

\subsection{Effect of the KDA prior}
\label{sec:test_kda_priors}

In this section we discuss the effect of the KDA prior on the BDC results.
For our tests we use the catalogues as detailed in Table~\ref{tbl:overview_kda}, so we include only sources that overlap with the GRS coverage and for which we could infer near of far KDA solutions.

First, we quantify the effect of the $\weight{CAT}{}$ weight on the distance estimation, for the two cases where the spiral arm priors are included ($\prob{SA}{} = 0.5$) or switched off ($\prob{SA}{} = 0$).
For this test we use only the KDA information from the ATLASGAL sample and do not consider any of the other catalogues. 
We perform different distance runs with the BDC for the ATLASGAL sample; for each run, we supply the BDC with the correct KDA solutions for the sources and just vary the weight $\weight{CAT}{}$, which determines the strength of the resulting $\prob{far}{}$ prior.
For example, for $\weight{CAT}{}=0.5$ a far KDA solution yields $\prob{far}{}=0.75$, whereas with $\weight{CAT}{}=0.75$ this increases to $\prob{far}{}=0.875$.
For $\weight{CAT}{}=1$ we would thus expect the highest correspondence between our calculated distances and the distances given in \citet{Urquhart2018}\footnote{However, since \citet{Urquhart2018} used an older version of the BDC (v$1$), we would not expect a perfect correspondence of the distance results even in this best case scenario.}.
To test how robust the BDC results are against wrong KDA solutions, we also perform distance calculations for which we intentionally supply the incorrect KDA solutions for the ATLASGAL sample.

\begin{table}
    \caption[BDC results for the ATLASGAL clump sample for different $\weight{CAT}{}$ and $\prob{SA}{}$ values and correct or incorrect KDA priors]{BDC results for the ATLASGAL clump sample for different $\weight{CAT}{}$ and $\prob{SA}{}$ values and correct or incorrect KDA priors.}
    \centering
    \footnotesize
    \renewcommand{\arraystretch}{1.3}
    \setlength{\tabcolsep}{5pt}
\begin{tabular}{cccccccc}
	\hline\hline
& & \multicolumn{3}{c}{correct KDA} & \multicolumn{3}{c}{incorrect KDA} \\\cmidrule(lr){3-5} \cmidrule(lr){6-8}
\multirow{2}{*}{$\prob{SA}{}$} & \multirow{2}{*}{w$_{\text{CAT}}$} & $\delta\left(0.5\right)$ & $\delta\left(1.0\right)$ & $\delta\left(1.5\right)$ & $\delta\left(0.5\right)$ & $\delta\left(1.0\right)$ & $\delta\left(1.5\right)$ \\
 & & [$\%$] & [$\%$] & [$\%$] & [$\%$] & [$\%$] & [$\%$] \\
	\hline
$0.5$ & $0$ & $36.2$ & $56.3$ & $62.7$ & $36.2$ & $56.3$ & $62.7$ \\
$0.5$ & $0.25$ & $40.9$ & $63.7$ & $70.1$ & $33.4$ & $49.7$ & $55.0$ \\
$0.5$ & $0.5$ & $46.6$ & $70.8$ & $77.5$ & $28.1$ & $43.6$ & $47.6$ \\
$0.5$ & $0.75$ & $50.1$ & $75.6$ & $83.3$ & $21.0$ & $32.0$ & $35.9$ \\
$0.5$ & $1$ & $55.1$ & $83.7$ & $92.6$ & $7.6$ & $10.6$ & $14.2$ \\
	\hline
$0$ & $0$ & $37.4$ & $52.3$ & $58.9$ & $37.4$ & $52.3$ & $58.9$ \\
$0$ & $0.25$ & $44.5$ & $62.6$ & $72.3$ & $23.8$ & $32.3$ & $36.7$ \\
$0$ & $0.5$ & $53.5$ & $77.0$ & $84.8$ & $16.6$ & $20.7$ & $23.7$ \\
$0$ & $0.75$ & $59.5$ & $82.7$ & $91.1$ & $8.9$ & $10.7$ & $12.8$ \\
$0$ & $1$ & $58.3$ & $81.9$ & $91.6$ & $2.4$ & $3.0$ & $5.4$ \\
	\hline
\end{tabular}
\label{tbl:weightcat}
\end{table}

Table~\ref{tbl:weightcat} lists the performance of the BDC results for $20$ runs, for which we vary $\weight{CAT}{}$ between the values $0, 0.25, 0.5, 0.75,\ \text{and}\ 1$, use or switch off the $\prob{SA}{}$ prior, and supply either the correct or incorrect KDA solutions.
The $\delta\left(x\right)$ parameter gives the percentage of calculated distance values whose absolute error is within $x$~kpc of the literature distances; we report $\delta\left(x\right)$ for $x$ intervals of $\pm 0.5$, $\pm 1.0$, and $\pm 1.5$~kpc. 
The runs with $\weight{CAT}{} = 0$ correspond to the default BDC distance estimations that do not consider any prior information on the KDA and thus serve as our benchmarks.

The runs for which we supplied the correct KDA solution show a clear increase in the fraction of matching distances with increasing $\weight{CAT}{}$ value.
This is expected, as KDA solutions are more enforced with increasing $\weight{CAT}{}$ values. 
In the runs where we use the prior for spiral arms ($\prob{SA}{} = 0.5$), the percentage of matching distance values is less; however, the vast majority of the estimated distances is still close to the literature values. 

Unsurprisingly, the supply of incorrect KDA solutions yields wrong distance estimates, especially the higher the $\weight{CAT}{}$ value, that means the more we enforce these KDA solutions. 
This effect is very strong for the runs where the spiral arm prior was switched off and the prior for the kinematic distances was the dominating factor for the BDC results. 
Using the prior for the spiral arms can mitigate the negative impact of the incorrect KDA solutions; for $\weight{CAT}{} = 0.5$ almost half the sources have distance results within $1.5$~kpc to the literature distances even though we intentionally forced the BDC to prioritise the wrong KDA solution.

This test demonstrated that the BDC run with $\prob{SA}{} = 0.5$ is more robust against priors using wrong KDA solutions than the run with $\prob{SA}{} = 0$.
Moreover, we find that $\weight{CAT}{}$ values of $0.5\ \text{to}\ 0.75$ are preferable values for the catalogue weights, as they offer a good balance between recovery of correct distances with the right KDA solutions for $\prob{SA}{} = 0$ and robustness against incorrect distances with the wrong KDA solutions for $\prob{SA}{} = 0.5$.

\begin{figure*}
    \centering
    \includegraphics[width=\textwidth]{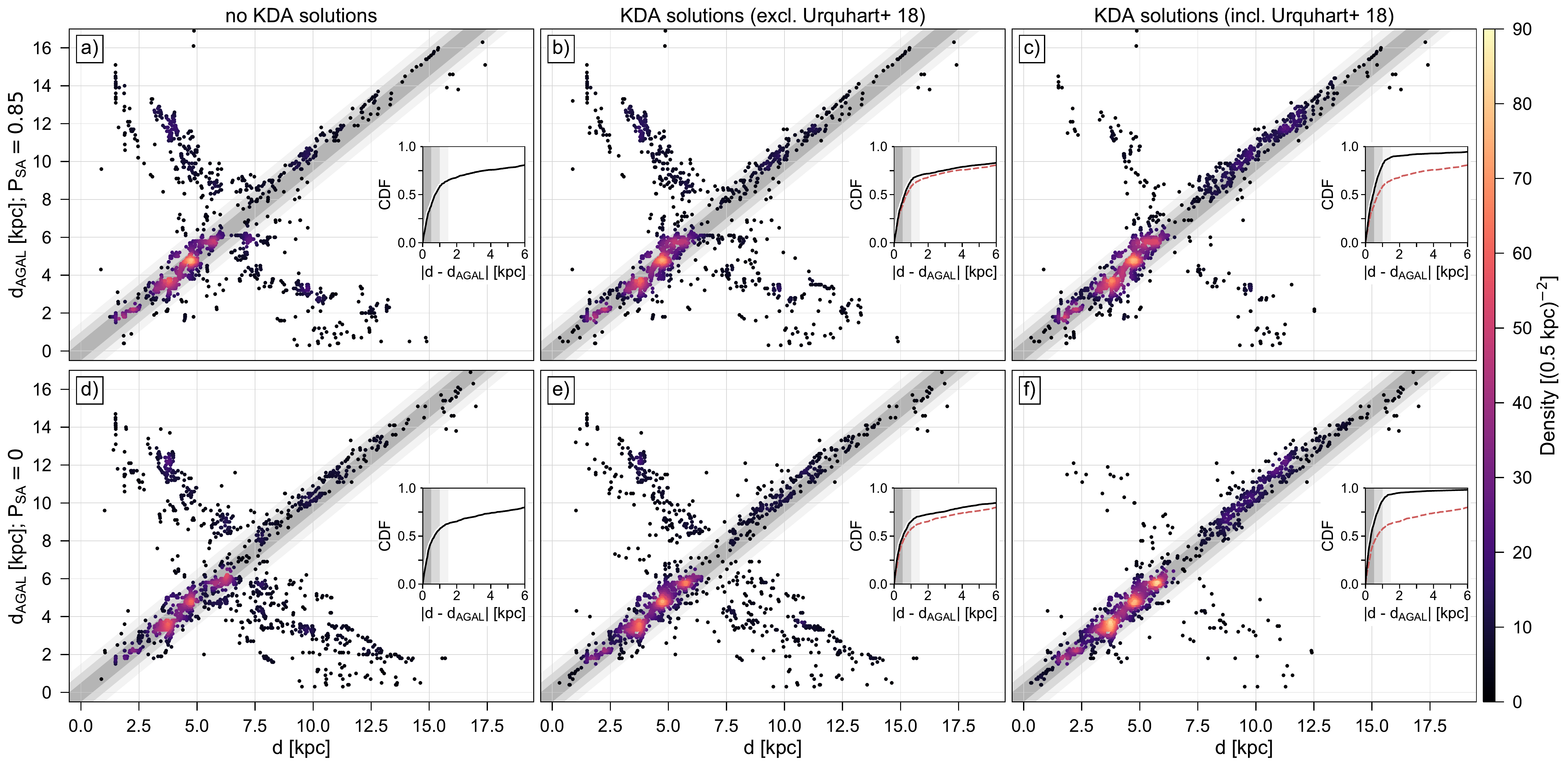}
    \caption[BDC results for the ATLASGAL clump sample compared to their literature values]{BDC results for the ATLASGAL clump sample plotted against their literature values (d$_{\text{AGAL}}$).
    The six panels correspond to different settings for the spiral arm and KDA priors.
    The points are colour-coded by their density.
    The insets show the cumulative distribution functions (CDFs) for the difference between the BDC results and the literature distances of the respective panels.
    The red dashed line in the insets of \textit{panels~b} and \textit{c} and \textit{panels~e} and \textit{f} correspond to the CDFs of \textit{panel~a} and \textit{d}, respectively.
    The grey-shaded areas in the main panels and insets correspond to $|\text{d} - \text{d}_{\text{AGAL}}|$ intervals of $0.5, 1.0,\ \text{and}\ 1.5$~kpc.
    See Appendix~\ref{sec:test_kda_priors} for more details.
    }
    \label{fig:atlasgal_crosscheck}
\end{figure*}

\begin{table}
    \caption[BDC results for ATLASGAL sample for different spiral arm and KDA priors]{BDC results for the ATLASGAL clump sample for different spiral arm and KDA priors.}
    \centering
    \footnotesize
    \renewcommand{\arraystretch}{1.3}
    \setlength{\tabcolsep}{5pt}
\begin{tabular}{ccccccc}
	\hline\hline
& \multicolumn{3}{c}{$\prob{SA}{}=0.5$} & \multicolumn{3}{c}{$\prob{SA}{}=0$} \\\cmidrule(lr){2-4} \cmidrule(lr){5-7}
\multirow{2}{*}{KDA} & $\delta \left(0.5\right)$ & $\delta \left(1.0\right)$ & $\delta \left(1.5\right)$ & $\delta\left(0.5\right)$ & $\delta\left(1.0\right)$ & $\delta\left(1.5\right)$ \\
 & [$\%$] & [$\%$] & [$\%$] & [$\%$] & [$\%$] & [$\%$] \\
	\hline
no & $39.1$ & $58.5$ & $65.1$ & $43.6$ & $57.4$ & $63.0$ \\
excl. U+18 & $43.7$ & $64.8$ & $70.1$ & $48.9$ & $65.3$ & $70.6$ \\
incl. U+18 & $53.7$ & $81.0$ & $89.1$ & $63.7$ & $86.8$ & $93.4$ \\
	\hline
\end{tabular}
\label{tbl:atlasgal_crosscheck}
\end{table}

Next, we want to quantify the effect of using all available KDA information with their corresponding weights (Table~\ref{tbl:overview_kda}).
Using the ATLASGAL sample, we again perform different BDC runs with $\prob{SA}{} = 0.5$ and $\prob{SA}{} = 0$.
For the KDA prior we either use none of the catalogues (giving us a benchmark for the default BDC performance), all KDA solutions from Table~\ref{tbl:overview_kda} excluding the ATLASGAL catalogue itself, and all KDA solutions including the ATLASGAL catalogue.
Figure \ref{fig:atlasgal_crosscheck} shows the BDC results plotted against the literature values for these six BDC runs. 
Table \ref{tbl:atlasgal_crosscheck} gives the corresponding percentages of matching distance values within ranges of $\delta\left(x\right)$ (with $x=0.5, 1,\ \text{and}\ 1.5$~kpc) that are highlighted with the grey-shaded areas in \fig\ref{fig:atlasgal_crosscheck}.
The runs which use no KDA solutions (panels~a and d) show a large dichotomy between near and far distance solutions. 
Including all KDA solutions apart from the \citet{Urquhart2018} catalogue itself  already manages to improve the correspondence between the distances and indicates that the sources of the remaining catalogues overlap with many of the ATLASGAL clumps.
Finally, as expected, the inclusion of the \citet{Urquhart2018} catalogue leads to the best correspondence of the distance results. 
However, for many sources the spiral arm priors lead to a preference of different KDA resolutions (panel c).
By switching the spiral arm prior off, we manage to drastically reduce the instances for which a different KDA solution was favoured. 
The remaining fraction of clumps for which a different KDA solution was chosen was due to mismatching KDA solutions from different catalogues. 
For example, \citet{Urquhart2018} give a near distance solution of $3.5$~kpc for the clump AGAL$020.662$-$00.139$, whereas spatially and spectrally overlapping sources in five other catalogues (A+09, RD+09, BH14, EB+15, E+17) favour a far distance solution, leading to a BDC value of $10.1$~kpc.

\begin{table}
    \caption[BDC results for the remaining catalogues]{BDC results for the remaining catalogues from Table~\ref{tbl:overview_kda} for different spiral arm and KDA priors.}
    \centering
    \footnotesize
    \renewcommand{\arraystretch}{1.3}
    \setlength{\tabcolsep}{5pt}
\begin{tabular}{ccccccc}
	\hline\hline
& \multicolumn{3}{c}{$\prob{SA}{}=0.5$} & \multicolumn{3}{c}{$\prob{SA}{}=0$} \\\cmidrule(lr){2-4} \cmidrule(lr){5-7}
\multirow{2}{*}{Abb.} & $\delta \left(1.5\right)$ & $\delta^{\,-} \left(1.5\right)$ & $\delta^{\,+} \left(1.5\right)$ & $\delta \left(1.5\right)$ & $\delta^{\,-} \left(1.5\right)$ & $\delta^{\,+} \left(1.5\right)$ \\
 & [$\%$] & [$\%$] & [$\%$] & [$\%$] & [$\%$] & [$\%$] \\
	\hline
E+12 & $21.2$ & $45.9$ & $80.8$ & $33.6$ & $67.8$ & $91.1$ \\
EB+15 & $56.6$ & $80.2$ & $88.5$ & $59.1$ & $87.1$ & $92.4$ \\
R+19 & $50.6$ & $79.8$ & $91.3$ & $63.5$ & $90.6$ & $95.7$ \\
E+17 & $51.3$ & $80.6$ & $91.4$ & $62.6$ & $89.3$ & $95.9$ \\
C+19 & $47.0$ & $69.4$ & $74.2$ & $52.3$ & $78.8$ & $84.6$ \\
RD+09 & $38.8$ & $69.3$ & $80.1$ & $52.5$ & $82.5$ & $92.0$ \\
BH14 & $29.8$ & $54.0$ & $65.0$ & $36.0$ & $66.1$ & $76.1$ \\
S+06 & $44.5$ & $74.1$ & $87.5$ & $51.0$ & $76.0$ & $89.4$ \\
A+09 & $25.4$ & $50.3$ & $63.3$ & $30.2$ & $56.8$ & $69.2$ \\
A+14 & $32.2$ & $58.7$ & $68.7$ & $41.9$ & $72.1$ & $83.2$ \\
R+18 & $43.5$ & $78.3$ & $78.3$ & $60.9$ & $91.3$ & $91.3$ \\
	\hline
\end{tabular}
\label{tbl:crosscheck}
\end{table}

Finally, we list the performance of the BDC results for the remaining catalogues  used for KDA information in Table~\ref{tbl:crosscheck}. 
We again perform different BDC runs with $\prob{SA}{}=0.5$ and $\prob{SA}{}=0$ and give the percentage of sources for which the distance was within a range of $1.5$~kpc to the literature distance for the cases where no literature KDA solution is used (labelled $\delta$), and KDA solutions from all catalogues excluding and including the one for which the distances are calculated (labelled $\delta^{\,-}$ and $\delta^{\,+}$, respectively).
We see already an improvement in matching distances for the $\delta^{\,-}$ runs, which indicates that there is a good overlap between sources from all catalogues.
As expected, we see the highest correspondence between the BDC and literature distance results for the runs in which the spiral arm prior is switched off. 
In conclusion, our tests showed that the BDC runs with supplied literature KDA solutions are able to match the vast majority of distance results from each of the individual catalogues used to infer KDA priors.
This result is not self-evident, given that many of these catalogues use different assumptions about the rotation curve parameters. 
We thus infer that our obtained distance results are consistent with the vast majority of the literature results.

\section{Effects of beam averaging on the observed linewidth}
\label{sec:beam_avg_veldisp}

We designed the following simplified experiment to test how fluctuations of the line centroids can broaden the linewidth via beam averaging effects.
We perform different runs for which we vary the spatial resolution of a given PPV cube to simulate observations of regions at different heliocentric distances.
For each run we construct PPV cubes with dimension ($100$, $100$, $30$) and populate each of the spectra with a single Gaussian component, whose velocity dispersion is two spectral channels.
For simplicity we do not assume any noise.
The mean position of the Gaussian is centred along the spectral axis with the centroid of the spectrum allowed to vary for each component; the standard deviation of this variation ($\Delta\text{v}_\text{cen}$) is set either to $0.5, 1,\ \text{or}\ 2$ times the velocity dispersion.
Assuming the pixel size to be equal to the FWHM of the resolution element or beam, we convolve this cube with a $2$D Gaussian kernel whose FWHM is set to either $2, 4, 8, 16, 32,\ \text{or}\ 64$ times the pixel size, thus simulating observations of regions at $2$ to $64$ times the distance of our original cube, which corresponds approximately to the variation present in the GRS.
We then determine the velocity dispersion of a Gaussian fit to the central spectrum of the spatially smoothed cube.

\begin{figure}
    \centering
    \includegraphics[width=0.75\columnwidth]{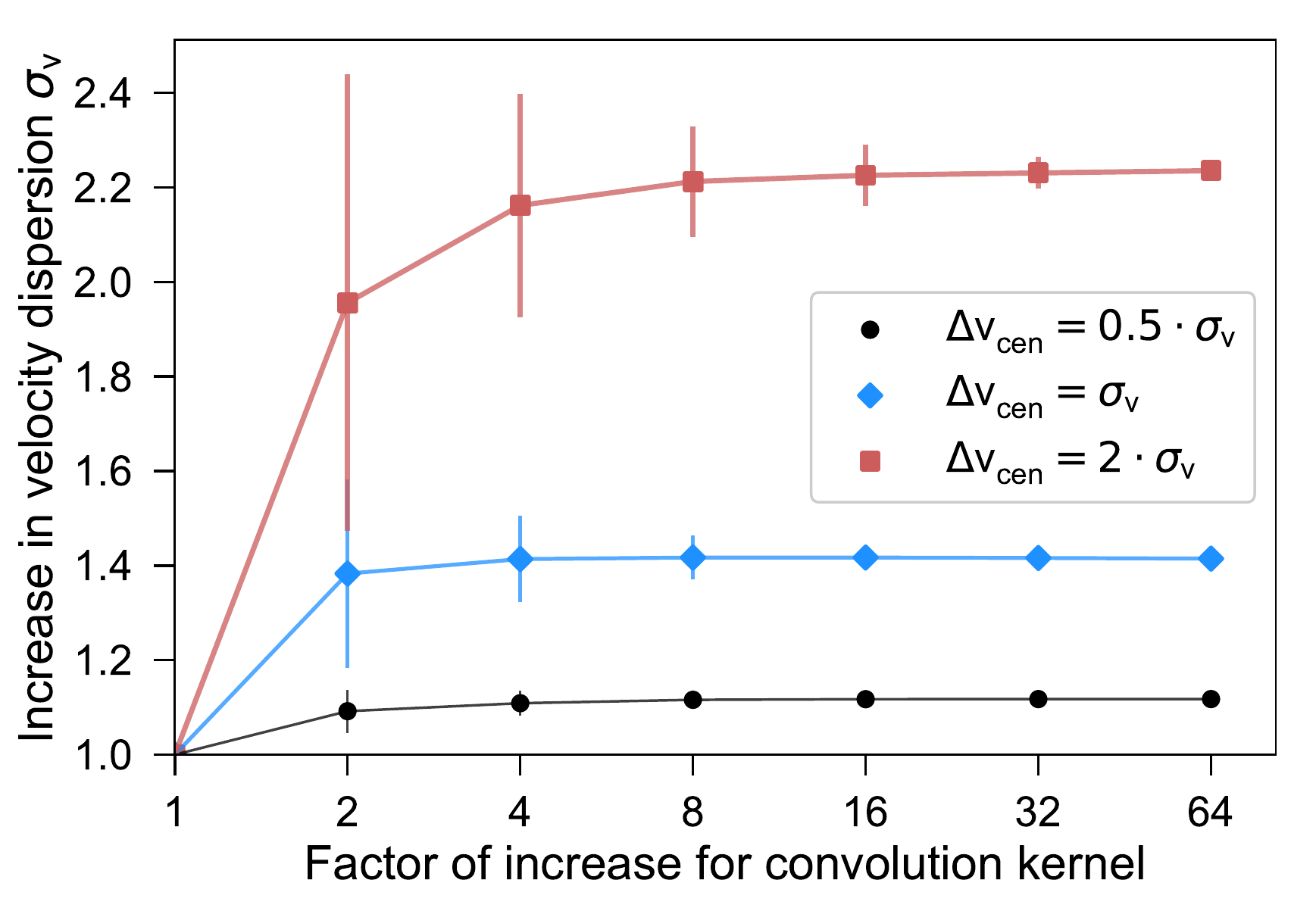}
    \caption[Increase in observed velocity dispersion with decreasing spatial resolution]{Increase in observed velocity dispersion with decreasing spatial resolution.
    The points and errorbars show the results of $100$ different realisations of a mock PPV cube containing emission lines with identical $\veldisp$ values; errorbars indicate $1\sigma$ intervals. 
    For each realisation, the standard deviation for the variation of emission line centroids ($\Delta\text{v}_\text{cen}$) was varied between $0.5, 1,\ \text{or}\ 2$ times $\veldisp$ (indicated in black, blue, and red, respectively).
    See Appendix~\ref{sec:beam_avg_veldisp} for more details.
    }
    \label{fig:beam_averaging_veldisp}
\end{figure}

Figure~\ref{fig:beam_averaging_veldisp} presents the results for $100$ different realisations of the PPV cube and shows that the measured $\veldisp$ in the spatially smoothed cubes increases significantly with increasing $\Delta\text{v}_\text{cen}$.
Variations of $\Delta\text{v}_\text{cen}$ on the order of the velocity dispersion of the emission line in the resolved cube lead to increases in $\veldisp$ in the spatially unresolved cubes by a factor of $\sim 1.4$.

In real observations $\Delta\text{v}_\text{cen}$ will not be distributed randomly.
Rather, the distribution of line centroids is observed to be highly structured \citep[][Henshaw et al., in press]{Riener2020}, with coherent gradients, which will result in similar effects as in our simplified case.
Moreover, variation in the non-thermal contribution to the linewidth can lead to an additional broadening of the lines at coarser spatial resolution.
We thus conclude that due to beam averaging effects it is very unlikely that we observe the same population of linewidths in regions located at far distances as in nearby regions, whose emission lines are spatially better resolved.

\section{Characterisation of the BDC performance}

Here we give some further details about the performance of the BDC. 
We also discuss the effect of the KDA and size-linewidth priors (Sects.~\ref{sec:prior_kda} and \ref{sec:prior_linewidth}) on our final distance results.

\subsection{Effect of \texorpdfstring{$\vlsr$}{VLSR} uncertainties}
\label{sec:effect_evel}

\begin{figure*}[!htbp]
    \centering
    \includegraphics[width=\textwidth]{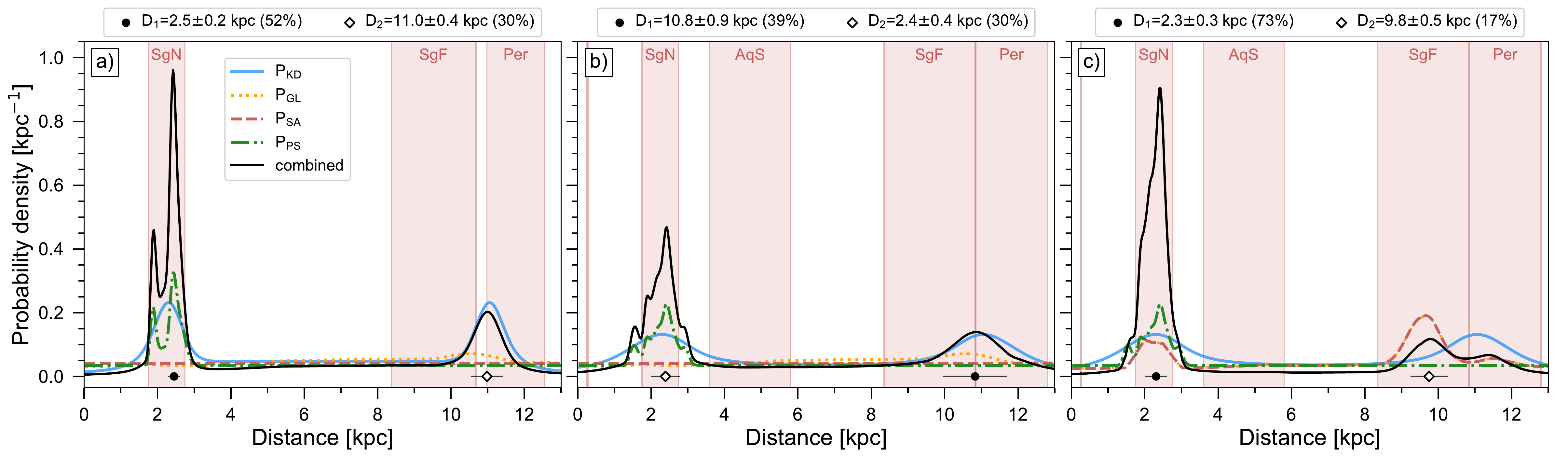}
    \caption[Effect of $\vlsr$ uncertainties on BDC results]{Effect of $\vlsr$ uncertainties on BDC results for a source located at $\ell=35\degr, b=0.1\degr,\ \text{and}\ \vlsr=40$~\kms.
    The BDC settings are the same for \textit{panels~a} and \textit{b}; for \textit{panel~c} we included the prior for spiral arms.
    We increase $\Delta\vlsr$ from $1$~\kms\ (\textit{panel~a}) to $10$~\kms\ (\textit{panels~b} and \textit{c}).
    The meaning of the lines and symbols is the same as in \fig\ref{fig:bdc_examples}.
See App.~\ref{sec:effect_evel} for more details.
    }
    \label{fig:effect_evel}
\end{figure*}

Version $2.4$ of the BDC allows to supply uncertainties for the $\vlsr$ measurement, which can have strong effects on the distance estimation (\fig\ref{fig:effect_evel}).
For each fit component, we chose either its estimated $\vlsr$ uncertainty or its $\veldisp$ value for $\Delta\vlsr$, whichever was the higher value.
The median uncertainty of the $\Delta\vlsr$ values for all fit components was $1.1$~\kms, with an IQR of $0.7$ to $1.9$~\kms.

Figure~\ref{fig:effect_evel} illustrates the effect of $\Delta\vlsr$ on the distance results.
The first two panels (a, b) show distance results obtained with identical BDC settings, where only the supplied uncertainty on the $\vlsr$ coordinate was different.
Increasing the $\vlsr$ uncertainty has multiple effects: the KD peaks get broadened and the association with parallax sources is increased.
In our example this causes a shift of the estimated most likely distance from the near to the far KD solution.
Finally, panel~(c) illustrates the effect of $\Delta\vlsr$ on the association with spiral arms.
Larger $\Delta\vlsr$ values lead to an increase in associations with Galactic features.
In our example it led to the consideration of the Aquila Spur and Aquila Rift as possible candidates for an association; however this had only a very limited effect on the combined distance PDF.

\subsection{Distance results without \texorpdfstring{P$_{\text{far}}$}{P\_far} priors}
\label{sec:runs_comp}

\begin{figure}
    \centering
    \includegraphics[width=\columnwidth]{grs_fov_bdc+kda+veldisp_inttot.pdf}%
    \hspace{-\columnwidth}%
    \begin{ocg}{fig:orig_off}{fig:orig_off}{0}%
    \end{ocg}%
    \begin{ocg}{fig:orig_on}{fig:orig_on}{1}%
    \includegraphics[width=\columnwidth]{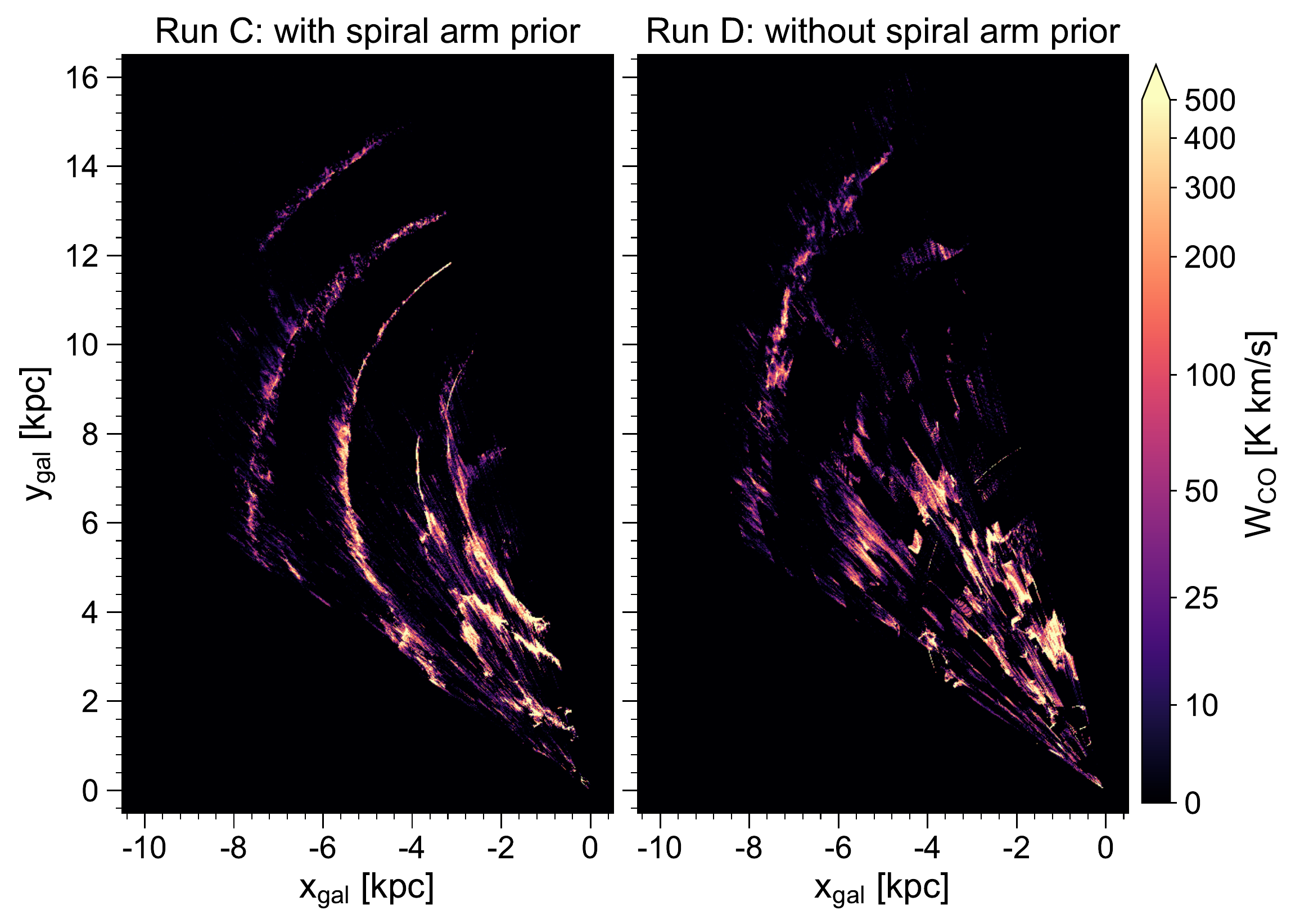}%
    \end{ocg}%
    \hspace{-\columnwidth}%
    \begin{ocg}{fig:grid_off_10}{fig:grid_off_10}{0}%
    \end{ocg}%
    \begin{ocg}{fig:grid_on_10}{fig:grid_on_10}{1}%
    \includegraphics[width=\columnwidth]{grs_fov_layer_grid.pdf}%
    \end{ocg}%
    \hspace{-\columnwidth}%
    \begin{ocg}{fig:arms_off_10}{fig:arms_off_10}{0}%
    \end{ocg}%
    \begin{ocg}{fig:arms_on_10}{fig:arms_on_10}{1}%
    \includegraphics[width=\columnwidth]{grs_fov_layer_spiral_arms_white.pdf}%
    \end{ocg}%
    \hspace{-\columnwidth}%
    \includegraphics[width=\columnwidth]{grs_fov_layer_sun+gc_white.pdf}%
    \caption[Face-on view of integrated emission obtained with BDC default settings]{Same as \fig\ref{fig:fov_inttot} but for the distance results obtained with the default settings of the BDC.
     When displayed in Adobe Acrobat, it is possible to switch to \ToggleLayer{fig:orig_off,fig:orig_on}{\protect\cdbox{the original map}} (\fig\ref{fig:fov_inttot}), hide the \ToggleLayer{fig:arms_on_10,fig:arms_off_10}{\protect\cdbox{spiral arm positions}} and the \ToggleLayer{fig:grid_on_10,fig:grid_off_10}{\protect\cdbox{grid}}.
    }
    \label{fig:bdc_default_inttot}
\end{figure}

Here we discuss the impact of the KDA and size-linewidth priors (see \sect\ref{sec:prior_kda} and \ref{sec:prior_linewidth}) on our distance results.
For this purpose we created four more distance runs with the BDC:
\begin{itemize}
	\item Run~C: Uses the default settings of the BDC, that means $\prob{SA}{}=0.85$, but does not include the KDA and size-linewidth priors.
	\item Run~D: Same as Run~C, but with the SA prior switched off ($\prob{SA}{}=0$), which also sets $\prob{GL}{}=0$ as these two priors are combined in the default version of the BDC.
	\item Run~E: Same as Run~A, but without the size-linewidth prior.
	\item Run~F: Same as Run~B, but without the size-linewidth prior.
\end{itemize}
Figure~\ref{fig:bdc_default_inttot} shows the map of $\wco$ values for Run~C and D.
A comparison with \fig\ref{fig:fov_inttot} reveals substantial differences to Run~A and B, respectively. 
The default BDC settings in Run~C lead to a much stronger association with the SA model and the results contain much less emission at close distances (d $< 2$~ kpc).
However, compared to Run~A, Run~C does not put emission in between the Perseus and Outer arm.
Run~D shows that without the KDA and GL priors the distance results contain a higher fraction of emission-free areas, which is especially notable around the far portion of the Sagittarius arm.
Without the size-linewidth prior, Run~D also puts significantly more emission from close distances towards the Perseus and Outer arm regions.
In terms of association with Galactic features: for Run~C $13.8\%$ of the emission (and $18.6\%$ of the fit components) were associated with interarm regions, which increased to $25.6\%$ ($34.6\%$ of $\Ncomp$) for Run~D. 

\begin{figure}
    \centering
    \includegraphics[width=\columnwidth]{grs_fov_bdc+kda+veldisp_veldisp.pdf}%
    \hspace{-\columnwidth}%
    \begin{ocg}{fig:orig_off_11}{fig:orig_off_11}{0}%
    \end{ocg}%
    \begin{ocg}{fig:orig_on_11}{fig:orig_on_11}{1}%
    \includegraphics[width=\columnwidth]{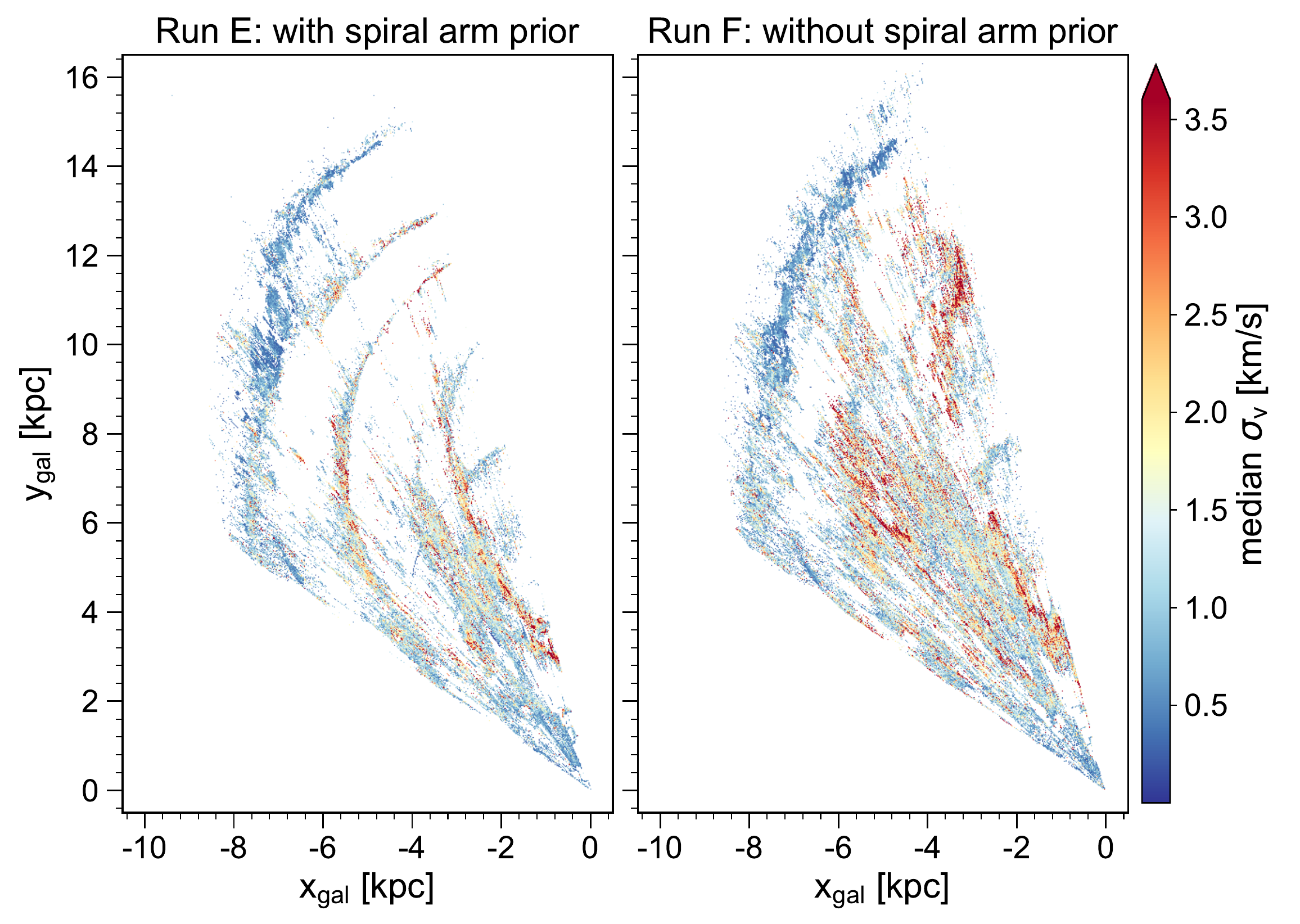}%
    \end{ocg}%
    \hspace{-\columnwidth}%
    \begin{ocg}{fig:grid_off_11}{fig:grid_off_11}{0}%
    \end{ocg}%
    \begin{ocg}{fig:grid_on_11}{fig:grid_on_11}{1}%
    \includegraphics[width=\columnwidth]{grs_fov_layer_grid.pdf}%
    \end{ocg}%
    \begin{ocg}{fig:arms_off_11}{fig:arms_off_11}{1}%
    \end{ocg}%
    \begin{ocg}{fig:arms_on_11}{fig:arms_on_11}{0}%
    \hspace{-\columnwidth}%
    \includegraphics[width=\columnwidth]{grs_fov_layer_spiral_arms_black.pdf}%
    \end{ocg}%
    \hspace{-\columnwidth}%
    \includegraphics[width=\columnwidth]{grs_fov_layer_sun+gc_black.pdf}%
    \caption[Face-on view of the median velocity dispersion values without the size-linewidth prior]{Face-on view of the median velocity dispersion values of Gaussian fit components for the BDC results obtained with (\textit{left}) and without (\textit{right}) the spiral arm prior.
    The values are binned in $10\times10$~pc cells and the median was calculated along the $\zgal$ axis.
    The position of the Sun and Galactic centre are indicated by the Sun symbol and black dot, respectively.
    When displayed in Adobe Acrobat, it is possible to switch to \ToggleLayer{fig:orig_off_11,fig:orig_on_11}{\protect\cdbox{the original map}} (\fig\ref{fig:fov_veldisp}), show the \ToggleLayer{fig:arms_off_11,fig:arms_on_11}{\protect\cdbox{spiral arm positions}} and hide the \ToggleLayer{fig:grid_on_11,fig:grid_off_11}{\protect\cdbox{grid}}.
    }
    \label{fig:fov_veldisp_comp}
\end{figure}

Figure~\ref{fig:fov_veldisp_comp} shows the resulting $\veldisp$ map if we had not used the size-linewidth prior.
Since we use this prior only for emission with $\vlsr < 20$~\kms, only positions at the largest $\rgal$ values are affected by it.
A comparison with Run~A and B shows that the size-linewidth prior helps in decreasing the confusion between near and far emission that causes the large fraction of emission lines with narrow linewidths at large distances in Run~E and F. 

\subsection{Estimated probabilities, \texorpdfstring{P$_{\text{far}}$}{P\_far} weights, and distance choices}
\label{app:bdc_prob_pfar_choice}

\begin{figure}
    \centering
    \includegraphics[width=\columnwidth]{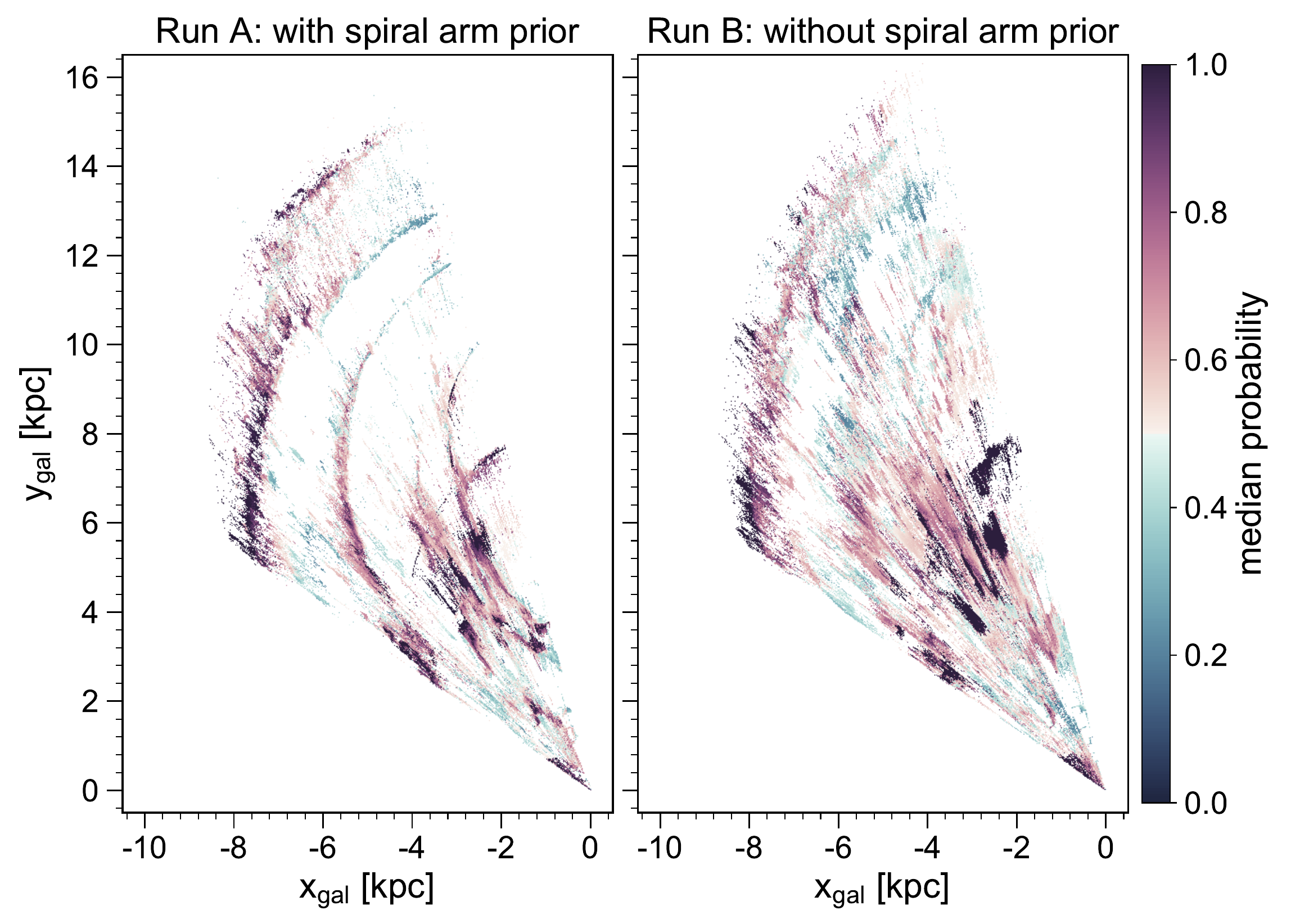}%
    \hspace{-\columnwidth}%
    \begin{ocg}{fig:grid_off_6}{fig:grid_off_6}{0}%
    \end{ocg}%
    \begin{ocg}{fig:grid_on_6}{fig:grid_on_6}{1}%
    \includegraphics[width=\columnwidth]{grs_fov_layer_grid.pdf}%
    \end{ocg}%
    \begin{ocg}{fig:arms_off_6}{fig:arms_off_6}{1}%
    \end{ocg}%
    \begin{ocg}{fig:arms_on_6}{fig:arms_on_6}{0}%
    \hspace{-\columnwidth}%
    \includegraphics[width=\columnwidth]{grs_fov_layer_spiral_arms_black.pdf}%
    \end{ocg}%
    \hspace{-\columnwidth}%
    \includegraphics[width=\columnwidth]{grs_fov_layer_sun+gc_black.pdf}%
    \caption[Face-on view of the median probability values]{Face-on view of the median probability values from the BDC results obtained with (\textit{left}) and without (\textit{right}) the spiral arm prior.
    The values are binned in $10\times10$~pc cells and the median was calculated along the $\zgal$ axis.
    The position of the Sun and Galactic centre are indicated by the Sun symbol and black dot, respectively.
     When displayed in Adobe Acrobat, it is possible to show the \ToggleLayer{fig:arms_off_6,fig:arms_on_6}{\protect\cdbox{spiral arm positions}} and hide the \ToggleLayer{fig:grid_on_6,fig:grid_off_6}{\protect\cdbox{grid}}.
    }
    \label{fig:fov_prob}
\end{figure}

\begin{figure}
    \centering
    \includegraphics[width=\columnwidth]{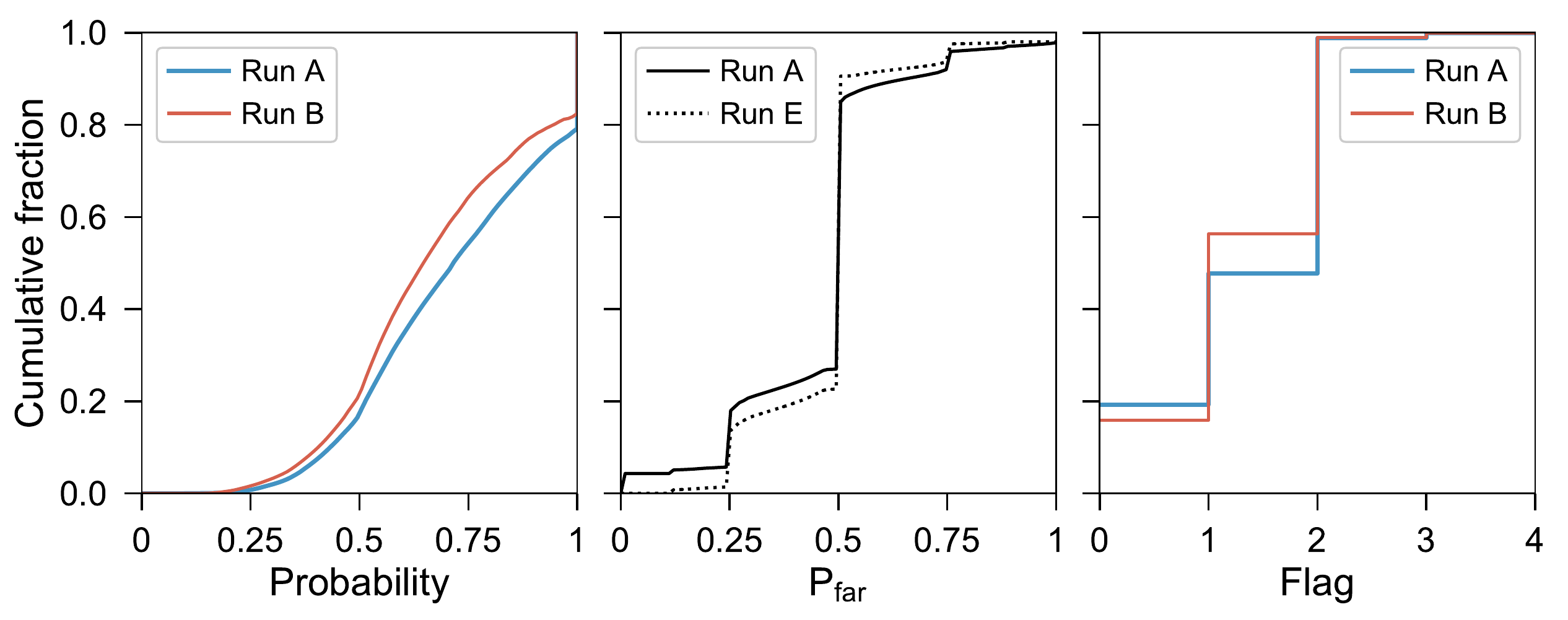}
    \caption[Cumulative distributions for estimated probabilities, $\prob{far}{}$ prior values, and distance choices]{Cumulative distributions for the estimated distance probabilities (\textit{left}), values for the $\prob{far}{}$ prior (\textit{middle}), and choice of the distance values (\textit{right}).
    See \sect\ref{sec:distresults} for more details.
    }
    \label{fig:cf_prob_pfar_flag}
\end{figure}

Figure~\ref{fig:fov_prob} shows a face-on view of the median estimated probability values and the left panel in \fig\ref{fig:cf_prob_pfar_flag} gives the cumulative distribution of all estimated probabilities.
These probabilities were estimated from the integrated areas of Gaussian fit components to the combined distance PDF (\sect\ref{sec:dist_choice}).
We thus get higher probabilities for regions where the combined distance PDF produced a dominant peak, which however could be caused by negative effects.
For example, the near distance solution of the KD prior is cut off for low $\vlsr$ values (cf. \fig\ref{fig:effect_lowvlsr}), thus yielding far distance solutions with high estimated probabilities. 
The KD prior is also down-weighted for lower $\rgal$ values, which could lead to strongly blended KD peaks.
This could result in broad Gaussian fits over both of these peaks, with high distance uncertainties as well as high estimated probability, which seems to occur at the lowest $\rgal$ values (cf. \fig\ref{fig:fov_edist}).

The cumulative distribution of the assigned probabilities for the distance values (left panel of \fig\ref{fig:cf_prob_pfar_flag}) shows that the chosen distance values from the BDC run with the SA prior have higher associated probabilities.
For Run~A, $43.8\%$ of the chosen distance values had probabilities $> 0.75$ and $16.7\%$ of the distance values had probabilities $< 0.5$; for Run~B, these fractions change to $33.6\%$ and $21.0\%$, respectively.  
Thus for Run~A the Gaussian fits to the combined distance PDF had higher integrated areas, confirming that the addition of the SA prior leads to more well-defined peaks.

The middle panel of \fig\ref{fig:cf_prob_pfar_flag} shows the cumulative distribution of chosen weights for the $\prob{far}{}$ prior for the case where the KDA prior was used (dotted line) and the case where the size-linewidth prior (\sect\ref{sec:prior_linewidth}) was used in addition to the KDA (solid line).
For the case in which $\prob{far}{}$ was only informed by the KDA prior, $23.1\%$ and $7.6\%$ of the fit components received a preference for the near and far KD solution, respectively. 
If the size-linewidth prior is used in addition to the KDA prior, these percentages increase to $27.5\%$ and $13.3\%$, respectively.

Finally, the right panel of \fig\ref{fig:cf_prob_pfar_flag} shows the cumulative distribution for the choice of distance values (\sect\ref{sec:dist_choice}).
The numbers indicate the following conditions: the distance assignment yielded only one distance solution ($0$); the associated Gaussian fit of one distance solution did not satisfy the criterion for the amplitude threshold ($1$); the distance solution with the highest probability (i.e. the highest integrated intensity of the associated Gaussian fit) was chosen ($2$); the distance solution with the lowest absolute distance error was chosen ($3$); and the near KD solution was picked randomly ($4$).
The cumulative distribution shows that conditions ($0$), ($1$), and ($2$) were responsible for the vast majority of final distance choices (contributing $19.2\%$, $28.5\%$,  and $51.2\%$ for Run~A and $15.9\%$, $40.4\%$, and $42.6\%$ for Run~B), whereas conditions ($3$) and ($4$) only contributed minimally.

\subsection{Deviation from Galactic rotation curve}
\label{sec:vlsr_deviation}

\begin{figure}
    \centering
    \includegraphics[width=\columnwidth]{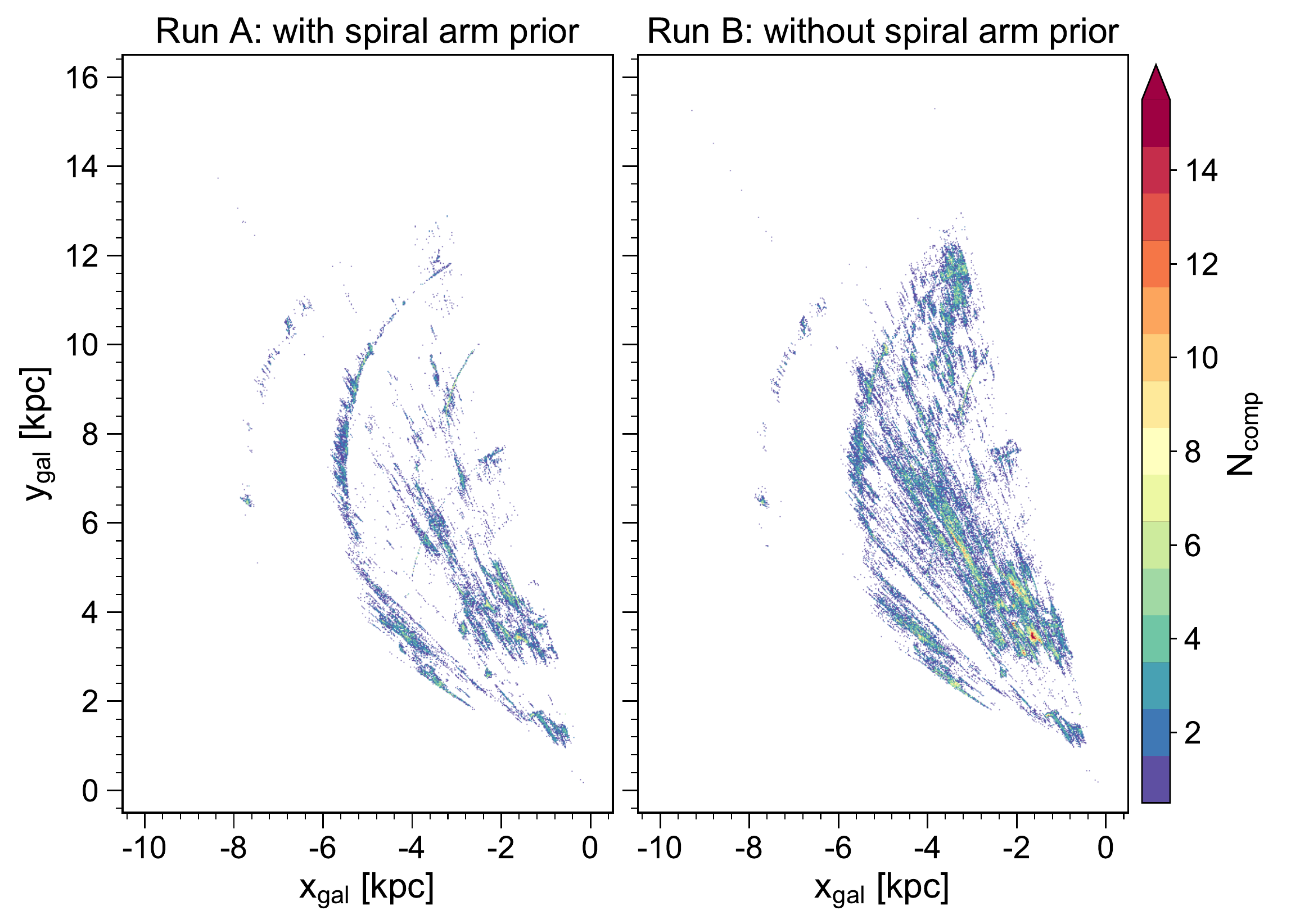}%
    \hspace{-\columnwidth}%
    \begin{ocg}{fig:grid_off_12}{fig:grid_off_12}{0}%
    \end{ocg}%
    \begin{ocg}{fig:grid_on_12}{fig:grid_on_12}{1}%
    \includegraphics[width=\columnwidth]{grs_fov_layer_grid.pdf}%
    \end{ocg}%
    \begin{ocg}{fig:arms_off_12}{fig:arms_off_12}{1}%
    \end{ocg}%
    \begin{ocg}{fig:arms_on_12}{fig:arms_on_12}{0}%
    \hspace{-\columnwidth}%
    \includegraphics[width=\columnwidth]{grs_fov_layer_spiral_arms_black.pdf}%
    \end{ocg}%
    \hspace{-\columnwidth}%
    \begin{ocg}{fig:vlsr_off_12}{fig:vlsr_off_12}{0}%
    \end{ocg}%
    \begin{ocg}{fig:vlsr_on_12}{fig:vlsr_on_12}{1}%
    \includegraphics[width=\columnwidth]{grs_fov_layer_vlsr_curves.pdf}%
    \end{ocg}%
    \hspace{-\columnwidth}%
    \includegraphics[width=\columnwidth]{grs_fov_layer_sun+gc_black.pdf}%
    \caption[Face-on view of components associated with $\vlsr$ deviations]{Face-on view of the number of components with distance results that cause an absolute $\vlsr$ deviation of more than $10$~\kms\ compared to the Galactic rotation curve model for the BDC results obtained with (\textit{left}) and without (\textit{right}) the spiral arm prior.
    The values are binned in $10\times10$~pc cells and the values were summed up along the $\zgal$ axis.
    The position of the Sun and Galactic centre are indicated by the Sun symbol and black dot, respectively.
     When displayed in Adobe Acrobat, it is possible to show the \ToggleLayer{fig:arms_off_12,fig:arms_on_12}{\protect\cdbox{spiral arm positions}}, hide the \ToggleLayer{fig:vlsr_on_12,fig:vlsr_off_12}{\protect\cdbox{curves of constant projected $\vlsr$}}, and hide the \ToggleLayer{fig:grid_on_12,fig:grid_off_12}{\protect\cdbox{grid}}.
    }
    \label{fig:fov_vlsr_deviation}
\end{figure}

Here we quantify the deviation of our distance results from the expected values from the Galactic rotation curve model.
These deviations are interesting as they can identify regions with high peculiar velocities.
Figure~\ref{fig:fov_vlsr_deviation} shows the number of fit components whose estimated distance values caused a deviation of more than $10$~\kms\ from the expected $\vlsr$ value based on the Galactic rotation curve.
In Run~A and B, $2.2\%$ and $6.2\%$ of the components showed such a large $\vlsr$ deviation. 
Since both panels show similar deviations occurring around the positions of the Sagittarius, Scutum, and Norma arms, it is likely that these differences from pure rotation curve velocities are to a large part due to the effect of the maser parallax sources (cf. \fig\ref{fig:schematic_spiral_arms+masers}).
For Run~B we can also identify an increase in the number of components with deviating $\vlsr$ values inside $\rgal \lesssim 5$~kpc.
For this region the BDC downweighted the KD prior, so these deviating components might to a large part simply be due to large associated distance uncertainties. 
A comparison with \fig\ref{fig:fov_edist} shows that these regions are indeed associated with increased uncertainty values.

\end{appendix} 

\end{document}